\documentclass{lmcs}
\pdfoutput=1

\usepackage{lastpage}
\lmcsdoi{21}{2}{11}
\lmcsheading{}{\pageref{LastPage}}{}{}%
{May~29,~2024}{Apr.~30,~2025}{}

\newif\ifignore 
\ignorefalse

\usepackage[utf8]{inputenc}
\usepackage{hyperref}
\usepackage{amssymb}
\usepackage{mathtools}
\usepackage{shadow}
\usepackage{pdfrender}
\usepackage{nicefrac}
\usepackage{xspace}

\usepackage{xypic}
\usepackage[all]{xy}
\xyoption{tips}
\SelectTips{cm}{}  

\newcommand*{\fatten}[1][.4pt]{%
  \textpdfrender{
    TextRenderingMode=FillStroke,
    LineWidth={\dimexpr(#1)\relax},
  }%
}

\ifdefined\mathsl\else
  \DeclareMathAlphabet{\mathsl}{\encodingdefault}{\rmdefault}{\mddefault}{\sldefault}
  \SetMathAlphabet{\mathsl}{bold}{\encodingdefault}{\rmdefault}{\bfdefault}{\sldefault}
\fi

\makeatletter
\newcommand{\mathoverlap}[2]{\mathpalette\mathoverlap@{{#1}{#2}}}
\newcommand{\mathoverlap@}[2]{\mathoverlap@@{#1}#2}
\newcommand{\mathoverlap@@}[3]{\ooalign{$\m@th#1#2$\crcr\hidewidth$\m@th#1#3$\hidewidth}}
\makeatother

\def\QEDbox{\qedsymbol} 
\def\QED{\hfill\QEDbox}

\newenvironment{myproof}[1][Proof]%
   { \proof
   }
   {
   }

\newcommand{\conglongrightarrow}{\mathrel{\smash{\stackrel{
           \raisebox{.5ex}{$\scriptstyle\cong$}}{
           \raisebox{0ex}[0ex][0ex]{$\longrightarrow$}}}}}

\newcommand{\after}{\mathrel{\circ}}
\newcommand{\klafter}{\mathbin{\mathoverlap{\circ}{\cdot}}}
\newcommand{\NNO}{\mathbb{N}}
\newcommand{\pNNO}{\mathbb{N}_{>0}}
\newcommand{\R}{\mathbb{R}}
\newcommand{\pR}{\mathbb{R}_{>0}}
\newcommand{\nnR}{\mathbb{R}_{\geq 0}}
\newcommand{\intd}{{\kern.2em}\mathrm{d}{\kern.03em}}

\newcommand{\Mlt}{\ensuremath{\mathcal{M}}}
\newcommand{\Dst}{\ensuremath{\mathcal{D}}}

\newcommand{\supp}{\ensuremath{\mathsl{supp}}}
\newcommand{\unit}{\ensuremath{\mathsl{unit}}}
\newcommand{\flatten}{\ensuremath{\mathsl{flat}}}
\newcommand{\iid}{\ensuremath{\mathsl{iid}}}
\newcommand{\Var}{\ensuremath{\mathsl{Var}}}
\newcommand{\expec}{\mathop{\mathbb{E}}}
\newcommand{\decouple}{\ensuremath{\mathsl{dcpl}}}
\newcommand{\projdelete}{\ensuremath{\mathsl{PD}}}
\newcommand{\drawdelete}{\ensuremath{\mathsl{DD}}}

\newcommand{\tvd}{\ensuremath{\mathsl{tvd}}}
\newcommand{\DKL}{\ensuremath{\mathsl{D}_{\mathsl{KL}}}}
\newcommand{\diameter}{\ensuremath{\mathsl{diam}}}
\newcommand{\Fact}{\ensuremath{\mathsl{Fact}}}
\newcommand{\FactSh}{\ensuremath{\mathsl{Fact}_{\mathsl{S}}}}

\newcommand{\Obs}{\ensuremath{\mathsl{Obs}}}

\newcommand{\concat}{\ensuremath{\mathbin{+{\kern-.5ex}+}}}
\newcommand{\one}{\ensuremath{\mathbf{1}}}
\newcommand{\zero}{\ensuremath{\mathbf{0}}}

\newcommand{\facto}[1]{\ensuremath{#1{\kern-2.5pt}\raisebox{-2.5pt}{\includegraphics[width=0.9em]{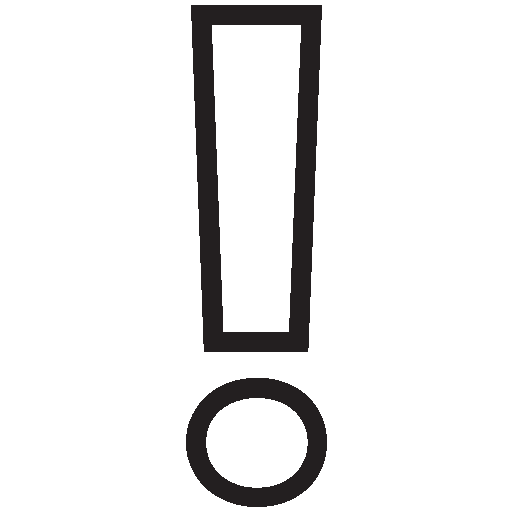}}{\kern-.2em}}}
\newcommand{\flrn}{\ensuremath{\mathsl{Flrn}}}
\newcommand{\indic}[1]{\mathbf{1}_{#1}}

\newcommand{\acc}{\ensuremath{\mathsl{acc}}}
\newcommand{\arr}{\ensuremath{\mathsl{arr}}}
\newcommand{\zip}{\ensuremath{\mathsl{zip}}}
\newcommand{\mzip}{\ensuremath{\mathsl{mzip}}}
\newcommand{\pml}{\ensuremath{\mathsl{pml}}}
\newcommand{\perm}{\ensuremath{\mathsl{prm}}}
\newcommand{\bibinom}[2]{\left({\kern-.5ex}\binom{#1}{#2}{\kern-.5ex}\right)}
\newcommand{\mulnom}{\ensuremath{\mathsl{mn}}}
\newcommand{\multinomial}[1][]{\ensuremath{\mulnom[#1]}}
\newcommand{\hypgeom}{\ensuremath{\mathsl{hg}}}
\newcommand{\hypergeometric}[1][]{\ensuremath{\hypgeom[#1]}}
\newcommand{\pol}{\ensuremath{\mathsl{pl}}}
\newcommand{\polya}[1][]{\ensuremath{\pol[#1]}}
\newcommand{\seqpolya}[1][]{\ensuremath{\mathsl{spol}[#1]}}

\newcommand{\projstoreadd}{\ensuremath{\mathsl{PSA}}}
\newcommand{\drawstoreadd}{\ensuremath{\mathsl{DSA}}}

\newcommand{\Dirichlet}{\ensuremath{\mathsl{Dir}}}
\newcommand{\poissonname}{\mathsl{Pois}}
\newcommand{\poisson}[1][]{\ensuremath{\poissonname[#1]}}
\newcommand{\Exp}{\ensuremath{\mathsl{Exp}}}
\newcommand{\tuple}[1]{\langle#1\rangle}
\newcommand{\set}[2]{\{#1\;|\;#2\}}
\newcommand{\setin}[3]{\{#1\in#2\;|\;#3\}}

\newcommand{\exin}[3]{\exists#1\in#2.\,#3}
\newcommand{\ketstrut}{\vrule height 10.5pt depth 4.5pt width 0pt}
\newcommand{\ket}[1]{\ensuremath{|{\kern.1em}#1{\kern.1em}\rangle}}
\newcommand{\bigket}[1]{\ensuremath{\big|{\kern.1em}#1{\kern.1em}\big\rangle}}
\newcommand{\Bigket}[1]{\ensuremath{\left|\ketstrut{\kern.1em}\right.{\kern-.2em}#1{\kern-.2em}\left.\ketstrut{\kern0em}\right>}}
\newcommand{\coefm}[1]{\ensuremath{\fatten[0.6pt]{(}{\kern1pt}#1{\kern1pt}\fatten[0.6pt]{)}}}

\newcommand{\pull}{\mathrel{\mathchoice%
   {\scalebox{-0.5}[1]{$\gg=$}}
   {\scalebox{-0.5}[1]{$\gg{\kern-1.5ex}=$}}
   {\scalebox{-0.5}[1]{${\kern.5ex}\scriptstyle\gg{\kern-0.2ex}={\kern.5ex}$}}
   {\scalebox{-0.5}[1]{$\scriptscriptstyle\gg=$}}}}
\newcommand{\push}{\mathrel{\mathchoice%
   {\scalebox{-0.5}[1]{$=\ll$}}
   {\scalebox{-0.5}[1]{$={\kern-1.5ex}\ll$}}
   {\scalebox{-0.5}[1]{${\kern.5ex}\scriptstyle={\kern-0.2ex}\ll{\kern.5ex}$}}
   {\scalebox{-0.5}[1]{$\scriptscriptstyle=\ll$}}}}
\newcommand{\pushing}[2]{{#1}_{*}(#2)}
\newcommand{\pushop}[1]{{#1}_{*}}

\newcommand{\upsum}[1]{\ensuremath{#1{\kern-.6ex}\uparrow}\xspace}
\newcommand{\downsum}[1]{\ensuremath{#1{\kern-.6ex}\downarrow}\xspace}

\newcommand{\Category}[1]{\ensuremath{\mathbf{#1}}\xspace}
\newcommand{\Sets}{\Category{Sets}}
\newcommand{\MetSh}{\Category{Met}_{\mathsl{S}}}
\newcommand{\MetLip}{\Category{Met}_{\mathsl{L}}}

\DeclareSymbolFont{T1op}{T1}{cmr}{m}{n}
\SetSymbolFont{T1op}{bold}{T1}{cmr}{bx}{n}
\DeclareMathSymbol{\mathguilsinglleft}{\mathopen}{T1op}{'016}
\DeclareMathSymbol{\mathguilsinglright}{\mathclose}{T1op}{'017}
\newcommand{\klin}[1]{\mathguilsinglleft#1\mathguilsinglright}

\newcommand{\msa}[1]{\!\!\;#1\;\!\!} 

\usepackage{tikzit}

\makeatletter
\newcommand{\monus}{\mathbin{\text{\@monus}}}
\newcommand{\@monus}{%
  \ooalign{\hidewidth\raise1ex\hbox{.}\hidewidth\cr$\m@th-$\cr}%
}
\makeatother

\newsavebox\sbpto
\savebox\sbpto{\begin{tikzpicture}[baseline=-2.5pt]
            \filldraw[fill=white,draw=white] circle (1.4pt);
            \filldraw[fill=white,draw=black,line width=0.2pt]circle(2pt);
                \end{tikzpicture}}
\newcommand\chanto{\mathrel{\ooalign{$\to$\cr
      \hfil\!$\usebox\sbpto$\hfil\cr}}}

\newcommand{\auxproof}[1]{
\ifignore\mbox{}\newline
\textbf{PROOF:} \dotfill\newline
{\it #1}\mbox{}\newline
\textbf{ENDPROOF}\dotfill
\fi}

\newcommand{\ignore}[1]{}

\title{Drawing with Distance}

\author[Bart Jacobs]{Bart Jacobs\lmcsorcid{0000-0002-0740-0336}}

\address{
iHub, Radboud University Nijmegen, The Netherlands. 
}
\email{bart@cs.ru.nl}



\begin{document}

\begin{abstract}
Drawing (a multiset of) coloured balls from an urn is one of the most
basic models in discrete probability theory. Three modes of drawing
are commonly distinguished: multinomial (draw-replace), hypergeometric
(draw-delete), and P\'olya (draw-add). These drawing operations are
represented as maps from urns to distributions over multisets of
draws.  The set of urns is a metric space via the Kantorovich
distance. The set of distributions over draws is also a metric space,
using Kantorovich-over-Kantorovich. It is shown that these three draw
operations are all isometries, that is, they exactly preserve the
Kantorovich distances. Further, drawing is studied in the limit, both
for large urns and for large draws. First it is shown that, as the urn
size increases, the Kantorovich distances go to zero between
hypergeometric and multinomial draws, and also between P\'olya and
multinomial draws.  Second, it is shown that, as the drawsize
increases, the Kantorovich distance goes to zero (in probability)
between an urn and (normalised) multinomial draws from the urn. These
results are known, but here, they are formulated in a novel metric
manner as limits of Kantorovich distances.  We call these two limit
results the law of large urns and the law of large draws.
\end{abstract}
\maketitle

\section{Introduction}\label{IntroSec}

Basic physical models in probability theory are flipping a coin,
rolling a dice, or drawing coloured balls from an
urn~\cite{JohnsonK77}. We start with an illustration of these urn
models. We consider a situation with a set $C = \{R,G,B\}$ of three
colours: red, green, blue. Assume that we have two urns $\upsilon_{1},
\upsilon_{2}$ with 10 coloured balls each. We describe these urns as
multisets of the form:
\[ \begin{array}{rclcrcl}
\upsilon_{1}
& = &
8\ket{G} + 2\ket{B}
& \qquad\mbox{and}\qquad &
\upsilon_{2}
& = &
5\ket{R} + 4\ket{G} + 1\ket{B}.
\end{array} \]

\noindent Recall that a multiset is like a set, except that elements
may occur multiple times. Here we describe urns as multisets using
`ket' notation $\ket{-}$. It separates multiplicities of elements
(before the ket) from the elements in the multiset (inside the
ket). Thus, urn $\upsilon_{1}$ contains $8$ green balls and $2$ blue
balls (and no red ones).  Similarly, urn $\upsilon_2$ contains $5$
red, $4$ green, and $1$ blue ball(s).

Below, we shall describe the Kantorovich distance between multisets
(of the same size). How this works does not matter for now; we simply
posit that the Kantorovich distance $d(\upsilon_{1}, \upsilon_{2})$
between these two urns is $\frac{1}{2}$ --- where we assume the discrete
distance on the set $C$ of colours.

We turn to draws from these two urns, in this introductory example of
size two. These draws are also described as multisets, with elements
from the set $C = \{R,G,B\}$ of colours. There are six multisets
(draws) of size $2$, namely:
\begin{equation}
\label{DrawsIntroEx}
2\ket{R} \qquad 
   1\ket{R} + 1\ket{G} \qquad
    2\ket{G} \qquad
    1\ket{R} + 1\ket{B} \qquad 
    2\ket{B} \qquad
    1\ket{G} + 1\ket{B}.
\end{equation}

\noindent As we see, there are three draws with 2 balls of the same
colour, and three draws with balls of different colours.

We consider the hypergeometric probabilities associated with these
draws, from the two urns. Let's illustrate this for the draw $1\ket{G}
+ 1\ket{B}$ of one green ball and one blue ball from the urn
$\upsilon_{1}$. The probability of drawing $1\ket{G} + 1\ket{B}$ is
$\frac{16}{45}$; it is obtained as sum of:
\begin{itemize}
\item first drawing-and-deleting a green ball from $\upsilon_{1} =
  8\ket{G} + 2\ket{B}$, with probability $\frac{8}{10}$.  It leaves an
  urn $7\ket{G} + 2\ket{B}$, from which we can draw a blue ball with
  probability $\frac{2}{9}$. Thus drawing ``first green then blue''
  happens with probability $\frac{8}{10} \cdot \frac{2}{9} =
  \frac{8}{45}$.

\item Similarly, the probability of drawing ``first blue then green''
  is $\frac{2}{10} \cdot \frac{8}{9} = \frac{8}{45}$.
\end{itemize}

\noindent We can similarly compute the probabilities for each of the
above six draws~\eqref{DrawsIntroEx} from urn $\upsilon_1$. This gives
the hypergeometric distribution, which we write using kets-over-kets
as:
\[ \begin{array}{rcl}
\hypergeometric[2](\upsilon_{1})
& = &
\frac{28}{45}\Bigket{2\ket{G}} + 
   \frac{16}{45}\Bigket{1\ket{G} + 1\ket{B}} + 
   \frac{1}{45}\Bigket{2\ket{B}}.
\end{array} \]

\noindent The fraction written before a big ket is the probability of
drawing the multiset (of size 2), written inside that big ket, from
the urn $\upsilon_1$.

Drawing from the second urn $\upsilon_2$ gives a different
distribution over these multisets~\eqref{DrawsIntroEx}. Since urn
$\upsilon_2$ contains red balls, they additionally appear in the
draws.
\[ \begin{array}{rcl}
\hypergeometric[2](\upsilon_{2})
& = &
\frac{2}{9}\Bigket{2\ket{R}} + 
   \frac{4}{9}\Bigket{1\ket{R} + 1\ket{G}} + 
   \frac{2}{15}\Bigket{2\ket{G}}
\\[+0.4em]
& & \qquad +\;
   \frac{1}{9}\Bigket{1\ket{R} + 1\ket{B}} + 
   \frac{4}{45}\Bigket{1\ket{G} + 1\ket{B}}.
\end{array} \]

\noindent We can also compute the distance between these two
hypergeometric distributions over multisets. It involves a Kantorovich
distance over the space of multisets (of size~2) with their own
Kantorovich distance. Again, details of the calculation are skipped at
this stage. The distance between the above two hypergeometric
draw-distributions is:
\[ \begin{array}{rcccl}
d\Big(\hypergeometric[2](\upsilon_{1}), \, \hypergeometric[2](\upsilon_{2})\Big)
& = &
\frac{1}{2}
& = &
d\big(\upsilon_{1}, \, \upsilon_{2}\big).
\end{array} \]

\noindent This coincidence of distances is non-trivial. It holds, in
general, for arbitrary urns (of the same size) over arbitrary metric
spaces of colours, for draws of arbitrary sizes. Moreover, the same
coincidence of distances holds for the multinomial and P\'olya modes
of drawing. These coincidences form key results of this paper, see
Theorems~\ref{MulnomIsomThm}, \ref{HypgeomIsomThm},
and~\ref{PolyaIsomThm} below.

In order to formulate and obtain these results, we describe
multinomial, hypergeometric and P\'olya distributions in the form of
(Kleisli) maps:
\begin{equation}
\label{DrawDiag}
\vcenter{\xymatrix@C+1pc{ \Dst(X)\ar[r]^-{\multinomial[K]} &
    \Dst\big(\Mlt[K](X)\big) &
    \Mlt[L](X)\ar@<-0.3pc>[l]_-{\hypergeometric[K]}
    \ar@<+0.3pc>[l]^-{\polya[K]} }}
\end{equation}

\noindent They all produce distributions (indicated by $\Dst$), in the
middle of this diagram, on multisets (draws) of size $K$, indicated by
$\Mlt[K]$, over a set $X$ of colours. Details will be provided
below. Using the maps in~\eqref{DrawDiag}, the coincidence of
distances that we saw above can be described as a preservation
property, in terms of distance preserving maps --- called
isometries. At this stage we wish to emphasise that the representation
of these different drawing operations as maps in~\eqref{DrawDiag} has
a categorical background. It makes it possible to formulate and prove
basic properties of drawing from an urn, such as naturality in the set
$X$ of colours. Also, as shown in~\cite{Jacobs21b} for the multinomial
and hypergeometric case, drawing forms a monoidal transformation (with
`zipping' for multisets as coherence map).  Below it will be shown
that the three draw maps~\eqref{DrawDiag} are even more well-behaved:
they are all isometries, that is, they exactly preserve Kantorovich
distances. These remarkable preservation results first appeared in the
conference publication~\cite{Jacobs24a}, which this paper extends. The
results are reproduced here, with more details, especially for the
P\'olya case.

This paper adds two more results about drawing and distance, to which
we refer as the law of large urns and the law of large draws. Recall
that multinomial and P\'olya drawing involves removal and addition of
the drawn balls from / to the urn. The effect of such removal /
addition is neglible when the urn is very large, in comparison to the
drawsize, so that one may expect that there is no difference with
multinomial drawing (where the urn remains unchanged). The law of
large urns make these intuitions precise in terms of limits of
distances going to zero:
\[ \begin{array}{rcccl}
\lim\limits_{\upsilon\rightarrow\infty}
   d\Big(\hypergeometric[K]\big(\upsilon\big), \;
   \multinomial[K]\big(\flrn(\upsilon)\big)\Big)
& = &
0
& = &
\lim\limits_{\upsilon\rightarrow\infty}
   d\Big(\polya[K]\big(\upsilon\big), \;
   \multinomial[K]\big(\flrn(\upsilon)\big)\Big).
\end{array} \]

\noindent The first equation expresses that the Kantorovich distances
between hypergeometric and multinomial distributions goes to zero as
the urns beccome large. The multinomial distribution acts on urns as
distributions, so that the normalisation operation $\flrn$ must be
inserted, see below for details. Similarly, the second equation makes
precise that P\'olya draws from large urns are close to multinomial
draws.

The second law in this paper is what we call the law of large draws.
It can be seen as a variation on the law of large numbers, in terms of
Kantorovich distances. This law takes the form:
\[ \begin{array}{rcl}
\lim\limits_{K\rightarrow\infty} \, \multinomial[K](\omega) \models
   d\big(\omega, \flrn(-)\big)
& \,=\, &
0.
\end{array} \]

\noindent Informally it says that (normalisations of) large
multinomial draws are close to the urn $\omega$. This closeness holds
``in probability'' as expressed by the validity sign $\models$ in the
above formulation. Details can be found in Theorem~\ref{LargeDrawThm}
below.

This paper concentrates on the mathematics behind these isometry and
large urn/draw results, and not on interpretations or applications. We
do like to briefly refer to interpretations in machine
learning~\cite{RubnerTG00} where the distance that we consider on
colours in an urn is called the \emph{ground distance}. Actual
distances between colours are used there, based on experiments in
psychophysics, using perceived differences~\cite{WyszeckiS82}.

The Kantorovich --- or Wasserstein-Kantorovich, or Monge-Kantorovich
--- distance is the standard distance on distributions and on
multisets, going back to~\cite{KantorovichR58}\footnote{The history on
this topic is not so clear; we refer to the bibliographic notes as the
end of Chapter~6 of~\cite{Villani09} for details.}.  After some
preliminaries on multisets and distributions, and on distances in
general, Sections~\ref{KantorovichDstSec} and~\ref{KantorovichMltSec}
of this paper recall the Kantorovich distance on distributions and on
multisets, together with several basic results. The appendix contains
some further background, especially about the dual formulations for
these distances. The three subsequent Sections~\ref{MulnomSec} --
\ref{PolyaSec} demonstrate that multinomial, hypergeometric and
P\'olya drawing are all isometries.  Distances occur on multiple
levels: on colours, on urns (as multisets or distributions) and on
draw-distributions. This may be confusing, but several illustrations
are included.

The so-called total variation distance is a special case of the
Kantorovich distance. The relation between these two distances is
subtle. We elaborate them in Section~\ref{TvdSec}, for convenience of
the reader. The large urn / draw results in
Theorem~\ref{DrawDistanceLimitThm} and~\ref{LargeDrawThm} are proven
first for the total variation distance, but can then be transferred to
the general Kantorovich distance.

\section{Preliminaries on multisets and distributions}\label{PrelimSec}

A \emph{multiset} over a set $X$ is a finite formal sum of the form
$\sum_{i} n_{i}\ket{x_i}$, for elements $x_{i} \in X$ and natural
numbers $n_{i}\in\NNO$ describing the multiplicities of these elements
$x_{i}$. We shall write $\Mlt(X)$ for the set of such multisets over
$X$. A multiset $\varphi\in\Mlt(X)$ may equivalently be described in
functional form, as a function $\varphi\colon X \rightarrow \NNO$ with
finite support: $\supp(\varphi) \coloneqq \setin{x}{X}{\varphi(x) \neq
  0}$. Such a function $\varphi\colon X \rightarrow \NNO$ can be
written in ket form as $\sum_{x\in X} \varphi(x)\ket{x}$. We switch
back-and-forth between the ket and functional form and use the
formulation that best suits a particular situation.

For a multiset $\varphi\in\Mlt(X)$ we write $\|\varphi\| \in \NNO$ for
the \emph{size} of the multiset. It is the total number of elements,
including multiplicities:
\[ \begin{array}{rcccl}
\|\varphi\|
& \coloneqq &
\displaystyle\sum_{x\in \supp(\varphi)} \varphi(x)
& = &
\displaystyle\sum_{x\in X} \varphi(x).
\end{array} \]

\noindent For a number $K\in\NNO$ we write $\Mlt[K](X) \subseteq
\Mlt(X)$ for the subset of multisets of size $K$. When the set $X$ has
$n\geq 1$ elements, the number of multisets of size $K$ in the set
$\Mlt[K](X)$ is given by the \emph{multichoose} coefficient:
\[ \begin{array}{rcccl}
\displaystyle\bibinom{n}{K} 
& \coloneqq &
\displaystyle\binom{n+K-1}{K}
& = &
\displaystyle\frac{(n+K-1)!}{K!\cdot (n-1)!}.
\end{array} \]

\noindent We refer to~\cite{Jacobs22a,Jacobs21g} for details, but we
do recall the analogy that the number of subsets of size $K$ of an
$n$-element set is given by the ordinary binomial coefficient
$\binom{n}{K} = \frac{n!}{K!\cdot (n-K)!}$.

For each set $X$ and number $K$ there is an `accumulation' map $\acc
\colon X^{K} \rightarrow \Mlt[K](X)$ that turns lists into multisets
via $\acc\big(x_{1}, \ldots, x_{K}\big) \coloneqq 1\ket{x_1} + \cdots
+ 1\ket{x_K}$. For instance $\acc\big(c,b,a,c,a,c) = 2\ket{a} +
1\ket{b} + 3\ket{c}$. A standard result (see~\cite{Jacobs21g}) is that
for a multiset $\varphi\in\Mlt[K](X)$ there are $\coefm{\varphi}
\coloneqq \frac{K!}{\facto{\varphi}}$ many sequences $\vec{x} \in
X^{K}$ with $\acc(\vec{x}) = \varphi$, where $\facto{\varphi} =
\prod_{x} \varphi(x)!$. This accumulation map is the coequaliser of
all transposition maps $X^{K} \conglongrightarrow X^{K}$ induced by
permutations of $K$, see~\cite{Jacobs22g,Jacobs21g}.

Multisets $\varphi,\psi\in\Mlt(X)$ can be added and compared
elementwise, so that $\big(\varphi + \psi\big)(x) = \varphi(x) +
\psi(x)$ and $\varphi\leq\psi$ means $\varphi(x) \leq \psi(x)$ for all
$x\in X$.  In the latter case, when $\varphi\leq\psi$, we can also
subtract $\psi-\varphi$ elementwise.\\[0.5\baselineskip]
\noindent 
The mapping $X \mapsto \Mlt(X)$ is functorial: for a function $f\colon
X\rightarrow Y$ we have $\Mlt(f) \colon \Mlt(X) \rightarrow \Mlt(Y)$
given by $\Mlt(f)(\varphi)(y) = \sum_{x\in f^{-1}(y)} \varphi(x)$.
When we view the function $f$ as mapping $X$-colours to $Y$-colours,
the associated map $\Mlt(f)$ is a repainting of balls in urns.  This
map $\Mlt(f)$ preserves sums and size. Clearly, repainting of balls in
an urn does not change the number of balls in the urn.  Preservation
of sums by $\Mlt(f)$ involves: first repainting the balls in two urns
separately and then combining their balls, is the same as first
combining the urns and then repainting.

For a multiset $\tau\in\Mlt(X\times Y)$ on a product set we can take
its two marginals $\Mlt(\pi_{1})(\tau) \in \Mlt(X)$ and
$\Mlt(\pi_{2})(\tau) \in \Mlt(Y)$ via functoriality, using the two
projection functions $\pi_{1}\colon X\times Y \rightarrow X$ and
$\pi_{2} \colon X\times Y \rightarrow Y$. Starting from
$\varphi\in\Mlt(X)$ and $\psi\in\Mlt(Y)$, we say that
$\tau\in\Mlt(X\times Y)$ is a \emph{coupling} of $\varphi,\psi$ if
$\varphi$ and $\psi$ are the two marginals of $\tau$. We define the
\emph{decoupling} map:
\begin{equation}
\label{MltDecoupleDiag}
\xymatrix@C+2pc{
\Mlt(X\times Y)\ar[rr]^-{\decouple \,\coloneqq\, \tuple{\Mlt(\pi_{1}), \Mlt(\pi_{2})}} & & \Mlt(X)\times\Mlt(Y)
}
\end{equation}

\noindent The inverse image $\decouple^{-1}(\varphi,\psi) \subseteq
\Mlt(X\times Y)$ is thus the subset of couplings of $\varphi,\psi$.

For two multisets $\varphi\in\Mlt(X)$ and $\psi\in\Mlt(Y)$ we can form
a tensor product $\varphi\otimes\psi\in\Mlt(X\times Y)$ via:
\[ \begin{array}{rcl}
\big(\varphi\otimes\psi\big)(x,y)
& \coloneqq &
\varphi(x) \cdot \psi(y).
\end{array} \]


\medskip

A \emph{distribution} over a set $X$ is a finite formal sum of the
form $\sum_{i} r_{i}\ket{x_i}$ with elements $x_{i}\in X$ and
multiplicities / probabilities $r_{i} \in [0,1]$ satisfying
$\sum_{i}r_{i} = 1$. Such a distribution can equivalently be described
as a function $\omega \colon X \rightarrow [0,1]$ with finite support,
satisfying $\sum_{x}\omega(x) = 1$. We write $\Dst(X)$ for the set of
distributions on $X$. This $\Dst$ is functorial, in the same way as
$\Mlt$. Both $\Dst$ and $\Mlt$ are monads on the category $\Sets$ of
sets and functions, but we only use this for $\Dst$. The unit and
multiplication / flatten maps $\unit \colon X \rightarrow \Dst(X)$ and
$\flatten \colon \Dst^{2}(X) \rightarrow \Dst(X)$ are given by:
\begin{equation}
\label{DstMonadEqn}
\begin{array}{rclcrcl}
\unit(x)
& \coloneqq &
1\ket{x}
& \qquad\qquad &
\flatten(\Omega)
& \coloneqq &
\displaystyle\sum_{x\in X} \left(\sum_{\omega\in\Dst(X)} 
   \Omega(\omega)\cdot \omega(x)\right)\ket{x}.
\end{array}
\end{equation}

\noindent Kleisli maps $c \colon X \rightarrow \Dst(Y)$ are also
called channels and written as $c\colon X \chanto Y$. The Kleisli
extension $\pushop{c} \colon \Dst(X) \rightarrow \Dst(Y)$ for such a
channel, is defined on $\omega\in\Dst(X)$ as:
\[ \begin{array}{rcccl}
\pushing{c}{\omega}
& \coloneqq &
\flatten\big(\Dst(c)(\omega)\big)
& = &
\displaystyle\sum_{y\in Y} \left(\sum_{x\in X} \omega(x)\cdot c(x)(y)\right)
   \ket{y}.
\end{array} \]

\noindent Channels $c \colon X \chanto Y$ and $d\colon Y \chanto Z$
can be composed to $d\klafter c \colon X \chanto Z$ via $(d\klafter
c)(x) \coloneqq \pushing{d}{c(x)}$. Each function $f\colon X
\rightarrow Y$ gives rise to a deterministic channel $\klin{f}
\coloneqq \unit \after f \colon X \chanto Y$, that is, via
$\klin{f}(x) = 1\bigket{f(x)}$.

An example of a channel is arrangement $\arr \colon \Mlt[K](X) \rightarrow
\Dst(X^{K})$. It maps a multiset $\varphi\in\Mlt[K](X)$ to the uniform
distribution of sequences that accumulate to $\varphi$.
\begin{equation}
\label{ArrEqn}
\begin{array}{rcccl}
\arr(\varphi)
& \coloneqq &
\displaystyle\sum_{\vec{x}\in\acc^{-1}(\varphi)} 
   \frac{1}{\coefm{\varphi}}\bigket{\vec{x}}
& = &
\displaystyle\sum_{\vec{x}\in\acc^{-1}(\varphi)} 
   \frac{\facto{\varphi}}{K!}\bigket{\vec{x}}.
\end{array} 
\end{equation}

\noindent One can show that $\klin{\acc} \klafter \arr =
\Dst(\acc) \after \arr = \unit \colon \Mlt[K](X) \rightarrow
\Dst\big(\Mlt[K](X)\big)$. The composite in the other direction produces
the uniform distribution of all permutations of a sequence:
\begin{equation}
\label{TranspositionEqn}
\begin{array}{rccclcrcl}
\arr \klafter \klin{\acc}
& = &
\arr \after \acc
& = &
\perm
& \qquad\mbox{where}\qquad &
\perm(\vec{x})
& \coloneqq &
\displaystyle\sum_{t\colon\! K\, \stackrel{\cong}{\rightarrow}\, K} \frac{1}{K!}\,
   \bigket{\underline{t}(\vec{x})},
\end{array}
\end{equation}

\noindent in which $\underline{t}\big(x_{1}, \ldots, x_{K}\big)
\coloneqq (x_{t(1)}, \ldots, x_{t(K)})$. In writing $t\colon K
\stackrel{\cong}{\rightarrow} K$ we implicitly identify the number $K$
with the set $\{1,\ldots,K\}$.

Each multiset $\varphi\in\Mlt(X)$ of non-zero size can be turned into
a distribution via normalisation. This operation is called frequentist
learning, since it involves learning a distribution from a multiset of
data, via counting. Explicitly:
\[ \begin{array}{rcl}
\flrn(\varphi)
& \coloneqq &
\displaystyle\sum_{x\in X} \, \frac{\varphi(x)}{\|\varphi\|} \, \bigket{x}.
\end{array} \]

\noindent For instance, if we learn from an urn with three red, two
green and five blue balls, we get the probability distribution for
drawing a ball of a particular colour from the urn:
\[ \begin{array}{rcl}
\flrn\Big(3\ket{R} + 2\ket{G} + 5\ket{B}\Big)
& = &
\frac{3}{10}\ket{R} + \frac{1}{5}\ket{G} + \frac{1}{2}\ket{B}.
\end{array} \]

\noindent This map $\flrn$ is a natural transformation (but not a map
of monads).

Given two distributions $\omega\in\Dst(X)$ and
$\rho\in\Dst(Y)$, we can form their parallel product
$\omega\otimes\rho\in\Dst(X\times Y)$, given in functional form as:
\[ \begin{array}{rcl}
\big(\omega\otimes\rho\big)(x,y)
& \coloneqq &
\omega(x) \cdot \rho(y).
\end{array} \]

\noindent The tensors of multisets and of distributions are related
via frequentist learning:
\begin{equation}
\label{FlrnTensorEqn}
\begin{array}{rcl}
\flrn\big(\varphi\otimes\psi\big)
& = &
\flrn(\varphi) \otimes \flrn(\psi).
\end{array} 
\end{equation}

Like for multisets, we call a joint distribution $\tau\in\Dst(X\times
Y)$ a \emph{coupling} of $\omega\in\Dst(X)$ and $\rho\in\Dst(Y)$ when
$\omega,\rho$ are the two marginals of $\tau$, that is, when
$\Dst(\pi_{1})(\tau) = \omega$ and $\Dst(\pi_{2}) = \rho$.  We can
express this also via a decouple map $\decouple =
\tuple{\Dst(\pi_{1}), \Dst(\pi_{2})}$ as in~\eqref{MltDecoupleDiag}.
The tensor $\omega\otimes\rho$ is an obvious coupling of the
distributions $\omega$ and $\rho$.

\medskip

An \emph{observation} on a set $X$ is a function of the form $p\colon
X \rightarrow \R$. Such a map $p$, together with a distribution
$\omega\in\Dst(X)$, is called a random variable --- but confusingly,
the distribution is often left implicit. The map $p\colon X
\rightarrow \R$ will be called a \emph{factor} if it restricts to
non-negative reals $X \rightarrow \nnR$. Each element $x\in X$ gives
rise to a point observation $\indic{x} \colon X \rightarrow \R$, with
$\indic{x}(x') = 1$ if $x = x'$ and $\indic{x}(x') = 0$ if $x \neq
x'$. 

For a distribution $\omega\in\Dst(X)$ and an observation $p\colon X
\rightarrow \R$ on the same set $X$ we write $\omega\models p$ for the
validity (expected value) of $p$ in $\omega$, defined as (finite) sum:
$\sum_{x\in X} \omega(x) \cdot p(x)$. For a function $f\colon X
\rightarrow Y$, a distribution $\omega\in\Dst(X)$ and an observation
$q\colon Y \rightarrow \R$ one has the following equality of
validities:
\begin{equation}
\label{TransformationValidityEqn}
\begin{array}{rcl}
\Dst(f)(\omega) \models q
& \,=\, &
\omega \models (q \after f).
\end{array}
\end{equation}

\noindent This equation is sometimes called `the law of the
unconscious statistician', see
\textit{e.g.}~\cite[\S2.2]{CasellaB02}. We shall write $\Obs(X) =
\R^{X}$ and $\Fact(X) = (\nnR)^{X}$ for the sets of observations and
factors on $X$.


\section{Preliminaries on metric spaces}\label{DistanceSec}

A metric space will be written as a pair $(X, d_{X})$, where $X$ is a
set and $d_{X} \colon X\times X \rightarrow \nnR$ is a distance
function, also called metric. This metric satisfies:
\begin{itemize}
\item separation: $d_{X}(x,x') = 0$ iff $x=x'$;

\item symmetry: $d_{X}(x,x') = d_{X}(x',x)$;

\item triangular inequality: $d_{X}(x,x'') \leq d_{X}(x,x') +
  d_{X}(x',x'')$.
\end{itemize}

\noindent Often, we drop the subscript $X$ in $d_X$ if it is clear from
the context. We use the standard distance $d(x,y) = |x-y|$ on real
and natural numbers.


\begin{defi}
\label{MetricMapDef}
Let $(X,d_{X})$, $(Y,d_{Y})$ be two metric spaces.
\begin{enumerate}
\item A function $f\colon X \rightarrow Y$ is called \emph{short}
(or also \emph{non-expansive}) if:
\[ \begin{array}{rcl}
d_{Y}\big(f(x), f(x')\big) 
& \leq &
d_{X}\big(x,x'\big), \qquad \mbox{for all }x,x'\in X.
\end{array} \]

\noindent Such a map is called an \emph{isometry} or an
\emph{isometric embedding} if the above inequality $\leq$ is an actual
equality $=$. This implies that the function $f$ is injective, and
thus an `embedding'.

We write $\MetSh$ for the category of metric spaces with short
maps between them. 

\item A function $f\colon X \rightarrow Y$ is \emph{Lipschitz} or
  $M$-\emph{Lipschitz}, if there is a number $M\in\pR$ such that:
\[ \begin{array}{rcl}
d_{Y}\big(f(x), f(x')\big) 
& \leq &
M\cdot d_{X}\big(x,x'\big), \qquad \mbox{for all }x,x'\in X.
\end{array} \]

\noindent The number $M$ is sometimes called the \emph{Lipschitz
constant}.  Thus, a short function is Lipschitz, with constant $1$.
We write $\MetLip$ for the category of metric spaces with Lipschitz
maps between them (with arbitrary Lipschitz constants).
\end{enumerate}
\end{defi}

\noindent 
It is easy to see that if $f\colon X \rightarrow Y$ is $M$-Lipschitz
and $g\colon Y \rightarrow Z$ is $L$-Lipschitz, then the composite $g
\after f\colon X \rightarrow Z$ is Lipschitz with constant $M\cdot L$.

\begin{lem}
\label{MetLipProdLem}
For two metric spaces $(X_{1},d_{1})$ and $(X_{2},d_{2})$ we equip the
cartesian product $X_{1}\times X_{2}$ of sets with the sum of the two
metrics:
\begin{equation}
\label{ProductMetricEqn}
\begin{array}{rcl}
d\Big((x_{1},x_{2}), (x'_{1},x'_{2})\Big)
& \coloneqq &
d_{1}(x_{1},x'_{1}) + d_{2}(x_{2},x'_{2}).
\end{array}
\end{equation}

\noindent We shall extend these product distances to $K$-ary form
$X_{1} \times \cdots \times X_{K}$ and $X^{K} = X \times \cdots \times
X$.

\begin{enumerate}
\item With the usual projections and tuples this forms a product in
  the category $\MetLip$. The definition yields a \emph{monoidal}
  product (tensor) in the category $\MetSh$ since the diagonal map $X
  \rightarrow X\times X$ is 2-Lipschitz and not short.

\item If $f_{i} \colon X_{i} \rightarrow Y_{i}$ is $M_{i}$-Lipschitz,
then $f_{1}\times f_{2} \colon X_{1} \times X_{2} \rightarrow Y_{1}\times Y_{2}$
is $\max(M_{1}, M_{2})$-Lipschitz.
\end{enumerate}
\end{lem}

\auxproof{
Projections are short:
\[ \begin{array}{rcl}
d_{X}\Big(\pi_{1}(x,y), \pi_{1}(x',y')\Big)
& = &
d_{X}(x,x')
\\
& \leq &
d_{X}(x,x') + d_{Y}(y,y')
\\
& = &
d_{X\times Y}\Big((x,y), (x',y')\Big).
\end{array} \]

When $f_{1}\colon Z \rightarrow X$ and $f_{2}\colon Z \rightarrow Y$
are $M_{1}-$ and $M_{2}$-Lipschitz, then for $M = M_{1} + M_{2}$ we
have that $\tuple{f_{1},f_{2}} \colon Z \rightarrow X\times Y$ is
$M$-Lipschitz, since:
\[ \begin{array}{rcl}
d_{X\times Y}\Big(\tuple{f_{1},f_{2}}(z), \tuple{f_{1},f_{2}}(z')\Big)
& = &
d_{X\times Y}\Big((f_{1}(z), f_{2}(z)), (f_{1}(z'), f_{2}(z')\Big)
\\
& = &
d_{X}\big(f_{1}(z), f_{1}(z')\big) + d_{Y}\big(f_{2}(z), f_{2}(z')\big)
\\
& \leq &
M_{1}\cdot d_{Z}(z,z') + M_{2}\cdot d_{Z}(z,z')
\\
& = &
\big(M_{1} + M_{2}\big)\cdot d_{Z}(z,z')
\\
& = &
M\cdot d_{Z}(z,z').
\end{array} \]

The same proof for the maximum: projections are short:
\[ \begin{array}{rcl}
d_{X}\Big(\pi_{1}(x,y), \pi_{1}(x',y')\Big)
& = &
d_{X}(x,x')
\\
& \leq &
\max\Big(d_{X}(x,x'), d_{Y}(y,y')\Big)
\\
& = &
d_{X\times Y}\Big((x,y), (x',y')\Big).
\end{array} \]

When $f_{1}\colon Z \rightarrow X$ and $f_{2}\colon Z \rightarrow Y$ are 
$M_{1}-$ and $M_{2}$-Lipschitz, then for $M = \max\big(M_{1}, M_{2}\big)$
we have that $\tuple{f_{1},f_{2}} \colon Z \rightarrow X\times Y$ is
$M$-Lipschitz, since:
\[ \begin{array}{rcl}
d_{X\times Y}\Big(\tuple{f_{1},f_{2}}(z), \tuple{f_{1},f_{2}}(z')\Big)
& = &
d_{X\times Y}\Big((f_{1}(z), f_{2}(z)), (f_{1}(z'), f_{2}(z')\Big)
\\
& = &
\max\Big(d_{X}\big(f_{1}(z), f_{1}(z')\big), d_{Y}\big(f_{2}(z), f_{2}(z')\big)\Big)
\\
& \leq &
\max\Big(M_{1}\cdot d_{Z}(z,z'), M_{2}\cdot d_{Z}(z,z')\Big)
\\
& = &
\max\big(M_{1},M_{2}\big)\cdot d_{Z}(z,z')
\\
& = &
M\cdot d_{Z}(z,z').
\end{array} \]

\noindent In particular, $\tuple{f_{1},f_{2}}$ is short when both
$f_{1},f_{2}$ are short.

For $f_{i} \colon X_{i} \rightarrow Y_{i}$ with Lipschitz constant $M_{i}$
we get:
\[ \begin{array}{rcl}
\lefteqn{d_{Y_{1}\times Y_{2}}\Big((f_{1}\times f_{2})(x_{1},x_{2}), \,
   (f_{1}\times f_{2})(x'_{1},x'_{2})\Big)}
\\[+0.2em]
& = &
d_{Y_{1}\times Y_{2}}\Big((f_{1}(x_{1}), f_{2}(x_{2})), \,
   (f_{1}(x'_{1}), f_{2}(x'_{2}))\Big)
\\[+0.2em]
& = &
d_{Y_{1}}\Big(f_{1}(x_{1}), f_{1}(x'_{1})\Big) + 
   d_{Y_{2}}\Big(f_{2}(x_{2}), f_{2}(x'_{2})\Big)
\\[+0.2em]
& \leq &
M_{1}\cdot d_{X_1}\big(x_{1}, x'_{1}\big) + M_{2}\cdot d_{X_2}\big(x_{2}, x'_{2}\big)
\\[+0.2em]
& \leq &
\max(M_{1},M_{2}) \cdot \Big(d_{X_1}\big(x_{1}, x'_{1}\big) + 
   d_{X_2}\big(x_{2}, x'_{2}\big)\Big)
\\[+0.2em]
& = &
\max(M_{1},M_{2}) \cdot d_{X_{1}\times X_{2}}\Big((x_{1},x_{2}), (x'_{1},x'_{2})\Big).
\end{array} \]
}

\noindent 
Products of metric spaces involve some subtleties, which are not
relevant for the main line of the paper, but which we briefly make
explicit as background information.

\begin{rem}
\label{ProductMetricRem}
In the above description~\eqref{ProductMetricEqn} we have used the sum
of the distances in the components. As stated, this formulation yields
a categorical product in the category of metric spaces with Lipschitz
maps, but not with short maps. 

Instead of the sum $+$ in~\eqref{ProductMetricEqn} one can use the
maximum. This yields a categorical product in the both the categories
$\MetLip$ and $\MetSh$ of metric spaces with Lipschitz and with short
maps. The maximum also works when the metric is 1-bounded, that is, of
the form $d\colon X\times X \rightarrow [0,1]$, restricted to the unit
interval $[0,1]$. This restriction is common in program semantics, see
\textit{e.g.}~\cite{deBakkerV96,Breugel01}. This 1-bounded case has
advantages since it admits infinite products and also coproducts of
metric spaces. We do not need such structure and we will work with
general $\nnR$-valued metrics.

Thus, in the category $\MetLip$ with Lipschitz maps one can
equivalently use the maximum or the sum of distances
in~\eqref{ProductMetricEqn}. The resulting products are isomorphic in
$\MetLip$. This works since for $r,s\in\nnR$ one has $\max(r,s) \leq
r+s$ and $r+s \leq 2\cdot\max(r,s)$. Here we use the sum of distances
in the product, since it is used for certain results, like
Lemma~\ref{DstMetricLem}~\eqref{DstMetricLemPerm}. A concrete example
where the sum of distances is used is for shortness of the addition
function $+\colon \R\times\R \rightarrow \R$, in:
\[ \begin{array}{rcl}
d_{\R}\big(r+s, r'+s'\big)
\hspace*{\arraycolsep}=\hspace*{\arraycolsep}
\big|\, (r+s) - (r'+s') \,\big|
& = &
\big|\, (r-r') + (s-s') \,\big|
\\
& \leq &
\big|r-r'\big| + \big|s-s'\big|
\\
& = &
d_{\R}\big(r,r'\big) + d_{\R}\big(s,s'\big)
\hspace*{\arraycolsep}\smash{\stackrel{\eqref{ProductMetricEqn}}{=}}\hspace*{\arraycolsep}
d_{\R\times\R}\big((r,s), (r',s')\big).
\end{array} \]
\end{rem}

Below we shall use the familiar zip function. It is in fact an
isometry.

\begin{lem}
\label{ZipLem}
For metric spaces $X,Y$ and for $K\in\NNO$ consider the zip function
$\zip \colon X^{K} \times Y^{K} \rightarrow (X\times Y)^{K}$ given
by $\zip\big(\vec{x},\vec{y}) = ((x_{1},y_{1}), \ldots, (x_{K},y_{K}))$.
This $\zip$ function is a bijective function and an isometry.

Similarly, concatenation $\concat \colon X^{K}\times X^{L} \rightarrow
X^{K+L}$ is a bijection and an isometry.
\end{lem}

\begin{myproof}
We only do the zip case. The zip function is obviously a
bijection. For $\vec{x}, \vec{x'} \in X^{K}$ and $\vec{y}, \vec{y'}
\in Y^{K}$ one has via repeated use of~\eqref{ProductMetricEqn},
\[ \begin{array}[b]{rcl}
\lefteqn{d_{(X\times Y)^{K}}\Big(\zip(\vec{x}, \vec{y}), 
   \zip(\vec{x'}, \vec{y'})\Big)}
\\[+0.2em]
& = &
d_{X\times Y}\Big((x_{1},y_{1}), (x'_{1},y'_{1})\Big) + \cdots +
   d_{X\times Y}\Big((x_{K},y_{K}), (x'_{K},y'_{K})\Big)
\\[+0.2em]
& = &
\Big(d_{X}(x_{1}, x'_{1}) + d_{Y}(y_{1},y'_{1})\Big) + \cdots + 
   \Big(d_{X}(x_{K}, x'_{K}) + d_{Y}(y_{K},y'_{K})\Big)
\\[+0.4em]
& = &
\Big(d_{X}(x_{1}, x'_{1}) + \cdots + d_{X}(x_{K}, x'_{K})\Big) + 
   \Big(d_{Y}(y_{1},y'_{1}) + \cdots + d_{Y}(y_{K},y'_{K})\Big)
\\[+0.4em]
& = &
d_{X^{K}}\big(\vec{x}, \vec{x'}\big) + d_{Y^{K}}\big(\vec{y}, \vec{y'}\big)
\hspace*{\arraycolsep}=\hspace*{\arraycolsep}
d_{X^{K}\times Y^{K}}\big((\vec{x}, \vec{y}), (\vec{x'}, \vec{y'})\big).
\end{array} \eqno{\QEDbox} \]

\auxproof{
For $\vec{x},\vec{x'}\in X^{K}$ and $\vec{y}, \vec{y'} \in X^{L}$ one
has:
\[ \begin{array}{rcl}
d_{X^{K+L}}\Big( \vec{x} \concat \vec{y}, \, \vec{x'} \concat \vec{y'}\Big)
& = &
\sum_{i} d_{X}\big(x_{i}, x'_{i}\big) + \sum_{j} d_{X}\big(y_{j}, y'_{j}\big)
\\[+0.2em]
& = &
d_{X^{K}}\big(\vec{x}, \vec{x'}\big) + d_{X^{L}}\big(\vec{y}, \vec{y'}\big)
\\[+0.2em]
& = &
d_{X^{K}\times X^{L}}\big((\vec{x},\vec{y}), \, (\vec{x'},\vec{y'})\big).
\end{array} \]
}
\end{myproof}

\section{The Kantorovich distance between distributions}\label{KantorovichDstSec}

This section introduces the Kantorovich distance between probability
distributions and recalls some basic results.  There are several
equivalent formulations for this distance.  We express them in terms
of validity and couplings, see also
\textit{e.g.}~\cite{Breugel01,Brezis18,DengD09,FritzP19,DesharnaisGJP04}.

\begin{defi}
\label{DstMetricDef}
Let $(X,d_{X})$ be a metric space. The \emph{Kantorovich} metric $d =
d_{\Dst(X)} \colon \Dst(X)\times \Dst(X) \rightarrow \nnR$ on
distributions over $X$ is defined by any of the three equivalent
formulas:
\begin{equation}
\label{DstMetricEqn}
\begin{array}{rcl}
d\big(\omega, \omega'\big)
& \,\coloneqq\, &
\displaystyle\bigwedge_{\tau\in\decouple^{-1}(\omega, \omega')} \tau \models d_{X}
\\[+1.4em]
& = &
\displaystyle\bigvee_{p, \, p'\in \Obs(X), \, p\oplus p' \,\leq\, d_{X}} 
   \, \omega\models p \,+\, \omega'\models p' 
\\[+1.4em]
& = &
\displaystyle\bigvee_{q\in \FactSh(X)} \big|\, \omega\models q \,-\,
   \omega'\models q \,\big|.
\end{array}
\end{equation}

\noindent This turns $\Dst(X)$ into a metric space. The operation
$\oplus$ in the second formulation is defined as $(p\oplus p')(x,x') =
p(x) + p'(x')$. The set $\FactSh(X) \subseteq \Fact(X)$ in the third
formulation is the subset of short factors $X \rightarrow \nnR$. To be
precise, we should write $\FactSh(X, d_{X})$ since the distance $d_X$
on $X$ is a parameter, but we leave it implicit for convenience. The
meet $\bigwedge$ and joins $\bigvee$ in~\eqref{DstMetricEqn} are
actually reached, by what are called the \emph{optimal} coupling and
the \emph{optimal} observations / factor.
\end{defi}

The equivalence of the first and second formulation
in~\eqref{DstMetricEqn} is an instance of strong duality in linear
programming, which can be obtained via Farkas' Lemma, see
\textit{e.g.}~\cite{MatouekG06}. The second formulation is commonly
associated with Monge. The single factor $q$ in the third formulation
can be obtained from the two observations $p,p'$ in the second
formulation, and vice-versa.  In~\cite{DesharnaisGJP04} it is shown
that instead of using all short factors $q$ in the third formulation
one can restrict to those factors that arise as interpretation of
formulas in a logic --- since these interpretations are dense.  What
we call the Kantorovich distance is also called the
Wasserstein-Kantorovich, or Monge-Kantorovich distance.

A proof of the equivalence of the three formulations for the
Kantorovich distance $d(\omega,\omega')$ between two distributions
$\omega,\omega'$ in~\eqref{DstMetricEqn} is given in the
appendix. These three formulations do not immediately suggest how to
calculate distances. What helps is that the minimum and maxima are
actually reached and can be computed. This is done via linear
programming, originally introduced by Kantorovich,
see~\cite{MatouekG06,Villani09,DengD09}.  In the sequel, we shall see
several examples of distances between distributions. They are obtained
via our (adapted) Python implementation of the linear optimisation,
which also produces an optimal coupling, observations or factor. This
implementation is used only for illustrations.

\begin{exa}
\label{DstMetricEx}
Consider the set $X$ containing the first eight natural numbers, so $X
= \{0,1,\ldots,7\} \subseteq \NNO$, with the usual distance, written
as $d_X$, between natural numbers: $d_{X}(n,m) = |n-m|$. We look at
the following two distributions on $X$.
\[ \begin{array}{rclcrcl}
\omega
& = &
\frac{1}{2}\ket{0} + \frac{1}{2}\ket{4}
& \qquad &
\omega'
& = &
\frac{1}{8}\ket{2} + \frac{1}{8}\ket{3} + 
   \frac{1}{8}\ket{6} + \frac{5}{8}\ket{7}.
\end{array} \]

\noindent We claim that the Kantorovich distance $d(\omega,\omega')$
is $\frac{15}{4}$. This will be illustrated for each of the three
formulations in Definition~\ref{DstMetricDef}.
\begin{itemize}
\item An optimal coupling $\tau\in\Dst(X\times X)$ of $\omega,\omega'$ is:
\[ \begin{array}{rcl}
\tau
& = &
\frac{1}{8}\bigket{0, 2} + 
   \frac{1}{8}\bigket{0, 3} + 
   \frac{1}{8}\bigket{0, 6} + 
   \frac{1}{8}\bigket{0, 7} + 
   \frac{1}{2}\bigket{4, 7}.
\end{array} \]

\noindent It is not hard to see that $\tau$'s first marginal is $\omega$,
and its second marginal is $\omega'$. We compute the distances as:
\[ \begin{array}{rcl}
\lefteqn{d(\omega,\omega')
\hspace*{\arraycolsep}=\hspace*{\arraycolsep}
\tau \models d_{X}}
\\[+0.2em]
& = &
\frac{1}{8}\cdot d_{X}(0, 2) + 
   \frac{1}{8}\cdot d_{X}(0, 3) + 
   \frac{1}{8}\cdot d_{X}(0, 6) + 
   \frac{1}{8}\cdot d_{X}(0, 7) + 
   \frac{1}{2}\cdot d_{X}(4, 7)
\\[+0.2em]
& = &
\frac{2}{8} + \frac{3}{8} + \frac{6}{8} + \frac{7}{8} + \frac{3}{2}
\hspace*{\arraycolsep}=\hspace*{\arraycolsep}
\frac{18}{8} + \frac{3}{2}
\hspace*{\arraycolsep}=\hspace*{\arraycolsep}
\frac{9}{4} + \frac{6}{4}
\hspace*{\arraycolsep}=\hspace*{\arraycolsep}
\frac{15}{4}.
\end{array} \]

\item There are the following two optimal observations $p,p' \colon X
  \rightarrow \R$, described as sums of weighted point predicates:
\[ \begin{array}{rcl}
p
& = &
-1\cdot\indic{1} - 2\cdot\indic{2} - 3\cdot\indic{3} - 4\cdot\indic{4} 
   - 5\cdot\indic{5} - 6\cdot\indic{6} - 7\cdot\indic{7} 
\\
p'
& = &
1\cdot\indic{1} + 2\cdot\indic{2} + 3\cdot\indic{3} + 4\cdot\indic{4} 
   + 5\cdot\indic{5} + 6\cdot\indic{6} + 7\cdot\indic{7}.
\end{array} \]

\noindent It is not hard to see that $(p\oplus p')(i,j) \coloneqq p(i)
+ p'(j) \leq d_{X}(i,j)$ holds for all $i,j\in X$. Using the second
formulation in~\eqref{DstMetricEqn} we get:
\[ \begin{array}{rcl}
\big(\omega\models p\big) + \big(\omega'\models p'\big)
& = &
\frac{1}{2}\cdot p(0) + \frac{1}{2}\cdot p(4) +
   \frac{1}{8}\cdot p'(2) + \frac{1}{8}\cdot p'(3) +
   \frac{1}{8}\cdot p'(6) + \frac{5}{8}\cdot p'(7) 
\\[+0.2em]
& = &
\frac{-4}{2} + \frac{2}{8} + \frac{3}{8} + \frac{6}{8} + \frac{35}{8}
\hspace*{\arraycolsep}=\hspace*{\arraycolsep}
-2 + \frac{46}{8}
\hspace*{\arraycolsep}=\hspace*{\arraycolsep}
\frac{30}{8}
\hspace*{\arraycolsep}=\hspace*{\arraycolsep}
\frac{15}{4}.
\end{array} \]

\item Finally, there is a (single) short factor $q \colon X
  \rightarrow \nnR$ given by:
\[ \begin{array}{rcl}
q
& = &
7\cdot\indic{0} + 6\cdot\indic{1} + 5\cdot\indic{2} + 
   4\cdot\indic{3} + 3\cdot\indic{4} + 2\cdot\indic{5} + 1\cdot\indic{6}.
\end{array} \]

\noindent Then:
\[ \begin{array}{rcl}
\big(\omega\models q\big) - \big(\omega'\models q\big)
& = &
\frac{1}{2}\cdot q(0) + \frac{1}{2}\cdot q(4) -
   \Big(\frac{1}{8}\cdot q(2) + \frac{1}{8}\cdot q(3) +
   \frac{1}{8}\cdot q(6) + \frac{5}{8}\cdot q(7)\Big)
\\[+0.2em]
& = &
\frac{7}{2} + \frac{3}{2} - \Big(\frac{5}{8} + \frac{4}{8} + \frac{1}{8}\Big)
\hspace*{\arraycolsep}=\hspace*{\arraycolsep}
\frac{10}{2} - \frac{10}{8}
\hspace*{\arraycolsep}=\hspace*{\arraycolsep}
\frac{20}{4} - \frac{5}{4}
\hspace*{\arraycolsep}=\hspace*{\arraycolsep}
\frac{15}{4}.
\end{array} \]
\end{itemize}

\noindent From the fact that the coupling $\tau$, the two observations
$p,p'$, and the single factor $q$ produce the same distance one can
deduce that they are optimal, using the formula~\eqref{DstMetricEqn}.
\end{exa}

We proceed with several standard properties of the Kantorovich
distance on distributions.

\begin{lem}
\label{DstMetricLem}
In the context of Definition~\ref{DstMetricDef}, the following
properties hold.
\begin{enumerate}
\item \label{DstMetricLemFun} For an $M$-Lipschitz function $f\colon
  X \rightarrow Y$, the pushforward map $\Dst(f) \colon \Dst(X)
  \rightarrow \Dst(Y)$ is also $M$-Lipschitz; as a result, $\Dst$
  lifts to a functor $\Dst \colon \MetLip \rightarrow \MetLip$, and
  also to $\Dst \colon \MetSh \rightarrow \MetSh$.

\item \label{DstMetricLemIsometric} If $f\colon X \rightarrow Y$ is
  an isometry, then so is $\Dst(f) \colon \Dst(X) \rightarrow
  \Dst(Y)$.

\item \label{DstMetricLemVal} For an $M$-Lipschitz factor $q\colon X
  \rightarrow \nnR$, the validity-of-$q$ factor $(-)\models q \colon
  \Dst(X) \rightarrow \nnR$ is also $M$-Lipschitz.


\item \label{DstMetricLemPoint} For each element $x\in X$ and
  distribution $\omega\in\Dst(X)$ one has: $d\big(1\ket{x},
  \omega\big) \,=\, \omega \models d_{X}(x, -)$; especially,
  $d\big(1\ket{x}, 1\ket{x'}\big) = d_{X}(x,x')$, making the map
  $\unit\colon X \rightarrow \Dst(X)$ an isometry.

\item \label{DstMetricLemFlatten} The monad multiplication $\flatten
  \colon \Dst^{2}(X) \rightarrow \Dst(X)$ is short, so that $\Dst$
  lifts from a monad on $\Sets$ to a monad on $\MetSh$ and on
  $\MetLip$.


\ignore{

X = range_sp(5)
v1 = random_distribution(X)
v2 = random_distribution(X)
W = random_distribution(Space(w1,w2))
V = random_distribution(Space(v1,v2))
print( wd(Sample >> W, Sample >> V) )
d = distribution_relationlift(W, V, WassersteinPred)
print( d[0] )

}

\item \label{DstMetricLemPush} If a channel $c \colon X \rightarrow
  \Dst(Y)$ is $M$-Lipschitz, then so is its Kleisli extension
  $\pushop{c} \coloneqq \flatten \after \Dst(c) \colon \Dst(X)
  \rightarrow \Dst(Y)$.

\item \label{DstMetricLemComp} If channel $c\colon X \chanto Y$ is
  $M$-Lipschitz and channel $d\colon Y \chanto Z$ is $L$-Lipschitz,
  then their (channel) composite $d\klafter c \colon X \chanto Z$ is
  $(M\cdot L)$-Lipschitz.

\item \label{DstMetricLemConv} For distributions
  $\omega_{i},\omega'_{i} \in \Dst(X)$ and numbers $r_{i}\in [0,1]$
  with $\sum_{i}r_{i} = 1$ one has:
\[ \begin{array}{rcl}
d\Big(\sum_{i}r_{i}\cdot\omega_{i}, \, 
   \sum_{i} r_{i}\cdot\omega'_{i}\Big)
& \leq &
\sum_{i}r_{i}\cdot d\big(\omega_{i}, \omega'_{i}\big).
\end{array} \]

\item \label{DstMetricLemPerm} The permutation channel $\perm \colon
  X^{K} \rightarrow \Dst(X^{K})$ from~\eqref{TranspositionEqn} is
  short.
\end{enumerate}
\end{lem}

\begin{myproof}\leavevmode
\begin{enumerate}
\item Let $f\colon X \rightarrow Y$ be $M$-Lipschitz. If
  $\tau\in\Dst(X\times X)$ is a coupling of
  $\omega,\omega'\in\Dst(X)$, then $\Dst(f\times f)(\tau) \in
  \Dst(Y\times Y)$ is a coupling of $\Dst(f)(\omega), \Dst(f)(\omega')
  \in \Dst(Y)$. Thus:
\[ \begin{array}{rcl}
d_{\Dst(Y)}\Big(\Dst(f)(\omega), \Dst(f)(\omega')\Big)
& \smash{\stackrel{\eqref{DstMetricEqn}}{=}} &
\displaystyle\bigwedge_{\sigma\in\decouple^{-1}(\Dst(f)(\omega), \Dst(f)(\omega'))} 
   \sigma \models d_{Y}
\\[+1.2em]
& \leq &
\displaystyle\bigwedge_{\tau\in\decouple^{-1}(\omega, \omega')} 
   \Dst(f\times f)(\tau) \models d_{Y}
\\[+1.2em]
& \smash{\stackrel{\eqref{TransformationValidityEqn}}{=}} &
\displaystyle\bigwedge_{\tau\in\decouple^{-1}(\omega, \omega')} 
   \tau \models d_{Y} \after (f\times f)
\\[+1.2em]
& \leq &
\displaystyle\bigwedge_{\tau\in\decouple^{-1}(\omega, \omega')} 
   \tau \models M\cdot d_{X}
\\[+1.2em]
& = &
\displaystyle M\cdot \left(\bigwedge_{\tau\in\decouple^{-1}(\omega, \omega')} 
   \tau \models d_{X}\right)
\hspace*{\arraycolsep}\smash{\stackrel{\eqref{DstMetricEqn}}{=}}\hspace*{\arraycolsep}
M\cdot d_{\Dst(X)}\big(\omega, \omega'\big).
\end{array} \]

\item Let $f\colon X \rightarrow Y$ be an isometry. Then it is
  $1$-Lipschitz, so $\Dst(f)$ is as well, by the previous item.  It thus
  suffices to prove: $d(\omega,\omega') \leq
  d\big(\Dst(f)(\omega), \Dst(f)(\omega')\big)$, for
  $\omega,\omega'\in\Dst(X)$. Let $p\colon X \rightarrow \nnR$ be short.
We turn it into a short factor $q\colon Y \rightarrow \nnR$ via the
definition:
\[ \begin{array}{rclcrcl}
q(y)
& \coloneqq &
\displaystyle\bigwedge_{x\in X} \, p(x) + d_{Y}(f(x),y)
& \qquad\mbox{satisfying, by construction,}\qquad &
q \after f
& = &
p.
\end{array} \]

\auxproof{
We need to prove the marked equation in:
\[ \begin{array}{rcccccl}
q(f(z))
& = &
\displaystyle\bigwedge_{x\in X} \, p(x) + d_{Y}(f(x),f(z))
& = &
\displaystyle\bigwedge_{x\in X} \, p(x) + d_{X}(x,z)
& \smash{\stackrel{(*)}{=}} &
p(z).
\end{array} \]

\noindent The equation $(\leq)$ obviously holds, by taking $x=z$. For
$(\geq)$ we distinguish two cases.
\begin{itemize}
\item When $p(x) \geq p(z)$ then also $p(x) + d_{X}(x,z) \geq p(z)$.

\item When $p(x) < p(z)$, then, using shortness of $p$ we have
$p(z) - p(x) \leq d_{X}(x,z)$, and thus $p(x) + d_{X}(x,z) \geq p(z)$.
\end{itemize}
}

\noindent This map $q$ is short, since for arbitrary $y,y'\in Y$ we have:
\[ \begin{array}{rcl}
\big|\, q(y) - q(y') \,\big|
& = &
\displaystyle\bigwedge_{x,x'\in X} \, \big|\, \big(p(x) + d_{Y}(f(x),y)\big)
   - \big(p(x') + d_{Y}(f(x'),y')\big) \,\big|
\\
& \leq &
\displaystyle\bigwedge_{x\in X} \, \big|\, p(x) + d_{Y}(f(x),y) 
   - p(x) - d_{Y}(f(x),y')\big) \,\big|
\\
& = &
\displaystyle\bigwedge_{x\in X} \, \big|\, d_{Y}(f(x),y) 
   - d_{Y}(f(x),y')\big) \,\big|
\hspace*{\arraycolsep}\leq\hspace*{\arraycolsep}
d_{Y}(y,y').
\end{array} \]

\noindent The latter inequality --- sometimes called the reverse triangle
inequality --- follows easily from the triangular inequality.

\auxproof{
We distinguish two cases:
\begin{itemize}
\item If $d(f(x),y) \leq d(f(x),y')$, then we use 
$d(y,y') + d(f(x),y) \geq d(f(x),y')$,
so that:
\[ \begin{array}{rcccl}
\big| \, d(f(x),y) - d(f(x),y') \,\big|
& = &
d(f(x),y') - d(f(x),y)
& \leq &
d(y,y').
\end{array} \]

\item If $d(f(x),y) > d(f(x),y')$, then we use $d(y,y') + d(f(x),y')
  \geq d(f(x),y)$, so that:
\[ \begin{array}{rcccl}
\big| \, d(f(x),y) - d(f(x),y') \,\big|
& = &
d(f(x),y) - d(f(x),y')
& \leq &
d(y,y').
\end{array} \]
\end{itemize}
}

Using $q \after f = p$ we get:
\[ \begin{array}{rcl}
\big|\, \omega\models p - \omega'\models p \,\big|
& = &
\big|\, \omega\models (q\after f) - \omega'\models (q\after f) \,\big|
\\[+0.4em]
& \smash{\stackrel{\eqref{TransformationValidityEqn}}{=}} &
\big|\, \Dst(f)(\omega)\models q - \Dst(f)(\omega')\models q \,\big|
\hspace*{\arraycolsep}\leq\hspace*{\arraycolsep}
d\big(\Dst(f)(\omega), \Dst(f)(\omega')\big).
\end{array} \]

\noindent Since this holds for all short factors $p$ we obtain, as
required:
\[ \begin{array}{rcccl}
d(\omega,\omega')
& = &
\displaystyle\bigvee_{p\in \FactSh(X)} \big|\, \omega\models p \,-\,
   \omega'\models p \,\big|
& \leq &
d\big(\Dst(f)(\omega), \Dst(f)(\omega')\big).
\end{array} \]

\item Let $q\colon X \rightarrow \nnR$ be $M$-Lipschitz, then
  $\frac{1}{M}\cdot q \colon X \rightarrow \nnR$ is short. The
  function $(-) \models q \colon \Dst(X) \rightarrow \nnR$ is then
  also $M$-Lipschitz, since for $\omega, \omega'\in \Dst(X)$,
\[ \begin{array}{rcl}
\big|\,\omega\models q - \omega' \models q\,\big|
& = &
M\cdot \big|\,\omega\models \frac{1}{M}\cdot q - 
   \omega' \models \frac{1}{M}\cdot q\,\big|
\\[+0.2em]
& \leq &
M \cdot \displaystyle\bigvee_{p\in \FactSh(X)} \big|\, \omega\models p -
   \omega'\models p \,\big|
\hspace*{\arraycolsep}=\hspace*{\arraycolsep}
M\cdot d\big(\omega, \omega'\big).
\end{array} \]


\item The only coupling of $1\ket{x}, \omega\in\Dst(X)$ is
  $1\ket{x}\otimes\omega\in\Dst(X\times X)$. Hence:
\[ \begin{array}{rcccccl}
d\big(1\ket{x}, \omega\big)
& = &
1\ket{x}\otimes\omega\models d_{X}
& = &
\displaystyle\sum_{x'\in X} \, \omega(x')\cdot d_{X}(x,x')
& = &
\omega \models d_{X}(x,-).
\end{array} \]

\item We first note that for a distribution of distributions
  $\Omega\in\Dst^{2}(X)$ and a short factor $p\colon X \rightarrow
  \nnR$ the validity in $\Omega$ of the short validity factor
  $(-)\models p \colon \Dst(X) \rightarrow \nnR$ from
  item~\eqref{DstMetricLemVal} satisfies:
\[ \begin{array}{rcl}
\Omega \models \big((-)\models p\big)
\hspace*{\arraycolsep}=\hspace*{\arraycolsep}
\displaystyle\sum_{\omega\in\Dst(X)}\, \Omega(\omega) \cdot 
   \big(\omega\models p\big)
& = &
\displaystyle\sum_{\omega\in\Dst(X)}\,\sum_{x\in X}\, \Omega(\omega) \cdot 
   \omega(x)\cdot p(x)
\\[+1.2em]
& \smash{\stackrel{\eqref{DstMonadEqn}}{=}} &
\displaystyle\sum_{x\in X}\, \flatten(\Omega)(x)\cdot p(x)
\hspace*{\arraycolsep}=\hspace*{\arraycolsep}
\flatten(\Omega)\models p.
\end{array} \]

\noindent Thus for $\Omega,\Omega'\in\Dst^{2}(X)$,
\[ \begin{array}{rcll}
\lefteqn{d_{\Dst(X)}\Big(\flatten(\Omega), \flatten(\Omega')\Big)}
\\[+0.2em]
& = &
\displaystyle \bigvee_{p\in \FactSh(X)} \, 
   \big|\, \flatten(\Omega) \models p \,-\, \flatten(\Omega') \models p \,\big|
\\[+1.2em]
& = &
\displaystyle \bigvee_{p\in \FactSh(X)} \, 
   \big|\, \Omega \models \big((-)\models p\big) \,-\, 
   \Omega' \models \big((-)\models p\big) \,\big| \quad
   & \mbox{as just shown}
\\[+1.2em]
& \leq &
\displaystyle \bigvee_{Q\in \FactSh(\Dst(X))} \, 
   \big|\, \Omega \models Q \,-\, 
   \Omega' \models Q \,\big|
   & \mbox{by item~\eqref{DstMetricLemVal}}
\\[+1.2em]
& = &
d_{\Dst^{2}(X)}\big(\Omega, \Omega'\big).
\end{array} \]

\item Directly by points~\eqref{DstMetricLemFun}
  and~\eqref{DstMetricLemFlatten}.

\item The channel composite $d \klafter c = \flatten \after \Dst(d)
  \after c$ consists of a functional composite of $M$-Lipschitz,
  $L$-Lipschitz, and $1$-Lipschitz maps, and is thus $(M\cdot L\cdot
  1)$-Lipschitz.  This uses items~\eqref{DstMetricLemFun}
  and~\eqref{DstMetricLemFlatten}.

\item If we have couplings $\tau_{i}$ for $\omega_{i},\omega'_{i}$,
  then $\sum_{i}r_{i}\cdot\tau_{i}$ is a coupling of
  $\sum_{i}r_{i}\cdot\omega_{i}$ and $\sum_{i}r_{i}\cdot\omega'_{i}$.
  Moreover:
\[ \begin{array}{rcccl}
d\Big(\sum_{i}r_{i}\cdot\omega_{i}, \, 
   \sum_{i} r_{i}\cdot\omega'_{i}\Big)
& \leq &
\Big(\sum_{i}r_{i}\cdot\tau_{i}\Big) \models d_{X}
& = &
\sum_{i}r_{i}\cdot \Big(\tau_{i} \models d_{X}\Big).
\end{array} \]

\noindent Since this holds for all $\tau_{i}$, we get:
$d\big(\sum_{i}r_{i}\cdot\omega_{i}, \, \sum_{i}
r_{i}\cdot\omega'_{i}\big) \leq \sum_{i}r_{i}\cdot d\big(\omega_{i},
\omega'_{i}\big)$.

\auxproof{
Alternatively:
\[ \begin{array}{rcl}
d\Big(\sum_{i}r_{i}\cdot\omega_{i}, \, 
   \sum_{i} r_{i}\cdot\omega'_{i}\Big)
& = &
\displaystyle\bigvee_{p\in \FactSh(X)} \big|\, \textstyle
   \big(\sum_{i}r_{i}\cdot\omega_{i}\big)\models p -
   \big(\sum_{i}r_{i}\cdot\omega'_{i}\big)\models p \,\big|
\\[+1.2em]
& = &
\displaystyle\bigvee_{p\in \FactSh(X)} \big|\, \textstyle
   \sum_{i}r_{i}\cdot\big(\omega_{i}\models p\big) -
   \sum_{i}r_{i}\cdot\big(\omega'_{i}\models p\big) \,\big|
\\[+1.2em]
& = &
\displaystyle\bigvee_{p\in \FactSh(X)} \big|\, \textstyle
   \sum_{i}r_{i}\cdot\Big(\omega_{i}\models p -
   \omega'_{i}\models p\Big) \,\big|
\\[+1.2em]
& \leq &
\displaystyle\bigvee_{p\in \FactSh(X)} \textstyle \sum_{i}r_{i}\cdot\big|\, 
   \omega_{i}\models p -
   \omega'_{i}\models p \,\big|
\\[+1.2em]
& = &
\sum_{i}r_{i}\cdot\displaystyle\bigvee_{p\in \FactSh(X)} \big|\, 
   \omega_{i}\models p -
   \omega'_{i}\models p \,\big|
\\[+1.2em]
& = &
\sum_{i}r_{i}\cdot d\big(\omega_{i}, \omega'_{i}\big).
\end{array} \]
}

\item We unfold the definition of the $\perm$ map
  from~\eqref{TranspositionEqn} and use the previous item in the first
  step below. We also use that the distance between two sequences is
  invariant under permutation (of both).
\[ \begin{array}[b]{rcl}
d_{\Dst(X^{K})}\big(\perm(\vec{x}), \perm(\vec{y})\big)
& \leq &
\displaystyle\sum_{t\colon\! K\, \stackrel{\cong}{\rightarrow}\, K}
   \frac{1}{K!} \cdot d_{\Dst(X^{K})}\big(1\bigket{\underline{t}(\vec{x})},
   1\bigket{\underline{t}(\vec{y})}\big)
\\[+1em]
& = &
\displaystyle\sum_{t\colon\! K\, \stackrel{\cong}{\rightarrow}\, K}
   \frac{1}{K!} \cdot d_{X^{K}}\big(\underline{t}(\vec{x}),
   \underline{t}(\vec{y})\big) \qquad\mbox{by item~\eqref{DstMetricLemPoint}}
\\[+1em]
& = &
\displaystyle\sum_{t\colon\! K\, \stackrel{\cong}{\rightarrow}\, K}
   \frac{1}{K!} \cdot d_{X^{K}}\big(\vec{x},\vec{y}\big)
\hspace*{\arraycolsep}=\hspace*{\arraycolsep}
d_{X^{K}}\big(\vec{x},\vec{y}\big).
\end{array} \eqno{\QEDbox} \]
\end{enumerate}
\end{myproof}

\noindent 
Later on we need the following facts about tensors of distributions.

\begin{prop}
\label{DstMetricTensorProp}
Let $X,Y$ be metric spaces, and $K$ be a positive natural number.
\begin{enumerate}
\item \label{DstMetricTensorPropTwo} The tensor map $\otimes \colon
  \Dst(X)\times\Dst(Y) \rightarrow \Dst(X\times Y)$ is an isometry.

\item \label{DstMetricTensorPropIID} The $K$-fold tensor map $\iid[K]
  \colon \Dst(X) \rightarrow \Dst(X^{K})$, given by $\iid[K](\omega)
  \coloneqq \omega^{K} = \omega \otimes \cdots \otimes \omega$, is
  $K$-Lipschitz. Actually, there is an equality: $d\big(\iid[K](\omega),
  \iid[K](\rho)\big) = K\cdot d(\omega,\rho)$.
\end{enumerate}
\end{prop}

\begin{myproof}
\begin{enumerate}
\item Let distributions $\omega,\omega'\in\Dst(X)$ and
  $\rho,\rho'\in\Dst(Y)$ be given. For the inequality
  $d_{\Dst(X)\times\Dst(Y)}\big((\omega,\rho), (\omega',\rho')\big)
  \leq d_{\Dst(X\times Y)}\big(\omega\otimes\rho,
  \omega'\otimes\rho'\big)$ one uses that a coupling $\tau \in
  \Dst\big((X\times Y)\times (X\times Y)\big)$ of $\omega\otimes\rho,
  \omega'\otimes\rho'\in \Dst(X\times Y)$ can be turned into two
  couplings $\tau_{1},\tau_{2}$ of $\omega,\omega'$ and of
  $\rho,\rho'$, namely as $\tau_{i} \coloneqq
  \Dst\big(\pi_{i}\times\pi_{i}\big)(\tau)$. For the reverse
  inequality one turns two couplings $\tau_{1},\tau_{2}$ of
  $\omega,\omega'$ and $\rho,\rho'$ into a coupling $\tau$ of
  $\omega\otimes\rho, \omega'\otimes\rho'$ via $\tau \coloneqq
  \Dst\big(\tuple{\pi_{1}\times\pi_{1}, \pi_{2}\times\pi_{2}}\big)
  \big(\tau_{1}\otimes\tau_{2}\big)$.

\auxproof{
Define:
\[ \begin{array}{rcl}
\tau_{1}
& \coloneqq &
\Dst\big(\pi_{1}\times\pi_{1}\big)(\tau) \,\in\, \Dst(X\times X)
\\
\tau_{2}
& \coloneqq &
\Dst\big(\pi_{2}\times\pi_{2}\big)(\tau) \,\in\, \Dst(Y\times Y).
\end{array} \]

\noindent It is not hard to see that $\tau_{1}$ is then a coupling
of $\omega,\omega'$ and $\tau_{2}$ of $\rho,\rho'$. We elaborate
one instance:
\[ \begin{array}{rcl}
\marg{\tau_{1}}{1,0}(x)
\hspace*{\arraycolsep}=\hspace*{\arraycolsep}
\displaystyle\sum_{x'\in X}\, \tau_{1}(x,x')
& = &
\displaystyle\sum_{x'\in X, \, y,y'\in Y} \tau\big((x,y), (x',y')\big)
\\[+1.2em]
& = &
\displaystyle\sum_{y\in Y}\, \marg{\tau}{1,0}(x,y)
\\[+1.2em]
& = &
\displaystyle\sum_{y\in Y}\, \big(\omega\otimes\rho\big)(x,y)
\\[+0.5em]
& = &
\displaystyle \omega(x) \cdot \left(\sum_{y\in Y}\, \rho(y)\right)
\hspace*{\arraycolsep}=\hspace*{\arraycolsep}
\omega(x).
\end{array} \]

\noindent Further,
\[ \begin{array}{rcl}
\lefteqn{d_{\Dst(X)\times\Dst(Y)}\Big((\omega,\rho), (\omega',\rho')\Big)}
\\[+0.2em]
& \smash{\stackrel{\eqref{ProductMetricEqn}}{=}} &
d_{\Dst(X)}(\omega,\omega') + d_{\Dst(Y)}(\rho,\rho')
\\[+0.2em]
& \leq &
\big(\tau_{1} \models d_{X}\big) + \big(\tau_{2} \models d_{Y}\big)
\\[+0.2em]
& = &
\displaystyle\sum_{x,x'\in X} \tau_{1}(x,x')\cdot d_{X}(x,x') \;+\;
   \sum_{y,y'\in Y} \tau_{2}(y,y') \cdot d_{Y}(y,y')
\\[+1.2em]
& = &
\displaystyle\sum_{x,x'\in X, \, y,y'\in Y} 
   \tau\big((x,y), (x',y')\big) \cdot d_{X}(x,x')
\\[+1em]
& & \qquad +\, \displaystyle
   \sum_{x,x'\in X, \, y,y'\in Y} \tau\big((x,y), (x',y')\big) \cdot d_{Y}(y,y')
\\[+1.2em]
& = &
\displaystyle\sum_{x,x'\in X, \, y,y'\in Y} 
   \tau\big((x,y), (x',y')\big) \cdot \Big(d_{X}(x,x') + d_{Y}(y,y')\Big)
\\[+1.2em]
& = &
\displaystyle\sum_{x,x'\in X, \, y,y'\in Y} 
   \tau\big((x,y), (x',y')\big) \cdot d_{X\times Y}\big((x,y), (x',y')\big)
\\[+0.8em]
& = &
\tau \models d_{X\times Y}.
\end{array} \]

\noindent Since this holds for all couplings $\tau$ of
$\omega\otimes\rho$ and $\omega'\otimes\rho'$, one gets:
\[ \begin{array}{rcl}
d_{\Dst(X)\times\Dst(Y)}\Big((\omega,\rho), (\omega',\rho')\Big) 
& \leq &
d_{\Dst(X\times Y)}\Big(\omega\otimes\rho, \omega'\otimes\rho'\Big).
\end{array} \]

For the reverse inequality $\geq$, one starts from two couplings,
$\tau_{1}\in\Dst(X\times X)$ of $\omega,\omega'\in\Dst(X)$ and
$\tau_{2}\in\Dst(Y\times Y)$ of $\rho,\rho'\in\Dst(Y)$. One then forms:
\[ \begin{array}{rcl}
\tau
& \coloneqq &
\Dst\big(\tuple{\pi_{1}\times\pi_{1}, \pi_{2}\times\pi_{2}}\big)
   \Big(\tau_{1}\otimes\tau_{2}\Big) 
   \,\in\, \Dst\big((X\times Y) \times (X\times Y)\big).
\end{array} \]

\noindent One then checks that $\tau$ is a coupling of
$\omega\otimes\rho$ and $\omega'\otimes\rho'$; the inequality $\geq$
is then obtained via reasoning as above. Details are left to the
reader. 

First, we check that $\tau$ is coupling of $\omega\otimes\rho$
and $\omega'\otimes\rho'$, in:
\[ \begin{array}{rcl}
\marg{\tau}{1,0}(x,y)
& = &
\displaystyle\sum_{x'\in X, \, y'\in Y} \tau\big((x,y), (x',y')\big)
\\[+1.2em]
& = &
\displaystyle\sum_{x'\in X, \, y'\in Y} 
   \big(\tau_{1}\otimes\tau_{2}\big)\big((x,x'), (y,y')\big)
\\[+1.2em]
& = &
\displaystyle\left(\sum_{x'\in X} \, \tau_{1}(x,x')\right) \cdot
   \left(\sum_{y'\in Y} \, \tau_{2}(y,y')\right)
\\[+1.2em]
& = &
\marg{\tau_{1}}{1,0}(x) \cdot \marg{\tau_{2}}{1,0}(y)
\\[+0.2em]
& = &
\big(\omega\otimes\rho\big)(x,y).
\end{array} \]

\noindent Further,
\[ \begin{array}{rcl}
\lefteqn{d_{\Dst(X\times Y)}\Big(\omega\otimes\rho, \omega'\otimes\rho'\Big).}
\\[+0.2em]
& = &
\tau \models d_{X\times Y}.
\\[+0.2em]
& = &
\displaystyle\sum_{x,x'\in X, \, y,y'\in Y} 
   \tau\big((x,y), (x',y')\big) \cdot d_{X\times Y}\big((x,x'), (y,y')\Big)
\\[+1.2em]
& = &
\displaystyle\sum_{x,x'\in X, \, y,y'\in Y} 
   \tau\big((x,y), (x',y')\big) \cdot d_{X}(x,x') \;+\;
   \sum_{x,x'\in X, \, y,y'\in Y} \tau\big((x,y), (x',y')\big) \cdot d_{Y}(y,y')
\\[+1.2em]
& = &
\displaystyle\sum_{x,x'\in X} \tau_{1}(x,x')\cdot d_{X}(x,x') \;+\;
   \sum_{y,y'\in Y} \tau_{2}(y,y') \cdot d_{Y}(y,y')
\\[+0.8em]
& = &
\big(\tau_{1} \models d_{X}\big) + \big(\tau_{2} \models d_{Y}\big).
\end{array} \]

\noindent Since this holds for all couplings $\tau_{1}$ and $\tau_{2}$ we
get:
\[ \begin{array}{rcl}
d_{\Dst(X)\times\Dst(Y)}\Big((\omega,\rho), (\omega',\rho')\Big) 
& \geq &
d_{\Dst(X\times Y)}\Big(\omega\otimes\rho, \omega'\otimes\rho'\Big).
\end{array} \]
}

\item For $\omega,\rho\in\Dst(X)$ and $K\in\NNO$, using the
previous item, we get:
\[ \begin{array}{rcccl}
d_{\Dst(X^{K})}\big(\omega^{K}, \rho^{K}\big)
& \smash{\stackrel{\eqref{DstMetricTensorPropTwo}}{=}} &
d_{\Dst(X)^{K}}\Big((\omega,\ldots,\omega), (\rho,\ldots,\rho)\Big)
& \smash{\stackrel{\eqref{ProductMetricEqn}}{=}} &
K\cdot d_{\Dst(X)}\big(\omega,\rho\big).
\end{array} \eqno{\QEDbox} \]
\end{enumerate}
\end{myproof}

\noindent 
For the next result we use for an arbitrary set $Y$ and natural number
$K > 0$ the set $\Dst[K](Y)$ of `fractional distributions' obtained
from multisets of size $K$:
\[ \begin{array}{rcl}
\Dst[K](Y)
& \coloneqq &
\set{\flrn(\varphi)}{\varphi\in\Mlt[K](Y)}.
\end{array} \]

\begin{fact}
\label{FractionalDstFact}
For fractional distributions $\omega, \omega'\in\Dst[K](X)$ over a
metric space $X$ there is an optimal coupling $\tau\in\Dst[K](X\times
X)$ with $d(\omega,\omega') = \tau\models d_{X}$.
\end{fact}

The fact that this coupling $\tau$ is also fractional, for the same
number $K$, can be seen by inspecting the simplex algorithm that is
standardly used in linear programming, see
\textit{e.g.}~\cite{MatouekG06}. The details of this go beyond the
scope of this paper. Since this fact is given without proof we shall
make explicit which results depend on it. 

We illustrate that these matters are subtle.

\begin{exa}
\label{CouplingFractionalEx}
Consider the set $X = \{1,2,3\}$ with the following two distributions
$\omega,\omega'\in\Dst[4](X)$, namely: $\omega = \frac{3}{4}\ket{1} +
\frac{1}{4}\ket{2}$ and $\omega' = \frac{1}{2}\ket{1} +
\frac{1}{4}\ket{2} + \frac{1}{4}\ket{3}$. We consider the following
two couplings of $\omega,\omega'$
\[ \begin{array}{rcl}
\tau
& = &
\frac{1}{2}\bigket{1,1} + \frac{1}{4}\bigket{1,2} + 
   \frac{1}{4}\bigket{2,3}
\\[+0.2em]
\sigma
& = &
\frac{1}{2}\bigket{1,1} + \frac{1}{8}\bigket{1,2} + 
   \frac{1}{8}\bigket{1,3} + \frac{1}{8}\bigket{2,2} + 
   \frac{1}{8}\bigket{2,3}
\end{array} \]

\noindent The first coupling $\tau$ is in $\Dst[4](X\times X)$ and
forms an optimal coupling, giving rise to $d(\omega,\omega') =
\frac{1}{2}$.  Interestingly, $\sigma$ is also an optimal coupling of
$\omega,\omega'$ giving rise to the same distance, but inhabits the
set $\Dst[8](X\times X)$.  Explicitly:
\[ \begin{array}{rcl}
\tau \models d_{X}
& = &
\frac{1}{4}\cdot d_{X}(1,2) + \frac{1}{4}\cdot d_{X}(2,3)
\hspace*{\arraycolsep}=\hspace*{\arraycolsep}
\frac{1}{4} \cdot 1 + \frac{1}{4} \cdot 1
\hspace*{\arraycolsep}=\hspace*{\arraycolsep}
\frac{1}{2}.
\\[+0.2em]
\sigma \models d_{X}
& = &
\frac{1}{8}\cdot d_{X}(1,2) + \frac{1}{8}\cdot d_{X}(1,3) +
   \frac{1}{8}\cdot d_{X}(2,3)
\hspace*{\arraycolsep}=\hspace*{\arraycolsep}
\frac{1}{8} \cdot 1 + \frac{1}{8} \cdot 2 + \frac{1}{8} \cdot 1
\hspace*{\arraycolsep}=\hspace*{\arraycolsep}
\frac{1}{2}.
\end{array} \]

\noindent The conclusion is that, in general, if $\tau\in\Dst(X\times
X)$ is an optimal coupling of $\omega, \omega'\in\Dst[K](X)$, then
$K\cdot\tau$ need not be a multiset of (size $K$).
Fact~\ref{FractionalDstFact} says that there is at least one optimal
coupling $\tau$ such that $K\cdot\tau$ does form a multiset.
\end{exa}

\section{Kantorovich distance and total variation distance}\label{TvdSec}

In this section we relate the so-called total variation distance
between distributions to the Kantorovich distance, in two different
ways.
\begin{itemize}
\item The total variation distance is an instance of the Kantorovich
  distance, for discrete metric spaces, see
  Proposition~\ref{TvdDiscreteProp}.

\item The total variation distance forms a (scaled) upperbound for the
  Kantorovich distance, see
  Proposition~\ref{KantorovichTotalVariationProp}.
\end{itemize}

\noindent Both facts are familiar from the literature, but proofs are
scattered around or left implicit for the discrete case. We include
the details for the record and for convenience of the reader. The
second point will be useful to prove further (limit) properties of the
Kantorovich distance, via the total variation distance. The latter
distance is easier to calculate and reason about.
\vskip0.5\baselineskip
\noindent 
We first recall the \emph{total variation distance} $\tvd$ for
distributions $\omega,\omega'\in\Dst(X)$, on a set $X$.
\begin{equation}
\label{TvdEqn}
\begin{array}{rcl}
\tvd(\omega, \omega')
& \coloneqq &
\frac{1}{2}\displaystyle\sum_{x\in X} \, \big|\,\omega(x) - \omega'(x)\,\big|.
\end{array}
\end{equation}

\noindent Since $\big|\,\omega(x) - \omega'(x)\,\big| \leq
\big|\omega(x)\big| + \big|\omega'(x)\big| = \omega(x) + \omega'(x)$ this
distance is 1-bounded, with values in the unit interval $[0,1]$. This
total variation distance is quite common, see
\emph{e.g.}~\cite{JacobsW20} and the references given there.

We first introduce a reformulation that is convenient to reason about
the total variation distance.

\begin{lem}
\label{TvdLem}
Fix two distributions $\omega,\omega'\in\Dst(X)$ and write:
\[ \left\{\begin{array}{rcl}
X_{>}
& \coloneqq &
\setin{x}{X}{\omega(x)>\omega'(x)}
\\
X_{=}
& \coloneqq &
\setin{x}{X}{\omega(x)=\omega'(x)}
\\
X_{<}
& \coloneqq &
\setin{x}{X}{\omega(x)<\omega'(x)}
\end{array}\right.
\qquad\qquad
\left\{\begin{array}{rcccl}
\upsum{X}
& \coloneqq &
\displaystyle\sum_{x\in X_{>}} \, \omega(x) - \omega'(x)
& \geq &
0
\\[+1.2em]
\downsum{X}
& = &
\displaystyle\sum_{x\in X_{<}} \, \omega'(x) - \omega(x)
& \geq &
0.
\end{array}\right. \]

\noindent Then: $\upsum{X} = \downsum{X} = \tvd(\omega,\omega')$.
\end{lem}

The sets $X_{>}, X_{=}, X_{<}$ and the numbers $\upsum{X},\downsum{X}$
depend on the distributions $\omega,\omega'$ but this is left implicit
for simplicity.

\begin{myproof}
We have both:
\[ \begin{array}{rcccl}
1
& = &
\displaystyle\sum_{x\in X}\, \omega(x)
& = &
\displaystyle\sum_{x\in X_{>}} \, \omega(x) +
\sum_{x\in X_{=}} \, \omega(x) +
\sum_{x\in X_{<}} \, \omega(x)
\\
1
& = &
\displaystyle\sum_{x\in X}\, \omega'(x)
& = &
\displaystyle\sum_{x\in X_{>}} \, \omega'(x) +
\sum_{x\in X_{=}} \, \omega'(x) +
\sum_{x\in X_{<}} \, \omega'(x).
\end{array} \]

\noindent Subtraction yields $0 = \upsum{X} - \downsum{X}$, so that
$\upsum{X} = \downsum{X}$. Further:
\[ \begin{array}[b]{rcl}
\tvd(\omega, \omega')
\hspace*{\arraycolsep}=\hspace*{\arraycolsep}
\frac{1}{2}\displaystyle\sum_{x\in X} \, \big|\,\omega(x) - \omega'(x)\,\big|
& = &
\frac{1}{2}\displaystyle\left(\sum_{x\in X_{>}} \, \omega(x) - \omega'(x)
   + \sum_{x\in X_{<}} \, \omega'(x) - \omega(x)\right)
\\[+1.4em]
& = &
\frac{1}{2}\Big(\upsum{X} + \downsum{X}\!\Big)
\hspace*{\arraycolsep}=\hspace*{\arraycolsep}
\frac{1}{2}\Big(\upsum{X} + \upsum{X}\!\Big)
\hspace*{\arraycolsep}=\hspace*{\arraycolsep}
\upsum{X}
\hspace*{\arraycolsep}=\hspace*{\arraycolsep}
\downsum{X}.
\end{array} \eqno{\QEDbox} \]
\end{myproof}

\noindent 
Each set $X$ forms a `discrete' metric with distance $d_{=} \colon X
\times X \rightarrow [0,1] \subseteq \nnR$ given by:
\begin{equation}
\label{DiscreteDistEqn}
\begin{array}{rcl}
d_{=}(x,x')
& \coloneqq &
\begin{cases}
0 & \mbox{if }x=x'
\\
1 & \mbox{if }x\neq x'.
\end{cases}
\end{array}
\end{equation}

\begin{prop}
\label{TvdDiscreteProp}
For each set $X$ with distributions $\omega,\omega'\in\Dst(X)$ one has:
\begin{equation}
\label{TvdDiscreteEqn}
\begin{array}{rcccl}
\tvd(\omega,\omega')
& \;= &
\displaystyle\bigwedge_{\tau\in\decouple^{-1}(\omega,\omega')} \, \tau\models d_{=}
& \;=\; &
d(\omega,\omega').
\end{array}
\end{equation}

\noindent The latter distance $d$ is the Kantorovich distance for
distributions on the discrete metric space $(X,d_{=})$.
\end{prop}

\begin{myproof}
Let $\tau$ be an arbitrary coupling of $\omega,\omega'$. Then:
\[ \begin{array}{rcl}
\omega(x)
\hspace*{\arraycolsep}=\hspace*{\arraycolsep}
\displaystyle\sum_{y\in X}\, \tau(x,y)
& = &
\displaystyle \tau(x,x) + \sum_{y\neq x}\, \tau(x,y)
\\
& \leq &
\displaystyle \sum_{z\in X} \, \tau(z,x) + \sum_{y\neq x}\, \tau(x,y)
\hspace*{\arraycolsep}=\hspace*{\arraycolsep}
\displaystyle \omega'(x) + \sum_{y\neq x}\, \tau(x,y).
\end{array} \]

\noindent Thus, using Lemma~\ref{TvdLem} we have:
\[ \begin{array}{rcl}
\tvd(\omega, \omega')
\hspace*{\arraycolsep}=\hspace*{\arraycolsep}
\upsum{X}
\hspace*{\arraycolsep}=\hspace*{\arraycolsep}
\displaystyle \sum_{x\in X_{>}} \, \omega(x) - \omega'(x)
& \leq &
\displaystyle \sum_{x\in X_{>}} \, \sum_{y\neq x}\, \tau(x,y)
\\[+1.2em]
& \leq &
\displaystyle \sum_{x\in X} \, \sum_{y\in X}\, \tau(x,y)\cdot d_{=}(x,y)
\hspace*{\arraycolsep}=\hspace*{\arraycolsep}
\tau \models d_{=}.
\end{array} \]

\noindent Since this holds for all couplings $\tau$ we get the inequality
$(\leq)$ in~\eqref{TvdDiscreteEqn}.

For the inequality $(\geq)$ we may assume $\omega\neq\omega'$ and so
$\tvd(\omega,\omega') \neq 0$. One uses the function $\rho\colon
X\times X \rightarrow \nnR$ defined as:
\begin{equation}
\label{OptimalCouplingEqn}
\begin{array}{rcl}
\rho(x,y)
& \coloneqq &
\begin{cases}
\min\big(\omega(x), \omega'(x)\big) & \mbox{if }x=y
\\[+0.5em]
\displaystyle\frac{(\omega(x) \monus \omega'(x)) \cdot
   (\omega'(y) \monus \omega(y))}{\tvd(\omega,\omega')} 
   & \mbox{otherwise,}
\end{cases}
\end{array}
\end{equation}

\noindent where $a \monus b \coloneqq \max(a - b, 0)$, for
$a,b\in\R$.  This formula occurs for instance in~\cite[proof of
  Thm.~6.15]{Villani09} (for the continuous case).

We first check that this $\rho$ is a coupling of $\omega,\omega'$.
Let $x\in X_{>}$ so that $\omega(x) > \omega'(x)$; then:
\[ \begin{array}{rcl}
\displaystyle\sum_{y\in X}\, \rho(x,y)
& = &
\min\big(\omega(x), \omega'(x)\big) + 
   (\omega(x) \monus \omega'(x)) \cdot \displaystyle\sum_{y\neq x} 
   \frac{\omega'(y) \monus \omega(y)}{\tvd(\omega,\omega')}
\\[+1.2em]
& = &
\omega'(x) + (\omega(x) - \omega'(x)) \cdot \displaystyle
   \frac{\sum_{y\in X_{<}} \omega'(y) - \omega(y)}
   {\tvd(\omega,\omega')}
\\[+0.8em]
& = &
\omega'(x) + (\omega(x) - \omega'(x)) \cdot \displaystyle
   \frac{X\downsum}{\tvd(\omega,\omega')}
\\[+0.6em]
& = &
\omega'(x) + (\omega(x) - \omega'(x)) \cdot 1
    \qquad\qquad \mbox{see Lemma~\ref{TvdLem}}
\\
& = &
\omega(x).
\end{array} \]

\noindent If $x\not\in X_{>}$, so that $\omega(x) \leq \omega'(x)$,
then it is obvious that $\sum_{y} \rho(x,y) = \min\big(\omega(x),
\omega'(x)\big) + 0 = \omega(x)$. This shows that the first marginal
$\Dst(\pi_{1})(\rho)$ is $\omega$. In a similar way one obtains
$\Dst(\pi_{2})(\rho) = \omega'$. In particular, we can conclude that
$\rho$ is a distribution.

Finally,
\[ \begin{array}{rcl}
1 - \big(\rho\models d_{=}\big)
\hspace*{\arraycolsep}=\hspace*{\arraycolsep}
\displaystyle\sum_{x,y} \rho(x,y) - \sum_{x\neq y} \rho(x,y)
& = &
\displaystyle\sum_{x\in X}\, \rho(x,x)
\\
& = &
\displaystyle\sum_{x\in X}\,\min\big(\omega(x), \omega'(x)\big)
\\[+1.0em]
& = &
\displaystyle\sum_{x\in X_{>}}\, \omega'(x) \;+\; 
   \sum_{x\not\in X_{>}} \omega(x)
\\
& = &
\displaystyle\sum_{x\in X_{>}}\, \omega'(x) + 1 -
   \displaystyle\sum_{x\in X_{>}}\, \omega(x) 
\\
& = &
\displaystyle 1 - \sum_{x\in X_{>}}\, \omega(x) - \omega'(x)
\\
& = &
1 - \upsum{X}
\hspace*{\arraycolsep}=\hspace*{\arraycolsep}
1 - \tvd(\omega,\omega').
\end{array} \]

\noindent Hence $\tvd(\omega, \omega') \,=\, \rho \models d_{=} \,\geq\,
\bigwedge_{\tau\in\decouple^{-1}(\omega,\omega')} \, \tau\models
d_{=}$.  In particular $\rho$ in~\eqref{OptimalCouplingEqn} is an
optimal coupling.
\end{myproof}\QED

\noindent 
For the next result we need the following notion. The diameter
$\diameter(U)$ of a subset $U\subseteq X$ of a metric space $X$, if it
exists, is the supremum:
\begin{equation}
\label{DiameterEqn}
\begin{array}{rcl}
\diameter(U)
& \coloneqq &
\displaystyle\bigvee_{x,x'\in U} d_{X}(x,x').
\end{array}
\end{equation}

\noindent The subset $U$ is called bounded if this join exists. This
is the case when $U$ is finite, as in the proposition below. This
diameter is used to show that the total variation distance forms a
scaled upperbound for the Kantorovich distance. The continuous version
of this result occurs in~\cite[6.16]{Villani09}.

\begin{prop}
\label{KantorovichTotalVariationProp}
Let $X$ be a metric space, with distributions $\omega,\omega'\in\Dst(X)$.
The Kantorovich distance $d$ and the total variation distance $\tvd$
are related via the following inequality.
\begin{equation}
\label{KantorovichTotalVariationIneq}
\begin{array}{rcl}
d\big(\omega,\omega'\big)
& \leq &
D\cdot \tvd\big(\omega,\omega'\big),
\end{array}
\end{equation}

\noindent where: 
\[ \begin{array}{rcl}
D 
& \coloneqq &
\max\set{d_{X}(x,x')}{x\in\supp(\omega), x'\in\supp(\omega')}
\end{array} \]

\end{prop}


\begin{myproof}
We use the coupling $\rho\in\Dst(X\times X)$ of $\omega,\omega'$
from~\eqref{OptimalCouplingEqn}. It satisfies, as we have
seen:
\begin{align*}
d\big(\omega,\omega'\big)
\leq \rho \models d_{X} &= \sum_{x,x'\in X} \, \rho(x,x')\cdot d_{X}(x,x') \\
&\leq \sum_{x \neq x'} \, \rho(x,x')\cdot D \\
&\quad \text{since $(x,x')\in\supp(\rho)$ implies
   $x\in\supp(\omega)$ and $x'\in\supp(\omega')$} \\
&= \big(\rho\models d_{=}\big)\cdot D \\
&= D\cdot \tvd\big(\omega,\omega'\big). \qedhere
\end{align*}


\end{myproof}

\section{The Kantorovich distance between multisets}\label{KantorovichMltSec}

There is also a Kantorovich distance between multisets of the same
size. This section recalls the definition and the main results.

\begin{defi}
\label{MltMetricDef}
Let $(X,d_{X})$ be a metric space and $K\in\pNNO$ a positive natural
number. We can turn the metric $d_{X}\colon X\times X \rightarrow
\nnR$ into the \emph{Kantorovich} metric $d\colon \Mlt[K](X)\times
\Mlt[K](X) \rightarrow \nnR$ on multisets (of the same size $K$), via:
\begin{equation}
\label{MltMetricEqn}
\begin{array}{rcl}
d\big(\varphi, \varphi'\big)
& \coloneqq &
\displaystyle\bigwedge_{\tau\in\decouple^{-1}(\varphi, \varphi')}
   \flrn(\tau) \models d_{X}
\\[+1.4em]
& = &
\displaystyle\bigwedge_{\vec{x}\in\acc^{-1}(\varphi), \,
   \vec{y}\in\acc^{-1}(\varphi')} \textstyle 
   \frac{1}{K}\cdot d_{X^{K}}\big(\vec{x}, \vec{y}\big)
\\[+1.4em]
& \smash{\stackrel{\eqref{ProductMetricEqn}}{=}} &
\displaystyle\bigwedge_{\vec{x}\in\acc^{-1}(\varphi), \,
   \vec{y}\in\acc^{-1}(\varphi')}\,
   \sum_{1\leq i\leq K} \textstyle\frac{1}{K}\cdot d_{X}(x_{i}, y_{i}).
\end{array}
\end{equation}

\noindent This distance is defined for $K > 0$. We can extend it
trivially to $K=0$ since there is only one multiset of size $K=0$,
namely the empty multiset, so $\Mlt[0](X) = \{\zero\}$. We set
$d(\zero, \zero) = 0$.
\end{defi}

All meets in~\eqref{MltMetricEqn} are finite and can be computed via
enumeration. Alternatively, one can use linear optimisation.  We give
an illustration below. The equality of the first two formulations is
standard; a proof is included in Appendix~\ref{MltApp}. In the above
formulation~\eqref{MltMetricEqn} the decouple map has type $\decouple
\colon \Mlt[K](X\times X) \rightarrow \Mlt[X]\times\Mlt[X]$.  Thus it
is required that the couplings $\tau$ in~\eqref{MltMetricEqn} are
inhabitants of the set $\Mlt[K](X\times X)$ and are thus multisets,
with natural numbers as multiplicities. The latter requirement is not
so explicit in formulations in the literature. It is justified by
Lemma~\ref{DecoupleLem} in the appendix. The requirement is relevant
and relates to Fact~\ref{FractionalDstFact}, see the proof of
Lemma~\ref{MltMetricLem}~\eqref{MltMetricLemFlrn} below.


\begin{exa}
\label{MltMetricEx}
Consider the following two multisets of size $4$ on the set $X =
\{1,2,3\} \subseteq \NNO$, with standard distance between natural
numbers.
\[ \begin{array}{rclcrcl}
\varphi
& = &
3\ket{1} + 1\ket{2}
& \qquad\qquad &
\varphi'
& = &
2\ket{1} + 1\ket{2} + 1\ket{3}.
\end{array} \]

\noindent An optimal coupling $\tau\in\Mlt[4](X\times X)$ is:
\[ \begin{array}{rcl}
\tau
& = &
2\bigket{1,1} + 1\bigket{1,2} + 1\bigket{2,3}.
\end{array} \]

\noindent The resulting Kantorovich distance $d(\varphi,\varphi')$ is:
\[ \begin{array}{rcccccl}
\flrn(\tau)\models d_{X}
& = &
\frac{1}{2}\cdot d_{X}(1,1) + \frac{1}{4}\cdot d_{X}(1,2) +
   \frac{1}{4}\cdot d_{X}(2,3)
& = &
\frac{1}{4} \cdot 1 + \frac{1}{4} \cdot 1
& = &
\frac{1}{2}.
\end{array} \]

\noindent Alternatively, we may proceed as follows. There are
$\coefm{\varphi} = \frac{4!}{3!\cdot 1!} = 4$ lists that accumulate to
$\varphi$, and $\coefm{\varphi'} = \frac{4!}{2!\cdot 1!\cdot 1!} = 12$
lists that accumulate to $\varphi'$. We can align them all and compute
the minimal distance. It is achieved for instance at:
\[ \begin{array}{rcccccl}
\frac{1}{4}\cdot d_{X^4}\Big((1,1,1,2), (1,1,2,3)\Big)
& \smash{\stackrel{\eqref{ProductMetricEqn}}{=}} &
\frac{1}{4}\cdot \big(0 + 0 + 1 + 1\big)
& = &
\frac{2}{4}
& = &
\frac{1}{2}.
\end{array} \]

\noindent As an aside: if we zip these two lists and then accumulate
we recover the coupling multiset $\tau$, in:
\[ \begin{array}{rcl}
\acc\Big(\zip\big((1,1,1,2), (1,1,2,3)\big)\Big)
& = &
\acc\Big((1,1), (1,1), (1,2), (2,3)\Big)
\\[+0.2em]
& = &
2\bigket{1,1} + 1\bigket{1,2} + 1\bigket{2,3}
\hspace*{\arraycolsep}=\hspace*{\arraycolsep}
\tau.
\end{array} \]

\noindent This is a more general phenomenon, see
Lemma~\ref{DecoupleLem}.
\end{exa}

\begin{lem}
\label{MltMetricLem}
We consider the situation in Definition~\ref{MltMetricDef}.
\begin{enumerate}
\item \label{MltMetricLemFlrn} Frequentist learning $\flrn \colon
  \Mlt[K](X) \rightarrow \Dst(X)$ is short, for $K > 0$. It is even an
  isometry by Fact~\ref{FractionalDstFact}.

\item \label{MltMetricLemScalar} For numbers $K, n\geq 1$ the scalar
  multiplication function $n\cdot (-) \colon \Mlt[K](X) \rightarrow
  \Mlt[n\cdot K](X)$ is also short, and an isometry by
  Fact~\ref{FractionalDstFact}.

\item \label{MltMetricLemSum} The sum of distributions $+ \colon
  \Mlt[K](X) \times \Mlt[L](X) \rightarrow \Mlt[K+L](X)$ is short.

\item \label{MltMetricLemFun} If $f\colon X \rightarrow Y$ is
  $M$-Lipschitz, then $\Mlt[K](f) \colon \Mlt[K](X) \rightarrow
  \Mlt[K](Y)$ is $M$-Lipschitz too.  Thus, the fixed size multiset
  functor $\Mlt[K]$ lifts to categories of metric spaces $\MetSh$ and
  $\MetLip$.

\item \label{MltMetricLemAcc} For $K > 0$ the accumulation map
  $\acc\colon X^{K} \rightarrow \Mlt[K](X)$ is
  $\frac{1}{K}$-Lipschitz, and thus short.

\item \label{MltMetricLemArr} The arrangement channel $\arr\colon
  \Mlt[K](X) \chanto X^{K}$ is $K$-Lipschitz; in fact there is an
  equality $d\big(\arr(\varphi), \arr(\varphi')\big) = K\cdot
  d(\varphi,\varphi')$.
\end{enumerate}
\end{lem}

\noindent 
We have already seen an illustration of item~\eqref{MltMetricLemFlrn}:
multisets $\varphi,\varphi'$ in Example~\ref{MltMetricEx} have the
same distance $\frac{1}{2}$ as the distributions $\omega =
\flrn(\varphi)$, $\omega' = \flrn(\varphi')$ in
Example~\ref{CouplingFractionalEx}.

\begin{myproof}
\begin{enumerate}


\item Via naturality of frequentist learning $\flrn \colon \Mlt[K]
  \Rightarrow \Dst$ we obtain that if the multiset
  $\tau\in\Mlt[K](X\times X)$ is a coupling of
  $\varphi,\varphi'\in\Mlt[K](X)$, then the distribution
  $\flrn(\tau)\in\Dst(X\times X)$ is a coupling of $\flrn(\varphi),
  \flrn(\varphi')\in\Dst(X)$. This gives:
\[ \begin{array}{rcccl}
d\big(\flrn(\varphi), \flrn(\varphi')\big)
& = &
\displaystyle\bigwedge_{\sigma\in\decouple^{-1}(\flrn(\varphi),\flrn(\varphi'))} \, 
   \sigma \models d_{X}
& \,\leq\, &
\flrn(\tau) \models d_{X}.
\end{array} \]

\noindent Since this holds for all $\tau$ we get shortness of $\flrn$ in:
\[ \begin{array}{rcccl}
d\big(\flrn(\varphi), \flrn(\varphi')\big)
& \leq &
\displaystyle\bigwedge_{\tau\in\decouple^{-1}(\varphi,\varphi')} \, 
   \flrn(\tau) \models d_{X}
& \,=\, &
d\big(\varphi,\varphi'\big).
\end{array} \]

For the reverse inequality $(\geq)$, let $\sigma\in\Dst(X\times X)$ be
an optimal coupling of $\flrn(\varphi), \flrn(\varphi') \in \Dst[K](X)$.
By Fact~\ref{FractionalDstFact} we may assume $\sigma\in\Dst[K](X\times X)$,
so that $K\cdot\sigma\in\Mlt[K](X\times X)$ This is then a coupling
of $\varphi,\varphi'\in\Mlt[K](X)$. We thus get:
\[ \begin{array}{rcl}
d\big(\flrn(\varphi), \flrn(\varphi')\big)
\hspace*{\arraycolsep}\,=\,\hspace*{\arraycolsep}
\sigma\models d_{X}
& \,=\, &
\flrn(K\cdot\sigma)\models d_{X}
\\
& \geq &
\displaystyle\bigwedge_{\tau\in\decouple^{-1}(\varphi,\varphi')} \, 
   \flrn(\tau) \models d_{X}
\hspace*{\arraycolsep}\,=\,\hspace*{\arraycolsep}
d\big(\varphi,\varphi'\big).
\end{array} \]

\auxproof{
\[ \begin{array}{rcccl}
\Dst\big(\pi_{1}\big)\Big(\flrn(\tau)\Big)
& = &
\flrn\Big(\Mlt(\pi_{1})(\tau)\Big)
& = &
\flrn(\varphi).
\end{array} \]
}
\item If $\tau\in\Mlt[K](X\times X)$ is a coupling of
  $\varphi,\varphi' \in\Mlt[K](X)$, then $n\cdot\tau\in\Mlt[n\cdot
  K](X\times X)$ is obviously a coupling of $n\cdot\varphi,
  n\cdot\varphi' \in \Mlt[n\cdot K](X)$. This gives:
\[ \qquad\begin{array}{rcccccl}
d(n\cdot\varphi, n\cdot\varphi')
& = \! &
\displaystyle\bigwedge_{\sigma\in\decouple^{-1}(n\cdot\varphi, n\cdot\varphi')} \!
   \flrn(\sigma) \models d_{X}
& \,\leq\, &
\flrn(n\cdot\tau) \models d_{X}
& \,=\, &
\flrn(\tau) \models d_{X}.
\end{array} \]

\noindent Since this holds for all $\tau$ we get $d(n\cdot\varphi,
n\cdot\varphi') \leq d(\varphi, \varphi')$. 

For the reverse inequality $(\geq)$, let multisets
$\varphi,\varphi'\in\Mlt[K](X)$ be given. Using that $\flrn$ is an
isometry, via Fact~\ref{FractionalDstFact}, we get:
\[ \begin{array}{rcl}
d_{\Mlt[K](X)}\big(\varphi, \varphi'\big)
& = &
d_{\Dst(X)}\big(\flrn(\varphi), \flrn(\varphi')\big)
\\[+0.2em]
& = &
d_{\Dst(X)}\big(\flrn(n\cdot\varphi), \flrn(n\cdot\varphi')\big)
\hspace*{\arraycolsep}=\hspace*{\arraycolsep}
d_{\Mlt[n\cdot K](X)}\big(n\cdot \varphi, n\cdot\varphi'\big).
\end{array} \]



\item Let $\sigma\in\Mlt[K](X\times X)$ and $\tau \in \Mlt[L](X\times
  X)$ be couplings of multisets $\varphi,\varphi'\in\Mlt[K](X)$ and
  $\psi,\psi'\in\Mlt[L](X)$. The multiset sum
  $\sigma+\tau\in\Mlt[K+L](X\times X)$ is then a coupling of
  $\varphi+\varphi'$ and $\psi+\psi'$. Hence:
\[ \begin{array}{rcl}
d\Big(\varphi+\psi, \, \varphi'+\psi'\Big)
& \leq &
\flrn(\sigma+\tau) \models d_{X}
\\[+0.2em]
& = &
\Big(\frac{K}{K+L}\cdot \flrn(\sigma) + \frac{L}{K+L}\cdot \flrn(\tau)\Big) 
   \models d_{X}
\\[+0.2em]
& = &
\frac{K}{K+L}\cdot \Big(\flrn(\sigma) \models d_{X}\Big) + 
   \frac{L}{K+L}\cdot \Big(\flrn(\tau) \models d_{X}\Big)
\\[+0.2em]
& \leq &
\flrn(\sigma) \models d_{X} \,+\, \flrn(\tau) \models d_{X}.
\end{array} \]

\noindent Since this holds for each coupling $\sigma$ and $\tau$ we
get:
\[ \begin{array}{rcl}
d\Big(\varphi+\psi, \, \varphi'+\psi'\Big)
& \leq &
\displaystyle\bigwedge_{\sigma\in\decouple^{-1}(\varphi,\varphi')} 
   \flrn(\sigma) \models d_{X} \;+
   \bigwedge_{\tau\in\decouple^{-1}(\psi,\psi')} 
   \flrn(\tau) \models d_{X}
\\[+1.4em]
& = &
d\big(\varphi, \varphi'\big) + d\big(\psi, \psi'\big)
\\[+0.2em]
& \smash{\stackrel{\eqref{ProductMetricEqn}}{=}} &
d\Big((\varphi,\psi), (\varphi',\psi')\Big).
\end{array} \]

\auxproof{
For multisets $\varphi,\varphi'\in\Mlt[K](X)$ and
  $\psi,\psi'\in\Mlt[L](X)$, using
  Lemma~\ref{DstMetricLem}~\eqref{DstMetricLemConv},
\[ \begin{array}{rcl}
\lefteqn{d\Big(\varphi+\psi, \, \varphi'+\psi'\Big)}
\\[+0.2em]
& \leq &
d\Big(\flrn(\varphi+\psi), \, \flrn(\varphi'+\psi')\Big)
    \qquad\mbox{by item~\eqref{MltMetricLemFlrn}}
\\[+0.4em]
& = &
d\Big(\frac{K}{K+L}\cdot\flrn(\varphi)+\frac{L}{K+L}\cdot\flrn(\psi),  \,
   \frac{K}{K+L}\cdot\flrn(\varphi')+\frac{L}{K+L}\cdot\flrn(\psi')\Big)
\\[+0.4em]
& \leq &
\frac{K}{K+L}\cdot d\big(\flrn(\varphi), \flrn(\varphi')\big) +
   \frac{L}{K+L}\cdot d\big(\flrn(\psi), \flrn(\psi')\big)
\\[+0.3em]
& \smash{\stackrel{\ref{MltMetricLemFlrn}}{=}} &
\frac{K}{K+L}\cdot d\big(\varphi, \varphi'\big) +
   \frac{L}{K+L}\cdot d\big(\psi, \psi'\big)
\\[+0.3em]
& \leq &
d\big(\varphi, \varphi'\big) + d\big(\psi, \psi'\big)
\\[+0.2em]
& \smash{\stackrel{\eqref{ProductMetricEqn}}{=}} &
d\Big((\varphi,\psi), (\varphi',\psi')\Big).
\end{array} \]
}

\item Let $f\colon X \rightarrow Y$ be $M$-Lipschitz. For sequences $\vec{x}\in\acc^{-1}(\varphi)$, $\vec{x'} \in
\acc^{-1}(\varphi')$ we have, by naturality of accumulation,
\[ \begin{array}{rcccl}
\acc\big(f^{K}(\vec{x})\big)
& = &
\Mlt(f)\big(\acc(\vec{x})\big)
& = &
\Mlt(f)\big(\varphi\big).
\end{array} \]

\noindent This means that $f^{K}(\vec{x}) \in
\acc^{-1}\big(\Mlt(f)\big(\varphi\big)\big)$. Thus:
\[ \begin{array}[b]{rcl}
d\Big(\Mlt(f)\big(\varphi\big), \Mlt(f)\big(\varphi'\big)\Big)
& = &
\frac{1}{K}\cdot\displaystyle
   \bigwedge_{\vec{y}\in\acc^{-1}(\Mlt(f)(\varphi)), \,
   \vec{y'}\in\acc^{-1}((\Mlt(f)(\varphi'))} 
   d_{Y^{K}}\big(\vec{y}, \vec{y'}\big)
\\[+1.4em]
& \leq &
\frac{1}{K}\cdot\displaystyle\bigwedge_{\vec{x}\in\acc^{-1}(\varphi), \,
   \vec{x'}\in\acc^{-1}(\varphi')} 
   d_{Y^{K}}\big(f^{K}(\vec{x}), f^{K}(\vec{x'})\big)
\\[+1.2em]
& \leq &
\frac{1}{K}\cdot\displaystyle\bigwedge_{\vec{x}\in\acc^{-1}(\varphi), \,
   \vec{x'}\in\acc^{-1}(\varphi')} 
   M\cdot d_{X^{K}}\big(\vec{x}, \vec{x'}\big)
\\[+1.4em]
& = &
M\cdot d\big(\varphi, \varphi').
\end{array} \]

\auxproof{
Let $f\colon X \rightarrow Y$ be $M$-Lipschitz. We use that
  frequentist learning $\flrn$ is an isometry and a natural
  transformation $\Mlt[K] \Rightarrow \Dst$. For multisets
  $\varphi,\varphi'\in\Mlt[K](X)$,
\[ \begin{array}{rcll}
\lefteqn{d_{\Mlt[K](Y)}\Big(\Mlt(f)\big(\varphi\big), 
   \Mlt(f)\big(\varphi'\big)\Big)}
\\[+0.2em]
& \smash{\stackrel{\ref{MltMetricLemFlrn}}{=}} &
d_{\Dst(Y)}\Big(\flrn\big(\Mlt(f)\big(\varphi\big)\big), 
   \flrn\big(\Mlt(f)\big(\varphi'\big)\big)\Big)
\\[+0.2em]
& = &
d_{\Dst(Y)}\Big(\Dst(f)\big(\flrn(\varphi)\big), 
   \Dst(f)\big(\flrn(\varphi')\big)\Big)
   \qquad & \mbox{by naturality of $\flrn$}
\\[+0.2em]
& \leq &
M\cdot d_{\Dst(X)}\big(\flrn(\varphi), \flrn(\varphi')\big)
   & \mbox{by Lemma~\ref{DstMetricLem}~\eqref{DstMetricLemFun}}
\\[+0.2em]
& \smash{\stackrel{\ref{MltMetricLemFlrn}}{=}} &
d_{\Mlt[K](X)}\big(\varphi, \varphi'\big).
\end{array} \]
}

\item The map $\acc \colon X^{K} \rightarrow \Mlt[K](X)$ is
  $\frac{1}{K}$-Lipschitz since for $\vec{y}, \vec{y'} \in X^{K}$,
\[ \begin{array}[b]{rcl}
d\Big(\acc(\vec{y}), \acc(\vec{y'})\Big)
& = &
\displaystyle\frac{1}{K}\cdot\bigwedge_{\vec{x}\in\acc^{-1}(\acc(\vec{y})), \,
   \vec{x'}\in\acc^{-1}(\acc(\vec{y'}))} d_{X^{K}}\big(\vec{x}, \vec{x'}\big)
\\
& \leq &
\displaystyle\frac{1}{K}\cdot d_{X^{K}}\big(\vec{y}, \vec{y'}\big).
\end{array} \]

\item For fixed $\varphi, \varphi'\in\Mlt[K](X)$, take arbitrary
  $\vec{x}\in\acc^{-1}(\varphi)$ and $\vec{x'} \in
  \acc^{-1}(\varphi')$. Then:
\[ \begin{array}{rcl}
d_{\Dst(X^{K})}\big(\arr(\varphi), \arr(\varphi')\big)
& = &
d_{\Dst(X^{K})}\big(\arr(\acc(\vec{x})), \arr(\acc(\vec{x'}))\big)
\\
& \smash{\stackrel{\eqref{TranspositionEqn}}{=}} &
d_{\Dst(X^{K})}\big(\perm(\vec{x}), \perm(\vec{x'})\big)
\\
& \leq &
d_{X^{K}}\big(\vec{x}, \vec{x'}\big) \qquad 
   \mbox{by Lemma~\ref{DstMetricLem}~\eqref{DstMetricLemPerm}.}
\end{array} \]

\noindent Since this holds for all sequences
$\vec{x}\in\acc^{-1}(\varphi)$, $\vec{x'} \in \acc^{-1}(\varphi')$ we
get an inequaltiy $d_{\Dst(X^{K})}\big(\arr(\varphi),
\arr(\varphi')\big) \leq K \cdot d_{\Mlt[K](X)}\big(\varphi,
\varphi'\big)$, see Definition~\ref{MltMetricDef}. This inequality is
an actual equality since $\acc$, and thus $\Dst(\acc)$, is
$\frac{1}{K}$-Lipschitz:
\[ \begin{array}[b]{rcl}
d_{\Mlt[K](X)}\big(\varphi, \varphi'\big)
& = &
d_{\Dst(\Mlt[K](X))}\big(1\ket{\varphi}, 1\ket{\varphi'}\big)
\\
& = &
d_{\Dst(\Mlt[K](X))}\Big(\Dst(\acc)\big(\arr(\varphi)\big), 
   \Dst(\acc)\big(\arr(\varphi')\big)\Big)
\\
& \leq &
\frac{1}{K}\cdot d_{\Dst(X^{K})}\big(\arr(\varphi), \arr(\varphi')\Big)
\end{array} \eqno{\QEDbox} \]
\end{enumerate}
\end{myproof}

\noindent 
We recall the multiset-zip operation $\mzip \colon
\Mlt[K](X)\times\Mlt[K](Y) \rightarrow \Dst\big(\Mlt[K](X\times
Y)\big)$ from~\cite{Jacobs21b}. It turns two multisets into a
distribution over couplings of these multisets, via the composite:
\begin{equation}
\label{MzipDiag}
\hspace*{-1em}\vcenter{\xymatrix@C-1.35pc@R-0.5pc{
\Mlt[K](X)\!\times\!\Mlt[K](Y)\ar[rr]^-{\arr\times\arr} & &
   \Dst\big(X^{K}\big)\!\times\! \Dst(Y^{K}\big)\ar[r]^-{\otimes} &
   \Dst\big(X^{K}\!\times\! Y^{K}\big)\ar[d]_-{\Dst(\zip)}^-{\cong}
\\
& & & \Dst\big((X\!\times\! Y)^{K}\big)\ar[rr]^-{\Dst(\acc)} & &
   \Dst\big(\Mlt[K](X\!\times\! Y)\big)
}}
\end{equation}

An example of how $\mzip$ works can be found in Example~\ref{MzipEx}
below, and also in~\cite[Ex.~16]{Jacobs21b}.

\begin{prop}
\label{MzipProp}
Consider the $\mzip$ operation~\eqref{MzipDiag} for metric spaces
$X,Y$.
\begin{enumerate}
\item \label{MzipPropShort} This $\mzip$ function is short.

\item \label{MzipPropIsom} It is even an isometry, via
  Fact~\ref{FractionalDstFact} (in an indirect way, through
  Lemma~\ref{MltMetricLem}~\eqref{MltMetricLemFlrn}).
\end{enumerate}
\end{prop}

\begin{myproof}
\begin{enumerate}
\item As we can see in Diagram~\eqref{MzipDiag}, the composite $\mzip
= \Dst(\acc) \after \Dst(\zip) \after \otimes \after (\arr\times\arr)$
is a composite of:
\begin{itemize}
\item a $K$-Lipschitz function $\arr\times\arr$, by
  Lemma~\ref{MltMetricLem}~\eqref{MltMetricLemArr} and
  Lemma~\ref{MetLipProdLem};

\item a $1$-Lipschitz function $\otimes$, by
  Proposition~\ref{DstMetricTensorProp}~\eqref{DstMetricTensorPropTwo};

\item a $1$-Lipschitz function $\Dst(\zip)$, by Lemma~\ref{ZipLem} and
  Lemma~\ref{DstMetricLem}~\eqref{DstMetricLemFun};

\item a $\frac{1}{K}$-Lipschitz function $\Dst(\acc)$, by
  Lemma~\ref{MltMetricLem}~\eqref{MltMetricLemAcc} and
  Lemma~\ref{DstMetricLem}~\eqref{DstMetricLemFun}.
\end{itemize}

\noindent The composite $\mzip$ is thus Lipschitz with constant
$K\cdot 1\cdot 1\cdot \frac{1}{K} = 1$, making it short.

\item We use the equation $\pushing{\flrn}{\mzip(\varphi,\psi)} =
  \flrn(\varphi\otimes\psi)$, from~\cite[Thm.~18]{Jacobs21b}. Then,
  for multisets $\varphi,\varphi'\in\Mlt[K](X)$ and
  $\psi,\psi'\in\Mlt[K](Y)$ one has:
\[ \begin{array}[b]{rcll}
\lefteqn{d\Big(\mzip(\varphi,\psi), \, \mzip(\varphi',\psi')\Big)}
\\[+0.2em]
& \geq &
d\Big(\pushing{\flrn}{\mzip(\varphi,\psi)}, \, 
   \pushing{\flrn}{\mzip(\varphi',\psi')}\Big)\quad
   & \mbox{by Lemma~\ref{DstMetricLem}~\eqref{DstMetricLemPush}}
\\[+0.4em]
& = &
d\Big(\flrn(\varphi\otimes\psi), \, \flrn(\varphi'\otimes\psi')\Big)
\\[+0.4em]
& = &
d\Big(\flrn(\varphi)\otimes\flrn(\psi), \, 
   \flrn(\varphi')\otimes\flrn(\psi')\Big)
   & \mbox{by~\eqref{FlrnTensorEqn}}
\\[+0.4em]
& = &
d\big(\flrn(\varphi), \flrn(\varphi')\big) + 
   d\big(\flrn(\psi), \flrn(\psi')\big) \quad
   & \mbox{by Proposition~\ref{DstMetricTensorProp}~\eqref{DstMetricTensorPropTwo}}
\\[+0.4em]
& = &
d\big(\varphi, \varphi'\big) + d\big(\psi, \psi'\big)
  & \mbox{by Lemma~\ref{MltMetricLem}~\eqref{MltMetricLemFlrn}}
\\[+0.2em]
& = &
d\Big((\varphi,\psi), \, (\varphi',\psi')\Big)
   & \mbox{by~\eqref{ProductMetricEqn}.}
\end{array} \eqno{\QEDbox} \]
\end{enumerate}
\end{myproof}

\noindent 
A central operation from~\cite{Jacobs21b} is the distributive law
$\pml \colon \Mlt[K]\big(\Dst(X)\big) \rightarrow
\Dst\big(\Mlt[K](X)\big)$, called parallel multinomial law. One way to
describe it is as composite $\pml = \Dst(\acc) \after \bigotimes
\after \arr \colon \Mlt[K]\big(\Dst(X)\big) \rightarrow \Dst(X)^{K}
\rightarrow \Dst\big(X^{K}\big) \rightarrow \Dst\big(\Mlt[K](X)\big)$.
The map in the middle is the $K$-ary tensor $\bigotimes \colon
\Dst(X)^{K} \rightarrow \Dst\big(X^{K}\big)$ of distributions ---
which is short, generalising
Proposition~\ref{DstMetricTensorProp}~\eqref{DstMetricTensorPropTwo}. Via
an argument as for $\mzip$ --- in the proof of
Proposition~\ref{MzipProp}~\eqref{MzipPropShort} --- one shows that
$\pml$ is short.

There is one more point we like to make about the multiset-zip
operation $\mzip$.

\begin{exa}
\label{MzipEx}
We shall now use the multiset-zip function~\eqref{MzipDiag} with $Y=X$
and check if the Kantorovich distance between multisets can be
described via this $\mzip$ operation (channel) --- assuming that $X$
is a metric space.

The obvious way to do this would be via the following composite.
\begin{equation}
\label{MzipDistAttempt}
\vcenter{\xymatrix@C+1.2pc{
\Mlt[K](X)\times\Mlt[K](X) \ar[r]^-{\mzip} &
   \Dst\big(\Mlt[K](X\times X)\big)\ar[r]^-{\Dst(-\models d_{X})} &
  \Dst(\nnR)\ar[r]^-{\mathsl{sum}} & \nnR
}}
\end{equation}

\noindent The sum operation $\mathsl{sum} \colon \Dst(\nnR)
\rightarrow \nnR$ is defined in the obvious way as
$\mathsl{sum}\big(\sum_{i} r_{i}\ket{s_i}\big) = \sum_{i} r_{i}\cdot
s_{i}$, exploiting that $\nnR$ is a convex set.

We demonstrate that the approach via~\eqref{MzipDistAttempt} does not
produce the Kantorovich distance, for $X = \{1,2,3\} \subseteq \NNO$
with usual distance and with multisets $\varphi = 3\ket{1} + 2\ket{3}$
and $\varphi' = 1\ket{1} + 3\ket{2} + 1\ket{3}$ of size $5$.  The
Kantorovich distance $d(\varphi,\varphi')$ is $3$. The multiset-zip
gives the following distribution over couplings of $\varphi,\varphi'$.
\[\begin{array}{rcl}
\mzip(\varphi, \varphi')
& \!= \!&
\frac{3}{10}\Bigket{2\ket{1, 2} \msa{+} 1\ket{1, 3} \msa{+} 1\ket{3, 1} \msa{+} 1\ket{3, 2}}
\\[+0.4em]
& & \quad \msa{+}\,
   \frac{3}{10}\Bigket{1\ket{1, 1} \msa{+} 1\ket{1, 2} \msa{+} 1\ket{1, 3} \msa{+} 2\ket{3, 2}}
\\[+0.4em]
& & \quad \msa{+}\,
   \frac{1}{10}\Bigket{3\ket{1, 2} \msa{+} 1\ket{3, 1} \msa{+} 1\ket{3, 3}} \msa{+}
   \frac{3}{10}\Bigket{1\ket{1, 1} \msa{+} 2\ket{1, 2} \msa{+} 1\ket{3, 2} \msa{+} 1\ket{3, 3}}.
\end{array} \]

\noindent By applying the function $(-) \models d_{X}$ to the
couplings inside the big ket's, as in~\eqref{MzipDistAttempt}, one
gets:
\[ \begin{array}{rcccl}
\Dst(\mathsl{sum})\Big(\frac{3}{10}\bigket{7} + \frac{3}{10}\bigket{5} + 
   \frac{1}{10}\bigket{5} + \frac{3}{10}\bigket{3}\Big)
& = &
\frac{50}{10}
& = &
5.
\end{array} \]

\noindent This outcome according to~\eqref{MzipDistAttempt} differs
from Kantorovich distance $d(\varphi,\varphi') = 3$.

\ignore{

X = nspace(3)
K = 5
m1 = random_multiset(K)(X)
m2 = random_multiset(K)(X)
print( m1 )
print( m2 )
print("")
print( distribution_relationlift(m1, m2, EuclideanPred)[0] )
print("")
DP = DPred(lambda m: m >= EuclideanPred, multiset_space(K)(X @ X))
print( Mzip(m1, m2, frac=True) )
print("")
print( Mzip(m1, m2) >= DP )

3|1> + 2|3>
1|1> + 3|2> + 1|3>

3.0

3/10|2|1, 2> + 1|1, 3> + 1|3, 1> + 1|3, 2>> + 
   3/10|1|1, 1> + 1|1, 2> + 1|1, 3> + 2|3, 2>> + 
   1/10|3|1, 2> + 1|3, 1> + 1|3, 3>> + 
   3/10|1|1, 1> + 2|1, 2> + 1|3, 2> + 1|3, 3>>

5.000000000000003

}

\end{exa}

\section{Multinomial drawing is isometric}\label{MulnomSec}

At this stage we have prepared the ground so that we are able to show
that the three draw channels --- multionomial $\multinomial[K]$,
hypergeometric $\hypergeometric[K]$ and P\'olya $\polya[K]$, for a
fixed drawsize $K$ --- are all isometric. The equality of metrics
involved in an isometry will be split in two inequalities, where one
is for shortness. The other inequality, for each of the drawing modes,
will be proven in basically the same manner. Each of the drawing
channels interacts well with frequentist learning, in the following
manner:
\begin{equation}
\label{FlrnDrawEqn}
\begin{array}{rclcrclcrcl}
\pushing{\flrn}{\multinomial[K](\omega)}
& = &
\omega
& \quad &
\pushing{\flrn}{\hypergeometric[K](\upsilon)}
& = &
\flrn(\upsilon)
& \quad &
\pushing{\flrn}{\polya[K](\upsilon)}
& = &
\flrn(\upsilon).
\end{array}
\end{equation}

\noindent With these equations we can already prove, without having
seen the details of $\multinomial[K]$, $\hypergeometric[K]$ or
$\polya[K]$, via Lemma~\ref{MltMetricLem}~\eqref{MltMetricLemFlrn},
including Fact~\ref{FractionalDstFact}, and
Lemma~\ref{DstMetricLem}~\eqref{DstMetricLemPush}:
\begin{equation}
\label{FlrnDrawProof}
\begin{array}{rcl}
d\big(\omega,\omega'\big)
& \smash{\stackrel{\eqref{FlrnDrawEqn}}{=}} &
d\big(\pushing{\flrn}{\multinomial[K](\omega)}, 
   \pushing{\flrn}{\multinomial[K](\omega')}\big)
\\[+0.2em]
& \leq &
d\big(\multinomial[K](\omega), \multinomial[K](\omega')\big)
\\[+0.2em]
d\big(\upsilon,\upsilon'\big)
& = &
d\big(\flrn(\upsilon), \flrn(\upsilon')\big)
\\[+0.2em]
& \smash{\stackrel{\eqref{FlrnDrawEqn}}{=}} &
d\big(\pushing{\flrn}{\hypergeometric[K](\upsilon)}, 
   \pushing{\flrn}{\hypergeometric[K](\upsilon')}\big)
\\[+0.2em]
& \leq &
d\big(\hypergeometric[K](\upsilon), \hypergeometric[K](\upsilon')\big)
\\[+0.2em]
d\big(\upsilon,\upsilon'\big)
& = &
d\big(\flrn(\upsilon), \flrn(\upsilon')\big)
\\[+0.2em]
& \smash{\stackrel{\eqref{FlrnDrawEqn}}{=}} &
d\big(\pushing{\flrn}{\polya[K](\upsilon)}, 
   \pushing{\flrn}{\polya[K](\upsilon')}\big)
\\[+0.2em]
& \leq &
d\big(\polya[K](\upsilon), \polya[K](\upsilon')\big).
\end{array}
\end{equation}

\noindent Hence in the next three sections, we shall prove for the
multinomial, hypergeometric, and P\'olya draw channel the
corresponding equation in~\eqref{FlrnDrawEqn}. This means that we only
have to prove shortness in order to get isometry, since the reverse
inequality is already handled in~\eqref{FlrnDrawProof}.

\medskip

We start with the multinomial case. Multinomial draws are of the
draw-and-replace kind. This means that a drawn ball is returned to the
urn, so that the urn remains unchanged.  Thus we may use a
distribution $\omega\in\Dst(X)$ as urn, giving for each colour $x\in
X$ the probability $\omega(x)$ of drawing a ball of that colour. For a
drawsize number $K\in\NNO$, the multinomial distribution
$\multinomial[K](\omega) \in \Dst\big(\Mlt[K](X)\big)$ on multisets /
draws of size $K$ can be defined via accumulated sequences of draws:
\begin{equation}
\label{MulnomEqn}
\begin{array}{rcl}
\multinomial[K](\omega)
\hspace*{\arraycolsep}\coloneqq\hspace*{\arraycolsep}
\Dst(\acc)\big(\iid[K](\omega)\big)
& = &
\Dst(\acc)\big(\omega^{K}\big)
\\[+0.2em]
& = &
\displaystyle\sum_{\vec{x}\in X^{K}} \, \omega^{K}\big(\vec{x}\big)
   \, \bigket{\acc(\vec{x})}
\\[+1.0em]
& = &
\displaystyle\sum_{\varphi\in\Mlt[K](X)} \, \sum_{\vec{x}\in \acc^{-1}(\varphi)} \, 
   \prod_{1\leq i\leq K} \omega(x_{i}) \, \bigket{\varphi}
\\[+1.4em]
& = &
\displaystyle\sum_{\varphi\in\Mlt[K](X)}\, \coefm{\varphi}\cdot
   \prod_{x\in X} \omega(x)^{\varphi(x)}\,\bigket{\varphi}.
\end{array}
\end{equation}

\noindent We recall that $\coefm{\varphi} =
\frac{K!}{\prod_{x}\varphi(x)!}$ is the number of sequences that
accumulate to a multiset / draw $\varphi\in\Mlt[K](X)$. 

\begin{exa}
\label{MultinomialIsometryEx}
Consider the following two distributions $\omega,\omega'\in\Dst(\NNO)$.
\[ \begin{array}{rclcrclcrcl}
\omega
& = &
\frac{1}{3}\ket{0} + \frac{2}{3}\ket{2}
& \quad\mbox{and}\quad &
\omega'
& = &
\frac{1}{2}\ket{1} + \frac{1}{2}\ket{2}
& \quad\mbox{with}\quad &
d(\omega,\omega')
& = &
\frac{1}{2}.
\end{array} \]

\noindent This distance $d(\omega,\omega')$ involves the standard
distance on $\NNO$, using as an optimal coupling $\frac{1}{3}\ket{0,
  1} + \frac{1}{6}\ket{2, 1} + \frac{1}{2}\ket{2, 2} \in
\Dst\big(\NNO\times\NNO\big)$.

We take draws of size $K=3$. There are 10 multisets of size $3$ over
$\{0,1,2\}$:
\[ \begin{array}{c}
\varphi_{1} 
=
3\ket{0}
\qquad
\varphi_{2}
=
2\ket{0} + 1\ket{1}
\qquad
\varphi_{3}
=
1\ket{0} + 2\ket{1}
\qquad
\varphi_{4}
=
3\ket{1}
\\
\varphi_{5}
=
2\ket{0} + 1\ket{2}
\qquad
\varphi_{6}
=
1\ket{0} + 1\ket{1} + 1\ket{2}
\qquad
\varphi_{7}
=
2\ket{1} + 1\ket{2}
\\
\varphi_{8}
=
1\ket{0} + 2\ket{2}
\qquad
\varphi_{9}
=
1\ket{1} + 2\ket{2}
\qquad
\varphi_{10}
=
3\ket{2}.
\end{array} \]

\noindent These multisets occur in the following multinomial
distributions of draws of size~3.
\begin{equation}
\label{MultinomialIsometryExDst}
\begin{array}{rcl}
\multinomial[3](\omega)
& = &
\frac{1}{27}\bigket{\varphi_{1}} + \frac{2}{9}\bigket{\varphi_{5}} + 
  \frac{4}{9}\bigket{\varphi_{8}} + \frac{8}{27}\bigket{\varphi_{10}}
\\[+0.2em]
\multinomial[3](\omega')
& = &
\frac{1}{8}\bigket{\varphi_{4}} + \frac{3}{8}\bigket{\varphi_{7}} + 
   \frac{3}{8}\bigket{\varphi_{9}} + \frac{1}{8}\bigket{\varphi_{10}}.
\end{array}
\end{equation}

\noindent An optimal coupling $\tau\in\Dst\big(\Mlt[3](\NNO) \times
\Mlt[3](\NNO)\big)$ between these two multinomial distributions is:
\[ \begin{array}{rcl}
\tau
& = &
\frac{1}{27}\Bigket{\varphi_{1}, \varphi_{4}} + 
  \frac{19}{216}\Bigket{\varphi_{5}, \varphi_{4}} +
  \frac{1}{8}\Bigket{\varphi_{10}, \varphi_{10}} +
  \frac{29}{216}\Bigket{\varphi_{5}, \varphi_{7}}
\\[+0.2em]
& & \;\; +\,
  \frac{5}{72}\Bigket{\varphi_{8}, \varphi_{7}} +
  \frac{3}{8}\Bigket{\varphi_{8}, \varphi_{9}} + 
  \frac{37}{216}\Bigket{\varphi_{10}, \varphi_{7}}.
\end{array} \]

\noindent We compute the distance between the multinomial
distributions, using $d_{\Mlt} = d_{\Mlt[3](\NNO)}$.
\[ \begin{array}{rcl}
\lefteqn{d\big(\multinomial[3](\omega), \multinomial[3](\omega')\big)
\hspace*{\arraycolsep}=\hspace*{\arraycolsep}
\tau \models d_{\Mlt}}
\\[+0.2em]
& = &
\frac{1}{27}\cdot d_{\Mlt}\big(\varphi_{1}, \varphi_{4}\big) + 
  \frac{19}{216}\cdot d_{\Mlt}\big(\varphi_{5}, \varphi_{4}\big) +
  \frac{1}{8}\cdot d_{\Mlt}\big(\varphi_{10}, \varphi_{10}\big) +
  \frac{29}{216}\cdot d_{\Mlt}\big(\varphi_{5}, \varphi_{7}\big)
  
\\[+0.2em]
& & \;\; +\,
  \frac{5}{72}\cdot d_{\Mlt}\big(\varphi_{8}, \varphi_{7}\big) +
  \frac{3}{8}\cdot d_{\Mlt}\big(\varphi_{8}, \varphi_{9}\big) + 
  \frac{37}{216}\cdot d_{\Mlt}\big(\varphi_{10}, \varphi_{7}\big)
\\[+0.2em]
& = &
\frac{1}{27}\cdot 1 + 
  \frac{19}{216}\cdot 1 +
  \frac{1}{8}\cdot 0 +
  \frac{29}{216}\cdot \frac{2}{3} +
  \frac{5}{72}\cdot \frac{2}{3} +
  \frac{3}{8}\cdot \frac{1}{3} + 
  \frac{37}{216}\cdot \frac{2}{3}
\hspace*{\arraycolsep}=\hspace*{\arraycolsep}
\frac{1}{2}.
\end{array} \]

\noindent This distance between multinomial distributions coincides
with the distance $d(\omega,\omega') = \frac{1}{2}$ between the
original urn distributions. This is a general phenomenon, see
Theorem~\ref{MulnomIsomThm} below. One sees here that the computation
of the distance between the multinomial distributions is more complex,
involving `Kantorovich-over-Kantorovich'.
\end{exa}

As shown in the beginning of this section, for one part of the
isometry of multinomial drawing we use the equation
$\pushing{\flrn}{\multinomial[K](\omega)} = \omega$
in~\eqref{FlrnDrawEqn} (originally from~\cite[Prop.~3]{Jacobs21b}). It
implies, for instance, that if we apply $\flrn$ to the multisets
$\varphi_i$ inside the kets in~\eqref{MultinomialIsometryExDst} and
then apply flattening, the original distributions $\omega$ and
$\omega'$ appear.

The proposition below contains a few more similar results, for future
use.

\begin{prop}
\label{FlrnMulnomProp}
Let $\omega\in\Dst(X)$ be a distribution with a drawsize number $K\in\NNO$.
\begin{enumerate}
\item \label{FlrnMulnomPropElt1} For each element $y\in X$,
\[ \begin{array}{rcl}
\displaystyle\sum_{\varphi\in\Mlt[K](X)}
   \multinomial[K](\omega)(\varphi) \cdot \varphi(y)
& \,=\, &
K \cdot \omega(y).
\end{array} \]

\noindent As a result, the validity $\multinomial[K](\omega) \models
\flrn(-)(y)$ equals $\omega(y)$.

\item \label{FlrnMulnomPropFlrn1} $\pushing{\flrn}{\multinomial[K](\omega)}
  = \omega$.

\item \label{FlrnMulnomPropElt2} For two elements $y\neq z$ in $X$,
\[ \begin{array}{rcl}
\displaystyle\sum_{\varphi\in\Mlt[K](X)}\, 
   \multinomial[K](\omega)(\varphi) \cdot \varphi(y) \cdot \varphi(z)
& = &
K \cdot (K- 1) \cdot \omega(y) \cdot \omega(z).
\end{array} \]

\item \label{FlrnMulnomPropElt3} For a single element $y\in X$,
\[ \begin{array}{rcl}
\displaystyle\sum_{\varphi\in\Mlt[K](X)}\, 
   \multinomial[K](\omega)(\varphi) \cdot \varphi(y) \cdot (\varphi(y)-1)
& = &
K \cdot (K - 1) \cdot \omega(y)^{2}.
\end{array} \]

\item \label{FlrnMulnomPropFlrn2} $\multinomial[K](\omega) \models
  \flrn(-)(y)^{2} = \displaystyle \frac{(K- 1) \cdot \omega(y)^{2} +
    \omega(y)}{K}$.
\end{enumerate}
\end{prop}

\begin{myproof}
\begin{enumerate}
\item The equation holds for $K=0$, since then $\varphi(y)=0$. Hence
we may assume $K > 0$. Then:
\[ \begin{array}[b]{rcl}
\lefteqn{\sum_{\varphi\in\Mlt[K](X)}\,  
   \multinomial[K](\omega)(\varphi) \cdot \varphi(y)}
\\
& \smash{\stackrel{\eqref{MulnomEqn}}{=}} &
\displaystyle\sum_{\varphi\in\Mlt[K](X),\,\varphi(y)\neq 0}\, \varphi(y) \cdot 
   \frac{K!}{\prod_{x} \varphi(x)!} \cdot 
   \textstyle {\displaystyle\prod}_{x}\, \omega(x)^{\varphi(x)}
\\[+1.2em]
& = &
\displaystyle\sum_{\varphi\in\Mlt[K](X),\,\varphi(y)\neq 0}
   K \cdot \frac{(K - 1)!}
     {\prod_{x} (\varphi - 1\ket{y})(x)!} 
   \cdot \omega(y) \cdot 
   \prod_{x}\, \omega(x)^{(\varphi-1\ket{y})(x)}
\\[+1.4em]
& = &
K\cdot\omega(y)\cdot \displaystyle\sum_{\varphi\in\Mlt[K-1](X)}\, 
   \coefm{\varphi} \cdot 
   \textstyle {\displaystyle\prod}_{x}\, \omega(x)^{\varphi(x)}
\\[+1.3em]
& = &
K\cdot\omega(y)\cdot \displaystyle\sum_{\varphi\in\Mlt[K-1](X)}\, 
   \multinomial[K - 1](\omega)(\varphi)
\\
& = &
K \cdot \omega(y).
\end{array} \]

\item By the previous point:
\[ \begin{array}[b]{rcl}
\pushing{\flrn}{\multinomial[K](\omega)}(y)
& = &
\displaystyle\sum_{\varphi\in\Mlt[K](X)} 
   \multinomial[K](\omega)(\varphi)\cdot \flrn(\varphi)(y)
\\[-0.4em]
& = &
\displaystyle\sum_{\varphi\in\Mlt[K](X)} 
   \multinomial[K](\omega)(\varphi)\cdot \frac{\varphi(y)}{K}
\hspace*{\arraycolsep}=\hspace*{\arraycolsep}
\displaystyle K \cdot\omega(y)\cdot \frac{1}{K}
\hspace*{\arraycolsep}=\hspace*{\arraycolsep}
\omega(y).
\end{array} \]

\item + \eqref{FlrnMulnomPropElt3} Essentially as for
  item~\eqref{FlrnMulnomPropElt1}

\auxproof{
Write $\varphi' = \varphi-1\ket{y} - 1\ket{z}$ in:
\[ \begin{array}[b]{rcl}
\lefteqn{\sum_{\varphi\in\Mlt[K](X)}\,  
   \multinomial[K](\omega)(\varphi) \cdot \varphi(y) \cdot \varphi(z)}
\\
& \smash{\stackrel{\eqref{MulnomEqn}}{=}} &
\displaystyle\sum_{\varphi\in\Mlt[K](X),\,\varphi(y)\neq 0,\,\varphi(z)\neq 0}\, 
   \frac{\varphi(y) \cdot \varphi(z) \cdot K!}
   {\prod_{x} \varphi(x)!} \cdot 
   \textstyle {\displaystyle\prod}_{x}\, \omega(x)^{\varphi(x)}
\\[+1.2em]
& = &
\displaystyle\sum_{\varphi\in\Mlt[K](X),\,\varphi(y)\neq 0,\,\varphi(z)\neq 0}
   \frac{K \cdot (K- 1) \cdot (K- 2)!}
    {(\varphi(y) \!-\! 1)! \cdot (\varphi(z)\!-\! 1)! \cdot 
   \prod_{x\neq y,z} \varphi(x)!} 
   \cdot \omega(y) \cdot \omega(z) \cdot 
   \prod_{x}\, \omega(x)^{\varphi'(x)}
\\[+1.2em]
& = &
K\cdot(K-1)\cdot\omega(y)\cdot\omega(z)\cdot 
   \displaystyle\sum_{\varphi\in\Mlt[K-2](X)}\, \coefm{\varphi} \cdot 
   \textstyle {\displaystyle\prod}_{x}\, \omega(x)^{\varphi(x)}
\\[+1.3em]
& = &
K\cdot(K-1)\cdot\omega(y)\cdot\omega(z)\cdot 
  \displaystyle\sum_{\varphi\in\Mlt[K-2](X)}\, 
   \multinomial[K- 2](\omega)(\varphi)
\\
& = &
K \cdot (K-1) \cdot \omega(y) \cdot \omega(z).
\end{array} \]

Write $\varphi' = \varphi - 2\ket{y}$ in:
\[ \begin{array}{rcl}
\lefteqn{\sum_{\varphi\in\Mlt[K](X)}\, 
   \multinomial[K](\omega)(\varphi) \cdot \varphi(y) \cdot (\varphi(y)-1)}
\\
& = &
\displaystyle\sum_{\varphi\in\Mlt[K](X), \varphi(y) > 1}\, 
   \frac{\varphi(y) \cdot (\varphi(y)-1) \cdot K!}
   {\prod_{x} \varphi(x)!} \cdot 
   \textstyle {\displaystyle\prod}_{x}\, \omega(x)^{\varphi(x)}
\\[+1.2em]
& = &
\displaystyle K \cdot (K- 1) \cdot \sum_{\varphi\in\Mlt[K](X), \varphi(y) > 1}\, 
   \frac{(K-2)!}{(\varphi(y)-2)!\cdot \prod_{x\neq y} \varphi(x)!} \cdot 
   \omega(y)^{2}\cdot \textstyle 
   {\displaystyle\prod}_{x}\, \omega(x)^{\varphi'(x)}
\\[+1.2em]
& = &
K \cdot (K- 1) \cdot \omega(y)^{2} \cdot \displaystyle
   \sum_{\varphi\in\Mlt[K-2](X)} \coefm{\varphi} \cdot \textstyle 
   {\displaystyle\prod}_{x}\, \omega(x)^{\varphi(x)} 
\\
& = &
K \cdot (K- 1) \cdot \omega(y)^{2}.
\end{array} \]
}
\item[\eqref{FlrnMulnomPropFlrn2}] We use the equation $a^{2} = a\cdot
  (a-1) + a$ in a combination of the previous
  items~\eqref{FlrnMulnomPropElt3} and~\eqref{FlrnMulnomPropElt1}:
   \[\begin{array}[b]{rcl}
\quad\!\!\;\multinomial[K](\omega) \models \flrn(-)(y)^{2}
&\! =\! &
\displaystyle\frac{1}{K^2}\sum_{\varphi\in\Mlt[K](X)}
   \multinomial[K](\omega)(\varphi) \cdot
   \Big(\varphi(y) \cdot (\varphi(y)-1) + \varphi(y)\Big)
\\[+1.2em]
&\! =\! &
\displaystyle \frac{K \cdot (K- 1) \cdot \omega(y)^{2} + K\cdot\omega(y)}
   {K^2}
\!=\!
\displaystyle \frac{(K- 1) \cdot \omega(y)^{2} \!+\! \omega(y)}{K}.
\end{array} \eqno{\QEDbox} \]
\end{enumerate}
\end{myproof}
\noindent 
We can now formulate and prove our first isometry result.

\begin{thm}
\label{MulnomIsomThm}
Let $X$ be an arbitrary metric space (of colours), and $K>0$ be a
positive natural (drawsize) number. The multinomial channel
\[ \xymatrix{
\Dst(X)\ar[rr]^-{\multinomial[K]} & & \Dst\big(\Mlt[K](X)\big)
} \]

\noindent is an isometry. This involves the Kantorovich
metric~\eqref{DstMetricEqn} for distributions over $X$ on the domain $\Dst(X)$,
and the Kantorovich metric for distributions over multisets of size
$K$, with their Kantorovich metric~\eqref{MltMetricEqn}, on the codomain
$\Dst\big(\Mlt[K](X)\big)$.
\end{thm}

\begin{myproof}
Let distributions $\omega,\omega'\in\Dst(X)$ be given. The map
$\multinomial[K]$ is short since:
\[ \begin{array}{rcll}
\lefteqn{d_{\Dst(\Mlt[K](X))}\Big(\multinomial[K](\omega), \, 
   \multinomial[K](\omega')\Big)}
\\[+0.4em]
& \smash{\stackrel{\eqref{MulnomEqn}}{=}} &
d_{\Dst(\Mlt[K](X))}\Big(\Dst(\acc)(\iid[K](\omega)), \, 
   \Dst(\acc)(\iid[K](\omega'))\Big) \quad
\\[+0.4em]
& \leq &
\frac{1}{K}\cdot d_{\Dst(X^{K})}\Big(\iid[K](\omega), \, \iid[K](\omega')\Big)
  & \mbox{by Lemma~\ref{MltMetricLem}~\eqref{MltMetricLemAcc}}
\\[+0.4em]
& = &
\frac{1}{K}\cdot K \cdot d_{\Dst(X)}\big(\omega, \, \omega'\big)
  & \mbox{by Proposition~\ref{DstMetricTensorProp}~\eqref{DstMetricTensorPropIID}}
\\[+0.4em]
& = &
d_{\Dst(X)}\big(\omega, \, \omega'\big).
\end{array} \]

\noindent The reverse inequality $(\geq)$ follows by combining
Proposition~\ref{FlrnMulnomProp}~\eqref{FlrnMulnomPropFlrn1} with the
argument in~\eqref{FlrnDrawProof}. \QED


\end{myproof}

\section{Hypergeometric drawing is isometric}\label{HypgeomSec}

We start with some preparatory observations on probabilistic
projection and drawing of single balls.

\begin{lem}
\label{ProbProjDrawLem}
For a metric space $X$ and a number $K$, consider the probabilistic
pro\-jec\-tion-delete $\projdelete$ and probabilistic draw-delete
$\drawdelete$ channels.
\[ \xymatrix{
X^{K+1}\ar[r]^-{\projdelete} & \Dst\big(X^{K}\big)
& &
\Mlt[K+1](X)\ar[r]^-{\drawdelete} & \Dst\big(\Mlt[K](X)\big)
} \]

\noindent They are defined via deletion of elements from sequences
and from multisets:
\[ \begin{array}{rcl}
\projdelete(x_{1}, \ldots, x_{K+1})
& \coloneqq &
\displaystyle\sum_{1\leq i\leq K+1} \frac{1}{K + 1}
   \bigket{x_{1}, \ldots, x_{i-1}, x_{i+1}, \ldots, x_{K+1}}
\\
\drawdelete(\psi)
& \coloneqq &
\displaystyle\sum_{x\in\supp(\psi)} \frac{\psi(x)}{K+1} \bigket{\psi - 1\ket{x}}
\\[+1.2em]
& = &
\displaystyle\sum_{x\in\supp(\psi)} \flrn(\psi)(x) \bigket{\psi - 1\ket{x}}.
\end{array} \]

\noindent We write $\psi-1\ket{x}$ for the multiset $\psi$ with one of
its elements $x\in\supp(\psi)$ removed. Then:
\begin{enumerate}
\item \label{ProbProjDrawLemAcc} $\klin{\acc} \klafter \projdelete =
  \drawdelete \klafter \klin{\acc}$;

\item \label{ProbProjDrawLemPush} $\pushing{\flrn}{\drawdelete(\psi)} =
  \flrn(\psi)$;

\item \label{ProbProjDrawLemProj} $\projdelete$ is
  $\frac{K}{K+1}$-Lipschitz, and thus short;

\item \label{ProbProjDrawLemDraw} $\drawdelete$ is short, and even an
  isometry using Fact~\ref{FractionalDstFact}.
\end{enumerate}
\end{lem}

\begin{myproof}
The first point is easy and the second one
is~\cite[Lem.~5~(ii)]{Jacobs21b}. 
\begin{enumerate}\setcounter{enumi}{2}
\item For $\vec{x}, \vec{y} \in X^{K+1}$, via
Lemma~\ref{DstMetricLem}~\eqref{DstMetricLemConv}
and~\eqref{DstMetricLemPoint},
\[ \begin{array}[b]{rcl}
d\Big(\projdelete(\vec{x}), \projdelete(\vec{y})\Big)
& = &
\displaystyle d\left(\sum_{1\leq i\leq K+1} \frac{1}{K+ 1}
   \bigket{x_{1}, \ldots, x_{i-1}, x_{i+1}, \ldots, x_{K+1}}, \right.
\\[-1.0em]
& & \qquad\quad
   \displaystyle\left.\sum_{1\leq i\leq K+1} \frac{1}{K+ 1}
   \bigket{y_{1}, \ldots, y_{i-1}, y_{i+1}, \ldots, y_{K+1}}\right)
\\[+1.0em]
& \leq &
\displaystyle\sum_{1\leq i\leq K+1} \frac{1}{K+ 1} \cdot
   d\Big(1\bigket{x_{1}, \ldots, x_{i-1}, x_{i+1}, \ldots, x_{K+1}},
\\[-1.0em]
& & \hspace*{10em}
    1\bigket{y_{1}, \ldots, y_{i-1}, y_{i+1}, \ldots, y_{K+1}}\Big)
\\[+0.6em]
& = &
\displaystyle\sum_{1\leq i\leq K+1} \frac{1}{K+ 1} \cdot
   d_{X^{K}}\Big((x_{1}, \ldots, x_{i-1}, x_{i+1}, \ldots, x_{K+1}),
\\[-1.0em]
& & \hspace*{11em}
    (y_{1}, \ldots, y_{i-1}, y_{i+1}, \ldots, y_{K+1})\Big)
\\[+0.2em]
& = &
\displaystyle\sum_{1\leq i\leq K+1} \frac{1}{K+1} \cdot
    \Big(d_{X^{K+1}}\big(\vec{x}, \vec{y}\big) - d_{X}(x_{i}, y_{i})\Big)
\\[+1.0em]
& = &
\displaystyle d_{X^{K+1}}\big(\vec{x}, \vec{y}\big) -
    \frac{1}{K+1} \cdot \sum_{1\leq i\leq K+1} d_{X}(x_{i}, y_{i})
\\[+1.0em]
& \smash{\stackrel{\eqref{ProductMetricEqn}}{=}} & 
\displaystyle d_{X^{K+1}}\big(\vec{x}, \vec{y}\big) -
    \frac{1}{K+1} \cdot d_{X^{K+1}}\big(\vec{x}, \vec{y}\big)
\\[+0.6em]
& = &
\displaystyle \frac{K}{K+1} \cdot d_{X^{K+1}}\big(\vec{x}, \vec{y}\big).
\end{array} \]

\item Via item~\ref{ProbProjDrawLemAcc} we get:
\[ \begin{array}{rcccccl}
\klin{\acc} \klafter \projdelete \klafter \arr
& = &
\drawdelete \klafter \klin{\acc} \klafter \arr
& = &
\drawdelete \klafter \unit
& = &
\drawdelete.
\end{array} \eqno{(*)} \]
\noindent Now we can show that $\drawdelete$ is short: for
$\psi,\psi'\in\Mlt[K+1](X)$
\[ \begin{array}{rcll}
\quad\lefteqn{d_{\Dst(\Mlt[K](X))}\big(\drawdelete(\psi), \,
   \drawdelete(\psi')\big)}
\\[+0.2em]
& = &
d_{\Dst(\Mlt[K](X))}\Big(\Dst(\acc)\big(\pushing{\projdelete}{\arr(\psi)}\big), \, 
   \Dst(\acc)\big(\pushing{\projdelete}{\arr(\psi')}\big)\Big)\quad
   & \mbox{by $(*)$}
\\[+0.3em]
& \leq &
\frac{1}{K}\cdot d_{\Dst(X^{K})}\Big(\pushing{\projdelete}{\arr(\psi)}, \, 
   \pushing{\projdelete}{\arr(\psi')}\Big)
   & \mbox{by Lemma~\ref{MltMetricLem}~\eqref{MltMetricLemAcc}}
\\[+0.3em]
& \leq &
\frac{1}{K}\cdot \frac{K}{K+1} \cdot 
   d_{\Dst(X^{K+1})}\big(\arr(\psi), \, \arr(\psi')\big)
   & \mbox{by item~\eqref{ProbProjDrawLemProj}}
\\[+0.3em]
& = &
\frac{1}{K+1}\cdot (K+1) \cdot d_{\Mlt[K+1](X))}\big(\psi, \, \psi'\big)
   & \mbox{by Lemma~\ref{MltMetricLem}~\eqref{MltMetricLemArr}}
\\[+0.2em]
& = &
d_{\Mlt[K+1](X))}\big(\psi, \, \psi'\big).
\end{array} \]

\noindent For the reverse inequality we use that $\flrn$ is an
isometry, via Fact~\ref{FractionalDstFact}, and (thus) that
$\pushop{\flrn}$ is short.
\[ \begin{array}[b]{rcll}
d_{\Dst(\Mlt[K](X))}\Big(\drawdelete(\psi), \drawdelete(\psi')\Big)
& \geq &
d_{\Dst(X)}\Big(\pushing{\flrn}{\drawdelete(\psi)}, 
   \pushing{\flrn}{\drawdelete(\psi')}\Big)
\\
& = &
d_{\Dst(X)}\big(\flrn(\psi), \flrn(\psi')\big)
   & \mbox{by item~\ref{ProbProjDrawLemPush}}
\\
& = &
d_{\Mlt[K+1](X)}\big(\psi, \psi'\big).
\end{array} \eqno{\QEDbox} \]
\end{enumerate}
\end{myproof}
\noindent 
Hypergeometric drawing uses the draw-delete mode, where a drawn ball
is removed from the urn. The urn thus changes with every draw. It will
be presented as a multiset, say initially of size $L$. The size $K$ of
draws (multisets) must thus be restricted as $K\leq L$, since one
cannot draw more balls than the urn contains --- unless one uses
negative probabilities, see~\cite{JacobsS23c}. In addition, we must
require that drawn multisets $\varphi$ are sub-multisets of the urn,
that is $\varphi \leq \upsilon$, if $\upsilon$ is the urn. This
requirement ensures that the number of drawn balls for each color is
below the number of balls of that colour in the urn.

The hypergeometric channel thus takes the form $\hypergeometric[K]
\colon \Mlt[L](X) \rightarrow \Dst\big(\Mlt[K](X)\big)$, for $L \geq
K$. It can be described as an iteration of draw-delete's,
see~\cite[Thm.~6]{Jacobs21b}:
\begin{equation}
\label{HypgeomEqn}
\begin{array}{rcccl}
\hypergeometric[K](\upsilon)
& \coloneqq &
\big(\underbrace{\drawdelete \klafter \cdots \klafter 
   \drawdelete}_{L-K\text{ times}}\big)(\upsilon)
& = &
\displaystyle\sum_{\varphi\in\Mlt[K](X), \, \varphi \leq \upsilon} 
   \frac{\binom{\upsilon}{\varphi}}{\binom{L}{K}}\,\bigket{\varphi},
\end{array}
\end{equation}

\noindent where $\binom{\upsilon}{\varphi} \coloneqq
\prod_{x\in X} \binom{\upsilon(x)}{\varphi(x)}$.

\begin{thm}
\label{HypgeomIsomThm}
The hypergeometric channel $\hypergeometric[K] \colon \Mlt[L](X)
\rightarrow \Dst\big(\Mlt[K](X)\big)$ defined in~\eqref{HypgeomEqn},
for $L\geq K$, is short and an isometry, via
Fact~\ref{FractionalDstFact}.
\end{thm}

\begin{myproof}
We see in~\eqref{HypgeomEqn} that $\hypergeometric[K]$ is a (channel)
iteration of short channels $\drawdelete$, and thus short itself. Via
iterated use of
Lemma~\ref{ProbProjDrawLem}~\eqref{ProbProjDrawLemPush} we get
$\pushing{\flrn}{\hypergeometric[K](\psi)} = \flrn(\psi)$. The proof
is then completed via the argumentation in~\eqref{FlrnDrawProof}.
\end{myproof}\QED
\\\par\noindent 
The introduction of this paper contains an illustration of this
hypergeometric isometry result, for urns over the set of colours $C =
\{R,G,B\}$, considered as a discrete metric space.

\section{P\'olya drawing is isometric}\label{PolyaSec}

Hypergeometric distributions use the draw-delete mode: a drawn ball is
removed from the urn. The less well-known P\'olya draws~\cite{Hoppe84}
use the draw-add mode. This means that a drawn ball is returned to the
urn, together with another ball of the same colour (as the drawn
ball). Thus, with hypergeometric draws the urn decreases in size, so
that only finitely many draws are possible, whereas with P\'olya draws
the urn grows in size, and the drawing may be repeated arbitrarily
many times. As a result, for P\'olya distributions we do not need to
impose restrictions on the size $K$ of draws. We do have to restrict
draws from urn $\upsilon$ to multisets $\varphi\in\Mlt[K](X)$ with
$\supp(\varphi) \subseteq \supp(\upsilon)$ since we can only draw
balls of colours that are in the urn. Following~\cite{Jacobs22a},
P\'olya distributions are formulated in terms of multichoose binomials
$\bibinom{n}{m} \coloneqq \binom{n+m-1}{m} = \frac{(n+m-1)!}{m!\cdot
  (n-1)!}$, for $n>0$.
\begin{equation}
\label{PolyaEqn}
\begin{array}{rcl}
\polya[K](\upsilon)
& \coloneqq &
\displaystyle\sum_{\varphi\in\Mlt[K](X), \, \supp(\varphi) \subseteq \supp(\upsilon)} 
   \frac{\big(\!\binom{\upsilon}{\varphi}\!\big)}
   {\big(\!\binom{L}{K}\!\big)}\,\bigket{\varphi},
\end{array}
\end{equation}

 where $\big(\!\binom{\upsilon}{\varphi}\!\big) \coloneqq
          {\displaystyle \prod_{x\in \supp(\upsilon)}}
          \big(\!\binom{\upsilon(x)}{\varphi(x)}\!\big)$.

\noindent The following result was already announced in the beginning of
Section~\ref{MulnomSec} and is used for part of the isometry proof.

\begin{prop}
\label{PolyaFlrnProp}
For a non-empty urn $\upsilon$ and drawsize $K>0$ one has
$\pushing{\flrn}{\polya[K](\upsilon)} = \flrn(\upsilon)$.
\end{prop}

\begin{myproof}
We use:
\[ \begin{array}{rcl}
\lefteqn{\sum_{\varphi\in\Mlt[K](\supp(\upsilon))} 
   \polya[K](\upsilon)(\varphi)\cdot\flrn(\varphi)(y)}
\\[+0.2em]
& \smash{\stackrel{\eqref{PolyaEqn}}{=}} &
\displaystyle\sum_{\varphi\in\Mlt[K](\supp(\upsilon))} 
   \frac{\big(\!\binom{\upsilon}{\varphi}\!\big)\cdot\varphi(y)}
   {\big(\!\binom{L}{K}\!\big)\cdot K}
\\[+1.2em]
& \smash{\stackrel{(*)}{=}} &
\displaystyle\sum_{\varphi\in\Mlt[K](\supp(\upsilon))} \, 
   \frac{\upsilon(y) \cdot 
      \big(\!\binom{\upsilon+1\ket{y}}{\varphi-1\ket{y}}\!\big)}
   {L \cdot \big(\!\binom{L+1}{K-1}\!\big)}
\\[+1.2em]
& = &
\displaystyle \frac{\upsilon(y)}{L} \cdot
   \sum_{\varphi\in\Mlt[K-1](\supp(\upsilon+1\ket{y}))} 
   \frac{\big(\!\binom{\upsilon+1\ket{y}}{\varphi}\!\big)}
      {\big(\!\binom{L+1}{K-1}\!\big)}
\\[+1.2em]
& = &
\displaystyle \flrn(\upsilon)(y) \cdot
    \sum_{\varphi\in\Mlt[K-1](\supp(\upsilon+1\ket{y}))} 
   \polya[K-1]\big(\upsilon+1\ket{y}\big)(\varphi)
\\[+0.2em]
& = &
\flrn(\upsilon)(y).
\end{array} \]

\noindent The marked equation $\smash{\stackrel{(*)}{=}}$ uses
the following property:
\[ \begin{array}{rcccccl}
\big(\!\binom{L}{K}\!\big)\cdot K
& = &
\displaystyle\frac{(L+K-1)!}{(L-1)!\cdot K!}\cdot K
& = &
\displaystyle L \cdot \frac{(L+K-1)!}{L!\cdot (K-1)!}
& = &
L \cdot \big(\!\binom{L+1}{K-1}\!\big).
\end{array} \eqno{\QEDbox} \]
\end{myproof}

\noindent 
In Equation~\eqref{HypgeomEqn} we have seen that the hypergeometric
channel can be expressed as an iteration of single-draw-delete's. It
is not the case that the P\'olya channel is an iteration of anaologous
single-draw-add channels.  But we do have the following result.

\begin{lem}
\label{ProjStoreAddLem}
Consider channel $\projstoreadd$, for `projection-store-add', of
the form:
\[ \xymatrix{
X^{L}\times X^{N} \ar[rr]^-{\projstoreadd} & & 
   \Dst\big(X^{L}\times X^{N+1}\big)
} \]

\noindent defined as:
\begin{equation}
\label{ProjStoreAddEqn}
\begin{array}{rcl}
\projstoreadd(\vec{x},\vec{y})
& \coloneqq &
\displaystyle \sum_{1\leq i\leq L+N} \sum_{1\leq j\leq N+1} \textstyle
   \frac{1}{(L+N)(N+1)}
   \bigket{\vec{x}, \; y_{1}, \ldots y_{j-1}, z_{i}, y_{j}, \ldots y_{N}}
\\
& & \qquad\mbox{where $z_{i}$ is the $i$-th element 
   of the concatenation $\vec{x} \concat \vec{y}$}
\end{array}
\end{equation}

\noindent Then:
\begin{enumerate}
\item \label{ProjStoreAddLemOne} $\projstoreadd$ has Lipschitz constant
  $\frac{L+N+1}{L+N}$;

\item \label{ProjStoreAddLemMany} The $K$-fold Kleisli iteration
$\projstoreadd^{K} = \projstoreadd \klafter \cdots \klafter \projstoreadd
\colon X^{L}\times X^{N} \rightarrow \Dst\big(X^{L}\times X^{N+K}\big)$ 
has Lipschitz constant $\frac{L+N+K}{L+N}$; 

\item \label{ProjStoreAddLemZero} Using the empty sequence
  $\tuple{}\in X^{0}$ as fixed second argument in the $K$-fold
  iteration gives a function $\projstoreadd^{K}(-,\tuple{}) \colon
  X^{L} \rightarrow \Dst(X^{L}\times X^{K}\big)$ with Lipschitz
  constant $\frac{L+K}{L}$.  This function is of the form:
\[ \begin{array}{rcl}
\projstoreadd^{K}(\vec{x},\tuple{})
& = &
1\bigket{\vec{x}} \otimes \seqpolya[K](\vec{x}),
\end{array} \]

\noindent where $\seqpolya[K] \coloneqq \Dst(\pi_{2}) \after
\projstoreadd^{K}(-,\tuple{}) \colon X^{L} \rightarrow
\Dst\big(X^{K}\big)$ is `sequence P\'olya'. We claim that has
Lipschitz constant $\frac{K}{L}$.

\item \label{ProjStoreAddLemIter} $K$-fold P\'olya drawing (in terms
  of multisets) can be described in terms of sequence P\'olya, namely
  as:
\[ \begin{array}{rcl}
\polya[K]
& = &
\klin{\acc} \klafter \seqpolya[K] \klafter \arr \;\colon\;
\Mlt[L](X) \rightarrow \Dst\big(\Mlt[K](X)\big).
\end{array} \]
\end{enumerate}
\end{lem}

\begin{myproof}
\begin{enumerate}
\item For arbitrary sequences $\vec{x},\vec{x'} \in X^{L}$ and
$\vec{y},\vec{y'} \in X^{N}$ we have, by immediately
using Lemma~\ref{DstMetricLem}~\eqref{DstMetricLemConv}:
\[ \begin{array}{rcl}
\lefteqn{d\Big(\projstoreadd(\vec{x},\vec{y}),
   \projstoreadd(\vec{x'},\vec{y'})\Big)}
\\[+0.2em]
& \leq &
\displaystyle \sum_{1\leq i\leq L+N} \sum_{1\leq j\leq N+1} 
   \textstyle\frac{1}{(L+N)(N+1)} \cdot d_{X^{L}\times X^{N+1}}
   \Big((\vec{x}, \; y_{1}, \ldots y_{j-1}, z_{i}, y_{j}, \ldots y_{N}),
\\[-1.0em]
& & \hspace*{18.5em}
    (\vec{x'}, \; y'_{1}, \ldots y'_{j-1}, z'_{i}, y'_{j}, \ldots y'_{N})\Big)
\\
& \smash{\stackrel{\eqref{ProductMetricEqn}}{=}} &
\displaystyle \sum_{1\leq i\leq L+N} \textstyle \frac{1}{L+N} \cdot 
   \Big(d_{X^L}(\vec{x}, \vec{x'}) + 
   d_{X}(z_{i}, z'_{i}) + d_{X^{N}}(\vec{y}, \vec{y'})\Big)
\\[+1.2em]
& = &
d_{X^L}(\vec{x}, \vec{x'}) + \frac{1}{L+N} \cdot 
   d_{X^{L}\times X^{N}}\big((\vec{x},\vec{y}), (\vec{x'},\vec{y'})\big) +
   d_{X^{N}}(\vec{y}, \vec{y'})
\\[+0.2em]
& = &
\frac{L+N+1}{L+N} \cdot 
   d_{X^{L}\times X^{N}}\big((\vec{x},\vec{y}), (\vec{x'},\vec{y'})\big).
\end{array} \]

\item By a simple iteration, using that function composition 
involves multiplication of Lipschitz constants:
\[ \begin{array}{rcl}
\displaystyle\frac{L+N+1}{L+N} \cdot \frac{L+N+2}{L+N+1} \cdot \ldots \cdot 
   \frac{L+N+K}{L+N+K-1}
& = &
\displaystyle\frac{L+N+K}{L+N}.
\end{array} \]

\item The distribution $\drawstoreadd(\vec{x},\vec{y})$
  in~\eqref{ProjStoreAddEqn} is defined as a sum of ket's that all
  have $\vec{x}$ as first component. This first component can be
  extracted via a tensor, as $\bigket{\vec{x}, -} = 1\bigket{\vec{x}}
  \otimes -$. Subsequently, this $1\bigket{\vec{x}}$ can be pulled
  outside the two sums in~\eqref{ProjStoreAddEqn}. This allows us in
  the calculations below to use that the tensor $\otimes$ is an
  isometry, see
  Proposition~\ref{DstMetricTensorProp}~\eqref{DstMetricTensorPropTwo}.

The previous point gives, for $N=0$, that the function
$\projstoreadd^{K}(-,\tuple{})$ has Lipschitz constant $\frac{L+K}{L}$,
since:
\[ \begin{array}{rcccl}
d\Big(\projstoreadd^{K}(\vec{x},\tuple{}), \,
   \projstoreadd^{K}(\vec{x'},\tuple{})\Big)
& \leq &
\frac{L+0+K}{L+0}\cdot \Big(d\big(\vec{x}, \vec{x'}\big) + 
   d(\tuple{}, \tuple{})\Big)
& = &
\frac{L+K}{L}\cdot d\big(\vec{x}, \vec{x'}\big).
\end{array} \]

\noindent At the same time, using
Proposition~\ref{DstMetricTensorProp}~\eqref{DstMetricTensorPropTwo}
and Lemma~\ref{DstMetricLem}~\eqref{DstMetricLemPoint},
\[ \begin{array}{rcl}
d\Big(\projstoreadd^{K}(\vec{x},\tuple{}), \,
   \projstoreadd^{K}(\vec{x'},\tuple{})\Big)
& = &
d\Big(1\bigket{\vec{x}} \otimes \seqpolya[K](\vec{x}), \,
   1\bigket{\vec{x'}} \otimes \seqpolya[K](\vec{x'})\Big)
\\[+0.4em]
& = &
d\big(1\bigket{\vec{x}}, 1\bigket{\vec{x'}}\big) +
   d\big(\seqpolya[K](\vec{x}), \seqpolya[K](\vec{x'})\big)
\\[+0.2em]
& = &
d\big(\vec{x}, \vec{x'}\big) +
   d\big(\seqpolya[K](\vec{x}), \seqpolya[K](\vec{x'})\big).
\end{array} \]

\noindent Combining these two facts gives Lipschitz constant $\frac{K}{L}$
for sequence P\'olya:
\[ \begin{array}{rcccl}
d\big(\seqpolya[K](\vec{x}), \seqpolya[K](\vec{x'})\big)
& \leq &
\frac{L+K}{L}\cdot d\big(\vec{x}, \vec{x'}\big) - d\big(\vec{x}, \vec{x'}\big)
& = &
\frac{K}{L}\cdot d\big(\vec{x}, \vec{x'}\big).
\end{array} \]

\item We prove $\klin{\acc} \klafter \seqpolya[K] \klafter \arr = \polya[K]$
  by induction on $K\geq 0$, where $\seqpolya[K] = \Dst(\pi_{2})
  \klafter \projstoreadd^{K}(-,\tuple{})$. The case $K=0$ is easy
  since $\projstoreadd^{0}(\vec{x}, \tuple{}) = 1\bigket{\vec{x},
    \tuple{}}$, so that:
\[ \begin{array}{rcl}
\big(\klin{\acc} \klafter \seqpolya[0] \klafter \arr\big)(\upsilon)
& = &
\Dst(\acc)\Big(\Dst(\pi_{2})\big(1\bigket{\vec{x}, \tuple{}}\big)\Big)
\\
& = &
\Dst(\acc)\big(1\bigket{\tuple{}}\big)
\hspace*{\arraycolsep}=\hspace*{\arraycolsep}
1\bigket{\acc(\tuple{})}
\hspace*{\arraycolsep}=\hspace*{\arraycolsep}
1\bigket{\zero}
\hspace*{\arraycolsep}=\hspace*{\arraycolsep}
\polya[0](\upsilon).
\end{array} \]

\noindent The latter equation holds because the only possible draw of
size $0$ is the empty multiset $\zero$.

The induction step is more work.
\[ \begin{array}{rcl}
\lefteqn{\big(\klin{\acc} \klafter \seqpolya[K+1] \klafter \arr\big)(\upsilon)}
\\[+0.2em]
& = &
\displaystyle \sum_{\vec{x}\in\acc^{-1}(\upsilon)}\! \frac{1}{\coefm{\upsilon}}
   \cdot \ \!\!\;\sum_{\vec{z}\in\supp(\upsilon)^{K+1}}
   \big(\projstoreadd \klafter \projstoreadd^{K}\big)
   (\vec{x}, \tuple{})(\vec{x}, \vec{z})
   \,\bigket{\acc(\vec{z})}
\\
& = &
\displaystyle \sum_{\vec{x}\in\acc^{-1}(\upsilon)}\! \frac{1}{\coefm{\upsilon}}
   \cdot \ \ \!\:\sum_{\vec{y}\in\supp(\upsilon)^{K}}\, \quad\ \;\sum_{z\in \vec{x} \concat \vec{y}}\,
   \frac{1}{L+K} \cdot \projstoreadd^{K}(\vec{x}, \tuple{})(\vec{x}, \vec{y})
   \,\bigket{\acc(\vec{y}) + 1\ket{z}}
\\
& = &
\displaystyle \sum_{\vec{x}\in\acc^{-1}(\upsilon)}\! \frac{1}{\coefm{\upsilon}} 
   \cdot\!\! \sum_{\varphi\in\Mlt[K](\supp(\upsilon))}\,\, \sum_{z\in \supp(\upsilon)}\!\!\!
   \frac{\upsilon(z) + \varphi(z)}{L+K} \cdot 
   \Dst(\acc)\big(\seqpolya[K](\vec{x})\big)(\varphi)
   \,\bigket{\varphi + 1\ket{z}}
\\[+1.4em]
& \smash{\stackrel{\text{(IH)}}{=}} &
\displaystyle\sum_{\varphi\in\Mlt[K](\supp(\upsilon))}\, \sum_{z\in \supp(\upsilon)}\,
   \frac{\upsilon(z) + \varphi(z)}{L+K} \cdot 
   \polya[K](\upsilon)(\varphi)
   \,\bigket{\varphi + 1\ket{z}}
\\
& \smash{\stackrel{\eqref{PolyaEqn}}{=}} &
\displaystyle\sum_{\varphi\in\Mlt[K](\supp(\upsilon))}\, \sum_{z\in \supp(\upsilon)}\,
   \frac{\upsilon(z) + \varphi(z)}{L+K} \cdot 
   \frac{\big(\!\binom{\upsilon}{\varphi}\!\big)}
   {\big(\!\binom{L}{K}\!\big)}
   \,\bigket{\varphi + 1\ket{z}}
\\
& \smash{\stackrel{(*)}{=}} &
\displaystyle\sum_{\varphi\in\Mlt[K](\supp(\upsilon))}\, \sum_{z\in \supp(\upsilon)}\,
   \frac{\varphi(z)+1}{K+1} \cdot 
   \frac{\big(\!\binom{\upsilon}{\varphi+1\ket{z}}\!\big)}
   {\big(\!\binom{L}{K+1}\!\big)}
   \,\bigket{\varphi + 1\ket{z}}
\\[+1.2em]
& = &
\displaystyle\sum_{\psi\in\Mlt[K+1](\supp(\upsilon))}\, \left(\sum_{z\in \supp(\upsilon)}\,
   \frac{\psi(z)}{K+1}\right) \cdot 
   \frac{\big(\!\binom{\upsilon}{\psi}\!\big)}
   {\big(\!\binom{L}{K+1}\!\big)}
   \,\bigket{\psi}
\\[+1.6em]
& \smash{\stackrel{\eqref{PolyaEqn}}{=}} &
\polya[K+1](\upsilon).
\end{array} \]

\noindent The marked equation $\stackrel{(*)}{=}$ follows from an easy
calculation. For instance:
\[ \begin{array}[b]{rcl}
(L+K)\cdot \big(\!\binom{L}{K}\!\big)
& = &
\displaystyle (L+K) \cdot \frac{(L+K-1)!}{(L-1)!\cdot K!}
\\[+0.8em]
& = &
\displaystyle (K+1) \cdot \frac{(L+K)!}{(L-1)!\cdot (K+1)!}
\hspace*{\arraycolsep}=\hspace*{\arraycolsep}
\textstyle(K+1)\cdot \big(\!\binom{L}{K+1}\!\big).
\end{array} \eqno{\QEDbox} \]
\end{enumerate}
\end{myproof}

\noindent 
We now obtain isometry for P\'olya.

\begin{thm}
\label{PolyaIsomThm}
Each P\'olya channel $\polya[K] \colon \Mlt[L](X) \rightarrow
\Dst\big(\Mlt[K](X)\big)$, for urn size $L > 0, K > 0$, is short and,
indirectly via Fact~\ref{FractionalDstFact}, also an isometry.
\end{thm}

\ignore{

X = Space(0,10,20,50)
for i in range(10):
    L = random.randint(2,10)
    K = random.randint(2,10)
    u1 = random_multiset(L)(X)
    u2 = random_multiset(L)(X)
    p1 = Polya(K)(u1)
    p2 = Polya(K)(u2)
    dr = distribution_relationlift(p1, p2, WassersteinPred)
    print( round(dr[0], 10) == round(wd(u1,u2),10) )

}

\begin{myproof}
The hard work has already been done in Lemma~\ref{ProjStoreAddLem}.
Shortness of P\'olya drawing is obtained from its description
$\polya[K] = \klin{\acc} \klafter \seqpolya[K] \klafter \arr \colon
\Mlt[L](X) \rightarrow \Dst\big(\Mlt[K](X)\big)$ in
Lemma~\ref{ProjStoreAddLem}~\eqref{ProjStoreAddLemIter}.  This allows
us to use the Lipschitz constants for accumulation and arrangement
from Lemma~\ref{MltMetricLem}~\eqref{MltMetricLemAcc},
\eqref{MltMetricLemArr}, and for sequence P\'olya from
Lemma~\ref{ProjStoreAddLem}~\eqref{ProjStoreAddLemZero}.  Thus, for
urns $\upsilon, \upsilon'\in\Mlt[L](X)$,
\[ \begin{array}{rcl}
d\Big(\polya[K](\upsilon), \, \polya[K](\upsilon')\Big)
& = &
d\Big(\Dst(\acc)\big(\pushing{\seqpolya[K]}{\arr(\upsilon)}\big), \,
      \Dst(\acc)\big(\pushing{\seqpolya[K]}{\arr(\upsilon')}\big)\Big)
\\[+0.2em]
& \leq &
\frac{1}{K}\cdot d\Big(\pushing{\seqpolya[K]}{\arr(\upsilon)}, \,
      \pushing{\seqpolya[K]}{\arr(\upsilon')}\Big)
\\[+0.4em]
& \leq &
\frac{1}{K}\cdot \frac{K}{L} \cdot d\big(\arr(\upsilon), \,
      \arr(\upsilon')\big)
\\[+0.3em]
& \leq &
\frac{1}{L} \cdot L \cdot d\big(\upsilon, \, \upsilon'\big)
\\[0.1em]
& = &
d\big(\upsilon, \, \upsilon'\big).
\end{array} \]
\noindent The reverse inequality $(\geq)$ follows from
Proposition~\ref{PolyaFlrnProp} and the argument
in~\eqref{FlrnDrawProof}.
\end{myproof}\hfill\QED

\noindent 
We illustrate that the P\'olya channel is an isometry.

\begin{exa}
\label{PolyaIsomEx}
We take as space of colours $X = \{0, 10, 50\} \subseteq \NNO$ with
two urns of size~$4$
\[ \begin{array}{rclcrcl}
\upsilon_{1}
& = &
3\ket{0} + 1\ket{10}
& \qquad &
\upsilon_{2}
& = &
1\ket{0} + 2\ket{10} + 1\ket{50}.
\end{array} \]

\noindent The Kantorovich distance between these urns is~$15$, via an
optimal coupling $1\ket{0, 0} + 2\ket{0, 10} + 1\ket{10, 50}$,
yielding the distance $\frac{1}{4}\cdot (0-0) + \frac{1}{2}\cdot
(10-0) + \frac{1}{4}\cdot (50-10) = 5 + 10 = 15$.

We look at P\'olya draws of size $K=2$. This gives distributions:
\[ \begin{array}{rcl}
\polya[2](\upsilon_{1})
& = &
\frac{3}{5}\Bigket{2\ket{0}} + 
   \frac{3}{10}\Bigket{1\ket{0} + 1\ket{10}} + 
   \frac{1}{10}\Bigket{2\ket{10}}
\\[+0.5em]
\polya[2](\upsilon_{2})
& = &
\frac{1}{10}\Bigket{2\ket{0}} + 
   \frac{1}{5}\Bigket{1\ket{0} + 1\ket{10}} + 
   \frac{3}{10}\Bigket{2\ket{10}} + 
   \frac{1}{10}\Bigket{1\ket{0} + 1\ket{50}}
\\[+0.4em]
& & \qquad +\;
   \frac{1}{5}\Bigket{1\ket{10} + 1\ket{50}} + 
   \frac{1}{10}\Bigket{2\ket{50}}
\end{array} \]

\noindent We compute the distance between these two distributions via
the last formulation in~\eqref{DstMetricEqn}, using an optimal short
factor $p\colon \Mlt[2](X) \rightarrow \nnR$ given by
$p(\varphi) = \sum_{x} \varphi(x)\cdot \frac{x}{2}$, that is:
\[ \begin{array}{rclcrclcrcl}
p\big(2\ket{0}\big)
& = &
0
& \qquad &
p\big(1\ket{0} + 1\ket{10}\big)
& = &
5
& \qquad &
p\big(2\ket{10}\big)
& = &
10
\\[+0.2em]
p\big(1\ket{0} + 1\ket{50}\big)
& = &
25
& &
p\big(1\ket{10} + 1\ket{50}\big)
& = &
30
& &
p\big(2\ket{50}\big)
& = &
50.
\end{array} \]

\noindent Then:
\[ \begin{array}{rcl}
\polya[2](\upsilon_{1}) \models p
& = &
\frac{3}{5}\cdot 0 +
   \frac{3}{10}\cdot 5 +
   \frac{1}{10}\cdot 10 
\hspace*{\arraycolsep}=\hspace*{\arraycolsep}
\frac{5}{2}
\\[+0.2em]
\polya[2](\upsilon_{2}) \models p
& = &
\frac{1}{10}\cdot 0 +
   \frac{1}{5}\cdot 5 + 
   \frac{3}{10}\cdot 10 +
   \frac{1}{10}\cdot 25 + 
   \frac{1}{5}\cdot 30 +
   \frac{1}{10}\cdot 50
\hspace*{\arraycolsep}=\hspace*{\arraycolsep}
\frac{35}{2}.
\end{array} \]

\noindent As predicted by Theorem~\ref{PolyaIsomThm}, the distance
between the P\'olya distributions then coincides with the distance
between the urns:
\[ \begin{array}{rcl}
d\Big(\polya[2](\upsilon_{1}), \, \polya[2](\upsilon_{2})\Big)
& = &
\Big|\,\polya[2](\upsilon_{1}) \models p - 
    \polya[2](\upsilon_{2}) \models p\,\Big|
\\[+0.5em]
& = &
\frac{35}{2} - \frac{5}{2}
\hspace*{\arraycolsep}=\hspace*{\arraycolsep}
15
\hspace*{\arraycolsep}=\hspace*{\arraycolsep}
d\big(\upsilon_{1}, \, \upsilon_{2}\big).
\end{array} \]
\end{exa}

\section{The law of large urns}\label{LargeUrnSec}

Hypergeometric and P\'olya draws involve removals from, and additions
to, the urn. When the urn $\upsilon$ is very large, in comparison to
the drawsize, the effects of such removals and additions are
relatively small. Thus, hypergeometric and P\'olya draws from large
urns look very much like multinomial draws --- which do not change the
urn. For the corresponding multinomial draws one uses the distribution
$\flrn(\upsilon)$ as urn, the normalised version of the original urn
$\upsilon$. This statement that hypergeometric and P\'olya draws from
large urns are like multinomial draws is intuitively clear and
well-known. We make it mathematically precise in terms of limits of
Kantorovich-over-Kantorovich distances going to zero.

The starting point is the following observation about binomial and
multichoose coeffients.

\begin{lem}
\label{CoefficientLimitLem}
Fix a number $m\in\NNO$. Then:
\[ \begin{array}{rcccl}
\displaystyle\lim\limits_{n\rightarrow\infty} \, \frac{\binom{n}{m}}{n^{m}} 
& \,=\, &
\displaystyle\frac{1}{m!}
& \,=\, &
\displaystyle\lim\limits_{n\rightarrow\infty} \, \frac{\bibinom{n}{m}}{n^{m}}.
\end{array} \]

\noindent We shall use these equations in the following alternative form.
\begin{equation}
\label{CoefficientLimitEqn}
\begin{array}{rcccl}
\displaystyle \lim\limits_{n\rightarrow\infty}\, \frac{n!}{(n-m)!\cdot n^{m}}
& \,=\, &
1
& \,=\, &
\displaystyle \lim\limits_{n\rightarrow\infty}\, \frac{(n+m-1)!}{(n-1)!\cdot n^{m}}.
\end{array}
\end{equation}
\end{lem}

\begin{myproof}
We may assume $n\geq m$. Then:
\[ \begin{array}{rcl}
\displaystyle\lim\limits_{n\rightarrow\infty} \, \frac{\binom{n}{m}}{n^{m}} 
& = &
\displaystyle \frac{1}{m!} \cdot 
    \lim\limits_{n\rightarrow\infty} \, \frac{n!}{(n-m)!\cdot n^{m}}
\\[+0.8em]
& = &
\displaystyle \frac{1}{m!} \cdot 
    \lim\limits_{n\rightarrow\infty} \, \frac{n}{n} \cdot \frac{n-1}{n} \cdot
    \ldots \cdot \frac{n-m+1}{n}
\\[+0.8em]
& = &
\displaystyle \frac{1}{m!} \cdot 
    \left(\lim\limits_{n\rightarrow\infty} \, \frac{n}{n}\right) \cdot 
    \left(\lim\limits_{n\rightarrow\infty} \, \frac{n-1}{n}\right) \cdot
    \ldots \cdot \left(\lim\limits_{n\rightarrow\infty} \, \frac{n-m+1}{n}\right)
\hspace*{\arraycolsep}=\hspace*{\arraycolsep}
\displaystyle\frac{1}{m!}.
\end{array} \]

\noindent Similarly, for the multichoose coefficient:
\[ \begin{array}[b]{rcl}
\displaystyle\lim\limits_{n\rightarrow\infty} \, \frac{\bibinom{n}{m}}{n^{m}} 
& = &
\displaystyle\lim\limits_{n\rightarrow\infty} \, 
   \frac{(n+m-1)!}{m! \cdot (n-1)! \cdot n^{m}} 
\\[+0.8em]
& = &
\displaystyle \frac{1}{m!} \cdot 
    \lim\limits_{n\rightarrow\infty} \, \frac{n}{n} \cdot \frac{n+1}{n} \cdot
    \ldots \cdot \frac{n+m-1}{n}
\hspace*{\arraycolsep}=\hspace*{\arraycolsep}
\displaystyle\frac{1}{m!}.
\end{array} \eqno{\QEDbox} \]
\end{myproof}

\begin{prop}
\label{DrawLimitProp}
Let $X = \{x_{1}, \ldots, x_{N}\}$ be a finite set with $N\geq 2$
elements and let $\varphi\in\Mlt[K](X)$ be a draw of size
$K\in\pNNO$.
\begin{enumerate}
\item \label{DrawLimitPropHyp} The hypergeometric probability of
  drawing $\varphi$ from an urn becomes equal to the multinomial
  probability of drawing $\varphi$ from the frequentist learning of
  the urn, when the number of balls in the urn for each color $x_i$
  goes to infinity:
\[ \begin{array}{rcl}
\displaystyle \lim\limits_{n_{1}, \ldots n_{N}\rightarrow\infty}\, 
  \frac{\multinomial[K]\big(\flrn(\sum_{i} n_{i}\ket{x_i})\big)(\varphi)}
       {\hypergeometric[K]\big(\sum_{i} n_{i}\ket{x_i}\big)(\varphi)}
& = &
1.
\end{array} \]

\item \label{DrawLimitPropPol} Similarly, P\'olya draw probabilities
  from a large urn are like multinomial probabilities:
\[ \begin{array}{rcl}
\displaystyle \lim\limits_{n_{1}, \ldots n_{N}\rightarrow\infty}\, 
  \frac{\multinomial[K]\big(\flrn(\sum_{i} n_{i}\ket{x_i})\big)(\varphi)}
       {\polya[K]\big(\sum_{i} n_{i}\ket{x_i}\big)(\varphi)}
& = &
1.
\end{array} \]
\end{enumerate}
\end{prop}

\begin{myproof}
\begin{enumerate}
\item By unpacking the relevant definitions:
\[ \begin{array}{rcl}
\lefteqn{\lim\limits_{n_{1}, \ldots n_{N}\rightarrow\infty}\, 
 \frac{\multinomial[K]\big(\flrn(\sum_{i} n_{i}\ket{x_i})\big)(\varphi)}
   {\hypergeometric[K]\big(\sum_{i} n_{i}\ket{x_i}\big)(\varphi)}}
\\
& = &
\displaystyle \lim\limits_{n_{1}, \ldots n_{N}\rightarrow\infty}\, 
   \frac{K!}{\prod_{i} \varphi(x_{i})!} \cdot
   \textstyle{\displaystyle\prod}_{i}\,
   \Big(\frac{n_{i}}{\sum_{j}n_{j}}\Big)^{\varphi(x_{i})} \cdot \displaystyle
   \frac{\binom{\sum_{j} n_{j}}{K}}{\prod_{i} \binom{n_{i}}{\varphi(x_{i})}}
\\[0.8em]
& = &
\displaystyle \lim\limits_{n_{1}, \ldots n_{N}\rightarrow\infty}\, 
    \frac{(\sum_{j}n_{j})!}{((\sum_{j}n_{j})-K)!\cdot (\sum_{j}n_{j})^{K}} \cdot
    \textstyle{\displaystyle\prod}_{i}\, \displaystyle
    \frac{(n_{i}-\varphi(x_{i}))!\cdot n_{i}^{\varphi(x_{i})}}{n_{i}!}
\\[0.8em]
& = &
\displaystyle\left(\lim\limits_{n\rightarrow\infty}\,
   \frac{n!}{(n-K)!\cdot n^{K}}\right) \cdot
   \left(\textstyle{\displaystyle\prod}_{i}\, \displaystyle
   \lim\limits_{n_{i}\rightarrow\infty}\, 
   \frac{(n_{i}-\varphi(x_{i}))!\cdot n_{i}^{\varphi(x_{i})}}{n_{i}!}\right)
\\[1.0em]
& \smash{\stackrel{\eqref{CoefficientLimitEqn}}{=}} &
1.
\end{array} \]

\item Similarly:
\[ \begin{array}[b]{rcl}
\lefteqn{\lim\limits_{n_{1}, \ldots n_{N}\rightarrow\infty}\, 
 \frac{\multinomial[K]\big(\flrn(\sum_{i} n_{i}\ket{x_i})\big)(\varphi)}
   {\polya[K]\big(\sum_{i} n_{i}\ket{x_i}\big)(\varphi)}}
\\
& = &
\displaystyle \lim\limits_{n_{1}, \ldots n_{N}\rightarrow\infty}\, 
   \frac{K!}{\prod_{i} \varphi(x_{i})!} \cdot
   \textstyle{\displaystyle\prod}_{i}\,
   \Big(\frac{n_{i}}{\sum_{j}n_{j}}\Big)^{\varphi(x_{i})} \cdot \displaystyle
   \frac{\big(\!\binom{\sum_{j} n_{j}}{K}\!\big)}
   {\prod_{i} \big(\!\binom{n_{i}}{\varphi(x_{i})}\!\big)}
\\[0.8em]
& = &
\displaystyle \lim\limits_{n_{1}, \ldots n_{N}\rightarrow\infty}\, 
    \frac{((\sum_{j}n_{j})+K-1)!}
      {((\sum_{j}n_{j})-1)!\cdot (\sum_{j}n_{j})^{K}} \cdot
    \textstyle{\displaystyle\prod}_{i}\, \displaystyle
    \frac{(n_{i}-1)!\cdot n_{i}^{\varphi(x_{i})}}{(n_{i}+\varphi(x_{i})-1)!}
\\[1.0em]
& = &
\displaystyle\left(\lim\limits_{n\rightarrow\infty}\,
   \frac{(n+K-1)!}{(n-1)!\cdot n^{K}}\right) \cdot
   \left(\textstyle{\displaystyle\prod}_{i}\, \displaystyle
   \lim\limits_{n_{i}\rightarrow\infty}\, 
   \frac{(n_{i}-1)!\cdot n_{i}^{\varphi(x_{i})}}{(n_{i}+\varphi(x_{i})-1)!}\right)
\\[1.0em]
& \smash{\stackrel{\eqref{CoefficientLimitEqn}}{=}} &
1.
\end{array} \eqno{\QEDbox} \]
\end{enumerate}
\end{myproof}

\noindent 
We translate these results into limits of Kantorovich distances going
to zero.

\begin{thm}
\label{DrawDistanceLimitThm}
Let $X = \{x_{1}, \ldots, x_{N}\} \subseteq X$ be a finite set of
colors, in a metric space $X$, of size $N\geq 2$, and let $K\in\NNO$
be a fixed drawsize.
\begin{enumerate}
\item When the urn size increases, the Kantorovich distance
  between hypergeometric and multinomial distributions goes to zero,
  as in:
\[ \begin{array}{rcl}
\lim\limits_{n_{1}, \ldots n_{N}\rightarrow\infty}\, 
   d\Big(\hypergeometric[K]\big(\sum_{i} n_{i}\ket{x_i}\big), \;
   \multinomial[K]\big(\flrn(\sum_{i} n_{i}\ket{x_i})\big)\Big)
& = &
0.
\end{array} \]

\item Similarly, for P\'olya distributions:
\[ \begin{array}{rcl}
\lim\limits_{n_{1}, \ldots n_{N}\rightarrow\infty}\, 
   d\Big(\polya[K]\big(\sum_{i} n_{i}\ket{x_i}\big), \;
   \multinomial[K]\big(\flrn(\sum_{i} n_{i}\ket{x_i})\big)\Big)
& = &
0.
\end{array} \]
\end{enumerate}
\end{thm}

\begin{myproof}
By Proposition~\ref{KantorovichTotalVariationProp} it suffices to
prove the result for the total variation distance, where we take as
constant $D = \diameter\big(\{x_{1}, \ldots, x_{n}\}\big)$.  We prove
the first point only, since the proof of the second one works
similarly. We use
Proposition~\ref{DrawLimitProp}~\eqref{DrawLimitPropHyp} in:
\[ \hspace*{-1em}\begin{array}[b]{rcl}
\lefteqn{\textstyle\lim\limits_{n_{1}, \ldots n_{N}\rightarrow\infty}\, 
   \tvd\Big(\hypergeometric[K]\big(\sum_{i} n_{i}\ket{x_i}\big), \;
   \multinomial[K]\big(\flrn\big(\sum_{i} n_{i}\ket{x_i}\big)\big)\Big)}
\\[+0.6em]
& \smash{\stackrel{\eqref{TvdEqn}}{=}} &
\lim\limits_{n_{1}, \ldots n_{N}\rightarrow\infty}\, 
   \frac{1}{2}\!\displaystyle\sum_{\varphi\in\Mlt[K](X)} 
   \left|\,\textstyle
   \hypergeometric[K]\big(\sum_{i} n_{i}\ket{x_i}\big)(\varphi) \,-\,
   \multinomial[K]\big(\flrn\big(\sum_{i} n_{i}\ket{x_i}\big)\big)(\varphi)
   \, \right|\hspace*{-1em}
\\[+1.2em]
& = &
\frac{1}{2}\!\displaystyle\sum_{\varphi\in\Mlt[K](X)}\,
   \lim\limits_{n_{1}, \ldots n_{N}\rightarrow\infty}\, \textstyle
   \hypergeometric[K]\big(\sum_{i} n_{i}\ket{x_i}\big)(\varphi) \cdot
   \displaystyle\left|\,
   \frac{\multinomial[K]\big(\flrn\big(\sum_{i} n_{i}\ket{x_i}\big)\big)(\varphi)}{\hypergeometric[K]\big(\sum_{i} n_{i}\ket{x_i}\big)(\varphi)}
   \;-\; 1 \, \right|\hspace*{-1em}
\\[+1.2em]
& \leq &
\frac{1}{2}\!\displaystyle\sum_{\varphi\in\Mlt[K](X)}\,
   \displaystyle\left|\, \lim\limits_{n_{1}, \ldots n_{N}\rightarrow\infty}\, 
   \frac{\multinomial[K]\big(\flrn\big(\sum_{i} n_{i}\ket{x_i}\big)\big)(\varphi)}{\hypergeometric[K]\big(\sum_{i} n_{i}\ket{x_i}\big)(\varphi)}
   \;-\; 1 \, \right|
\\[+1.2em]
& = &
0.
\end{array} \eqno{\QEDbox} \]

\auxproof{
Similarly:
\[ \begin{array}{rcl}
\lefteqn{\textstyle\lim\limits_{n_{1}, \ldots n_{N}\rightarrow\infty}\, 
   d\Big(\polya[K]\big(\sum_{i} n_{i}\ket{x_i}\big), \;
   \multinomial[K]\big(\flrn\big(\sum_{i} n_{i}\ket{x_i}\big)\big)\Big)}
\\[+0.2em]
& = &
\lim\limits_{n_{1}, \ldots n_{N}\rightarrow\infty}\, 
   \frac{1}{2}\!\displaystyle\sum_{\varphi\in\Mlt[K](X)} 
   \left|\,\textstyle
   \polya[K]\Big(\sum_{i} n_{i}\ket{x_i}\Big)(\varphi) \,-\,
   \multinomial[K]\Big(\flrn\big(\sum_{i} n_{i}\ket{x_i}\big)\Big)(\varphi)
   \, \right|
\\[+1.2em]
& = &
\frac{1}{2}\!\displaystyle\sum_{\varphi\in\Mlt[K](X)}\,
   \lim\limits_{n_{1}, \ldots n_{N}\rightarrow\infty}\, \textstyle
   \polya[K]\big(\sum_{i} n_{i}\ket{x_i}\big)(\varphi)
\\[+1em]
& & \hspace*{10em} \cdot\,
   \displaystyle\left|\,
   \frac{\multinomial[K]\big(\flrn\big(\sum_{i} n_{i}\ket{x_i}\big)\big)(\varphi)}{\polya[K]\big(\sum_{i} n_{i}\ket{x_i}\big)(\varphi)}
   \;-\; 1 \, \right|
\\[+1.2em]
& \leq &
\frac{1}{2}\!\displaystyle\sum_{\varphi\in\Mlt[K](X)}\,
   \displaystyle\left|\, \lim\limits_{n_{1}, \ldots n_{N}\rightarrow\infty}\, 
   \frac{\multinomial[K]\big(\flrn\big(\sum_{i} n_{i}\ket{x_i}\big)\big)(\varphi)}{\polya[K]\big(\sum_{i} n_{i}\ket{x_i}\big)(\varphi)}
   \;-\; 1 \, \right|
\\[+1.2em]
& = &
0.
\end{array} \]
}
\end{myproof}

\noindent 
We may reformulate the limit results in
Theorem~\ref{DrawDistanceLimitThm} as `urn limits' of the form:
\[ \begin{array}{rcccl}
\lim\limits_{\upsilon\rightarrow\infty}
   d\Big(\hypergeometric[K]\big(\upsilon\big), \;
   \multinomial[K]\big(\flrn(\upsilon)\big)\Big)
& = &
0
& = &
\lim\limits_{\upsilon\rightarrow\infty}
   d\Big(\polya[K]\big(\upsilon\big), \;
   \multinomial[K]\big(\flrn(\upsilon)\big)\Big).
\end{array} \]

\noindent The meaning of $\upsilon\rightarrow\infty$ for
$\upsilon\in\Mlt(X)$ is that each color multiplicity $\upsilon(x)$
goes to infinity, for each element $x$ of the finite set of colours
$X$. We refer to the above equations as the law(s) of large urns.

\section{The law of large draws}\label{LargeDrawSec}

This section contains another limit property, not for large urns but
for large draws. It may be seen as a metric reformulation of the law
of large numbers.

But first some preliminaries. For a distribution $\omega\in\Dst(X)$
and an observation $p\colon X \rightarrow \R$ the variance
$\Var(\omega,p) \in \nnR$ describes how much the observable is
spread-out, in relation to the validity, or expected value,
$\omega\models p$. Explicitly:
\begin{equation}
\label{VarEqn}
\begin{array}{rcl}
\Var(\omega,p)
& \,\coloneqq\, &
\omega \models \big(p - (\omega\models p)\cdot \one\big)^{2}.
\end{array}
\end{equation}

\noindent We recall a number of basic facts about this variance.

\begin{lem}
\label{VarLem}
Let $\omega\in\Dst(X)$ and $p\colon X \rightarrow \R$ be given.
\begin{enumerate}
\item \label{VarLemVar} The variance~\eqref{VarEqn} can also be described as:
\[ \begin{array}{rcl}
\Var(\omega, p)
& \,=\, &
\big(\omega\models p^{2}\big) - \big(\omega\models p\big)^{2}.
\end{array} \]

\item As a result, $\big(\omega\models p^{2}\big) \geq
  \big(\omega\models p\big)^{2}$.

\item \label{VarLemAbs} There is also an inequality:
\[ \begin{array}{rcl}
\omega \models \big|\, p - (\omega\models p)\cdot\one \,\big|
& \,\leq\, &
\sqrt{\Var(\omega,p)}.
\end{array} \]
\end{enumerate}
\end{lem}

\begin{myproof}
\begin{enumerate}
\item Via a standard argument:
\[ \begin{array}{rcl}
\lefteqn{\Var(\omega, p)}
\\[+0.2em]
& \smash{\stackrel{\eqref{VarEqn}}{=}} &
\omega \models \big(p - (\omega\models p)\cdot\one\big)^{2}
\\[+0.2em]
& = &
\displaystyle\sum_{x\in X}\, \omega(x)\cdot\big(p(x) - (\omega\models p)\big)^{2}
\\[+1.0em]
& = &
\displaystyle\sum_{x\in X}\, \omega(x) \cdot
   \Big(p(x)^{2} - 2(\omega\models p) \cdot p(x) + (\omega\models p)^{2}\Big)
\\[+1.2em]
& = &
\displaystyle\left(\sum_{x\in X}\, \omega(x) \cdot p^{2}(x)\right) - 
   2(\omega\models p) \cdot \left(\sum_{x\in X}\, \omega(x) \cdot p(x)\right) + 
   \left(\sum_{x\in X}\, \omega(x) \cdot (\omega\models p)^{2}\right)
\\[+1.2em]
& = &
\big(\omega\models p^{2}\big) - 2\big(\omega\models p)\cdot(\omega\models p) +
    (\omega\models p)^{2}
\\[+0.2em]
& = &
\big(\omega\models p^{2}\big) - (\omega\models p)^{2}.
\end{array} \]

\item Obviously, by the previous point, since $\Var(\omega,p) \geq 0$.

\item We abbreviate $q \coloneqq \big|\, p - (\omega\models
  p)\cdot\one \,\big|$, so that $q(x) = \big|\, p(x) - (\omega\models
  p)\,\big|$. Then:
\[ \begin{array}{rcccl}
\Var(\omega,p)
& \,=\, &
\omega \models q^{2}
& \,\geq\, &
\big(\omega \models q\big)^{2}, 
   \qquad\mbox{by the previous point.}
\end{array} \]

\noindent As a result,
\[ \begin{array}{rcccl}
\omega \models \big|\, p - (\omega\models p)\cdot\one \,\big|
& \,=\, &
\omega \models q
& \,\leq\, &
\sqrt{\Var(\omega,p)}.
\end{array} \eqno{\QEDbox} \]
\end{enumerate}
\end{myproof}

\noindent 
Informally, the next result expresses that large multinomial draws,
when normalised, are close to the urn distribution. This is
intuitively clear: when we draw very many balls (in draw-replace
mode), then the frequencies of the colours in the draw resemble the
urn. The more formal aspect of this statement is below. It expresses
that this closeness works ``in probability'', that is, the
probabilities of the draws must be taken into account. Indeed, a very
large draw may differ significantly from the urn, for instance when
all drawn balls have the same colour, but the likelihood of such draws
is low.

\begin{thm}
\label{LargeDrawThm}
For each distribution $\omega\in\Dst(X)$ on a metric space $X$,
\[ \begin{array}{rcl}
\lim\limits_{K\rightarrow\infty} \, \multinomial[K](\omega) \models
   d\big(\omega, \flrn(-)\big)
& \,=\, &
0.
\end{array} \]
\end{thm}

\begin{myproof}
Since Proposition~\ref{KantorovichTotalVariationProp} applies, with $D
= \diameter\big(\supp(\omega)\big)$, it suffices to prove the limit
equation for the total variation distance $\tvd$. 
\[ \begin{array}[b]{rcl}
\lefteqn{\lim\limits_{K\rightarrow\infty} \, \multinomial[K](\omega) \models
   \tvd\big(\omega, \flrn(-)\big)}
\\[+0.6em]
& \smash{\stackrel{\eqref{TvdEqn}}{=}} &
\lim\limits_{K\rightarrow\infty} \, \displaystyle \sum_{\varphi\in\Mlt[K](X)} \,
   \multinomial[K](\omega)(\varphi) \cdot \textstyle
   \frac{1}{2}\cdot \displaystyle \sum_{y\in\supp(\omega)}
   \big|\, \flrn(\varphi)(y) - \omega(y) \, \big|
\\[+1.2em]
& = &
\frac{1}{2}\! \displaystyle \sum_{y\in\supp(\omega)}
   \lim\limits_{K\rightarrow\infty} \, \multinomial[K](\omega) \models
   \big|\, \flrn(-)(y) - 
   \big(\multinomial[K](\omega) \models \flrn(-)(y)\big) \, \big|
\\
& & \qquad\mbox{by Proposition~\ref{FlrnMulnomProp}~\eqref{FlrnMulnomPropElt1}}
\\
& \leq &
\frac{1}{2}\! \displaystyle \sum_{y\in\supp(\omega)}
   \lim\limits_{K\rightarrow\infty} \!
   \sqrt{\Var\big(\multinomial[K](\omega), \flrn(-)(y)\big)}
   \qquad\mbox{by Lemma~\ref{VarLem}~\eqref{VarLemAbs}}
\\[+0.8em]
& = &
\frac{1}{2}\! \displaystyle \sum_{y\in\supp(\omega)}
   \lim\limits_{K\rightarrow\infty} \!
   \displaystyle
      \sqrt{\Big(\multinomial[K](\omega) \models \flrn(-)(y)^{2}\Big) -
   \Big(\multinomial[K](\omega) \models \flrn(-)(y)\Big)^{2}}
\\
& & \qquad\mbox{by Lemma~\ref{VarLem}~\eqref{VarLemVar}}
\\[+0.2em]
& = &
\frac{1}{2}\! \displaystyle \sum_{y\in\supp(\omega)}
   \lim\limits_{K\rightarrow\infty} \!
   \sqrt{\frac{(K - 1) \cdot \omega(y)^{2} + \omega(y)}{K} - \omega(y)^{2}}
  \qquad\mbox{by Proposition~\ref{FlrnMulnomProp}~\eqref{FlrnMulnomPropFlrn2},
  \eqref{FlrnMulnomPropFlrn1}}
\\[+0.8em]
& = &
\frac{1}{2}\! \displaystyle \sum_{y\in\supp(\omega)}
   \lim\limits_{K\rightarrow\infty} 
   \frac{\sqrt{\omega(y)\cdot (1 - \omega(y))}}{\sqrt{K}}
\\[+0.6em]
& = &
0.
\end{array} \eqno{\QEDbox} \]
\end{myproof}

\begin{cor}
\label{LargeDrawCor}
For two distributions $\omega,\rho\in\Dst(X)$ on a metric space $X$,
\[ \begin{array}{rcl}
\lim\limits_{K\rightarrow\infty} \, \multinomial[K](\omega) \models
   d\big(\flrn(-), \rho\big)
& \,=\, &
d\big(\omega, \rho\big).
\end{array} \]
\end{cor}

\begin{myproof}
We use the triangular inequality, both for $(\leq)$ and $(\geq)$.
First, for each multiset $\varphi$ we have $d\big(\flrn(\varphi),
\rho\big) \leq d\big(\flrn(\varphi), \omega\big) + d\big(\omega,
\rho\big)$. Hence:
\[ \begin{array}{rcl}
\lim\limits_{K\rightarrow\infty} \, \multinomial[K](\omega) \models
   d\big(\flrn(-), \rho\big)
& \leq &
\lim\limits_{K\rightarrow\infty} \, \multinomial[K](\omega) \models 
   \Big(d\big(\flrn(-), \omega\big) + d\big(\omega, \rho\big)\Big)
\\[+0.6em]
& = &
\Big(\lim\limits_{K\rightarrow\infty} \, \multinomial[K](\omega) \models 
   d\big(\flrn(-), \omega\big)\Big) + d\big(\omega, \rho\big)
\\[+0.6em]
& = &
d\big(\omega, \rho\big).
\end{array} \]

\noindent In the other direction, we use $d\big(\omega, \rho\big) \leq
d\big(\omega, \flrn(\varphi)\big) + d\big(\flrn(\varphi),
\rho\big)$. Hence:
\[ \begin{array}[b]{rcl}
d\big(\omega, \rho\big)
& = &
\lim\limits_{K\rightarrow\infty} \, \multinomial[K](\omega) \models 
   d\big(\omega, \rho\big)
\\[+0.2em]
& \leq &
\lim\limits_{K\rightarrow\infty} \, \multinomial[K](\omega) \models \Big(
  d\big(\omega, \flrn(-)\big) + d\big(\flrn(-), \rho\big)\Big)
\\[+0.6em]
& = &
\Big(\lim\limits_{K\rightarrow\infty} \, \multinomial[K](\omega) \models 
  d\big(\omega, \flrn(-)\big)\Big) +
  \Big(\lim\limits_{K\rightarrow\infty} \, \multinomial[K](\omega) \models 
  d\big(\flrn(-), \rho\big)\Big)
\\[+0.4em]
& = &
\lim\limits_{K\rightarrow\infty} \, \multinomial[K](\omega) \models 
  d\big(\flrn(-), \rho\big).
\end{array} \eqno{\QEDbox} \]
\end{myproof}

\noindent 
We conclude this section with some observations about large P\'olya
draws. Earlier in this section we have looked at large multinomial
draw. Large hypergeometric draws, when the drawsize goes to infinity,
do not make sense, since the urn is finite and thus empty at some
stage\footnote{In a theory with negative probabilities one can have
hypergeometric overdrawing, see~\cite{JacobsS23c}.}. But one can ask:
what happens with large P\'olya draws? Does the distance between the
(normalised) urn and (normalised) draws also go to zero, when the
drawsize goes to infinity? That is, do large P\'olya draws look very
much like the urn, in probability? Recall that P\'olya draws are used
to capture clustering effects. Hence it is unclear what happens.  The
distance could go to infinity. In fact, the limit goes to a specific
number.

We sketch the background of this observation without going into all
details, since it involves continuous probability theory and thus
leads beyond the scope of this paper. It is well-known, see standard
textbooks,
like~\cite{BernardoS00,Billingsley95,Feng10,Kallenberg21,Wilks62},
that the P\'olya distribution can be obtained via a pushforward of the
multinomial channel over the Dirichlet distribution. This takes the
following form, for an urn $\upsilon\in\Mlt(X)$ with $\supp(\upsilon)
= X$, for a finite set $X$.
\begin{equation}
\label{PolyaViaMulnomEqn}
\begin{array}{rcl}
\polya[K](\upsilon)(\varphi)
& = &
\displaystyle\int_{\omega\in\Dst(X)} \Dirichlet(\upsilon)(\omega)
   \cdot \multinomial[K](\omega)(\varphi) \intd\omega.
\end{array}
\end{equation}

\noindent We can now formulate a P\'olya analogue of Theorem~\ref{LargeDrawThm}.

\begin{prop}
\label{LargePolyaDrawProp}
For a non-empty urn $\upsilon\in\Mlt(X)$,
\begin{equation}
\label{LargePolyaDrawEqn}
\begin{array}{rcl}
\lim\limits_{K\rightarrow\infty} \, \polya[K](\upsilon) \models
   d\big(\flrn(-), \flrn(\upsilon)\big)
& \,=\, &
\displaystyle\int_{\omega\in\Dst(X)} \Dirichlet(\upsilon)(\omega)
   \cdot d\big(\omega, \flrn(\upsilon)\big)\intd\omega
\\[+1.0em]
& \,=\, &
\Dirichlet(\upsilon) \models d\big(-, \flrn(\upsilon)\big).
\end{array}
\end{equation}

\noindent The latter formulation uses validity $\models$ in a
continuous setting.
\end{prop}

\noindent 
The integral in~\eqref{LargePolyaDrawEqn} is a special number
associated with the urn / multiset $\upsilon$. It remains an open
question to compute this integral more concretely.

\begin{myproof}
We apply the formula~\eqref{PolyaViaMulnomEqn} in:
\[ \begin{array}[b]{rcl}
\lefteqn{\lim\limits_{K\rightarrow\infty} \, \polya[K](\upsilon) \models
   d\big(\flrn(-), \flrn(\upsilon)\big)}
\\[+0.2em]
& = &
\displaystyle\lim\limits_{K\rightarrow\infty} \, 
   \sum_{\varphi\in\Mlt[K](X)} \, \polya[K](\upsilon)(\varphi) \cdot
   d\big(\flrn(\varphi), \flrn(\upsilon)\big)
\\
& = &
\displaystyle\lim\limits_{K\rightarrow\infty} \, 
   \sum_{\varphi\in\Mlt[K](X)} \, 
   \left(\int_{\omega\in\Dst(X)} \Dirichlet(\upsilon)(\omega)
   \cdot \multinomial[K](\omega)(\varphi) \intd\omega\right) \cdot
   d\big(\flrn(\varphi), \flrn(\upsilon)\big)
\\
& = &
\displaystyle\int_{\omega\in\Dst(X)} \Dirichlet(\upsilon)(\omega)
   \cdot \lim\limits_{K\rightarrow\infty} \, 
   \sum_{\varphi\in\Mlt[K](X)} \, 
   \multinomial[K](\omega)(\varphi) \cdot
   d\big(\flrn(\varphi), \flrn(\upsilon)\big)\intd\omega
\\
& = &
\displaystyle\int_{\omega\in\Dst(X)} \Dirichlet(\upsilon)(\omega)
   \cdot \lim\limits_{K\rightarrow\infty} \, 
   \Big(\multinomial[K](\omega) \models 
   d\big(\flrn(-), \flrn(\upsilon)\big)\Big)\intd\omega
\\[+1em]
& = &
\displaystyle\int_{\omega\in\Dst(X)} \Dirichlet(\upsilon)(\omega)
   \cdot d\big(\omega, \flrn(\upsilon)\big)\intd\omega
   \qquad\mbox{by Corollary~\ref{LargeDrawCor}}
\\[+1em]
& = &
\Dirichlet(\upsilon) \models d\big(-, \flrn(\upsilon)\big).
\end{array} \eqno{\QEDbox} \]
\end{myproof}

\noindent 
The plot below gives an impression of the
numbers~\eqref{LargePolyaDrawEqn} in the binary case with total
variation distance, for urns $i\ket{a} + j\ket{b}$ for the 400 cases
$1 \leq i,j \leq 20$.
\[ \includegraphics[scale=0.6]{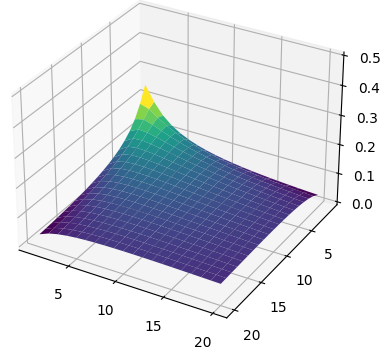} \]

\section{Concluding remarks}

Category theory provides a fresh look at the area of probability
theory, see \textit{e.g.}~\cite{Fritz20} or~\cite{Jacobs21g} for an
overview. This paper demonstrates that draw operations, viewed as
(Kleisli) maps, are remarkably well-behaved: they exactly preserve
Kantorovich distances.  Such distances on urns filled with coloured
balls are relatively simple, starting from a `ground' metric on the
set of colours. But on draw distributions, the distances involve
Kantorovich-over-Kantorovich. In addition, well-known limit
behaviour, for large urns and large draws, is formulated in terms of
Kantorovich distances.

This paper arose from a fresh categorical perspective on classical urn
models, in line with earlier work~\cite{Jacobs21b,Jacobs22c,Jacobs22a}
of the author. Is category theory necessary to obtain (or prove) these
results? No, one can strip all categorical language from what has been
described above and formulate and prove the isometry results in
traditional probabilistic language. But it is a fact that these
results have not appeared within the traditional, non-categorical
setting. We do not argue that category theory is necessary, but we do
believe that its new, fresh perspective is helpful and leads to new
ideas and results. More generally, category theory often helps in a
particular mathematical setting to organise the material in a
structured manner and to ask relevant questions: does this category
have limits or colimits? Is this operation functorial, and if so, what
does it preserve, does it have adjoints, \textit{etc}. This paper uses
category theory in a light-weight manner and is not a hard-core paper
in categorical probability theory, using for instance Markov
categories~\cite{Fritz20}.  Hopefully it works as invitation to invest
efforts to learn the new categorically-inspired approach and what it
brings.

This paper concentrates on drawing from an urn using finite discrete
probability distributions. A natural question is whether other
operations, especially involving infinite discrete and continuous
probability, also preserve distance. This question is solved
in~\cite[Prop.~3.2]{AngelisG21} for parameterised distributions
$\omega[\theta]$ on (subsets of) the reals: the Kantorovich distance
equals the (real-number) distance between the expected values:
\begin{equation}
\label{ContinuousCase}
\begin{array}{rcl}
d\big(\omega[\theta_{1}], \omega[\theta_{2}]\big)
& = &
\big| \, \expec(\omega[\theta_{1}]) -  \expec(\omega[\theta_{2}]) \,\big|.
\end{array} 
\end{equation}

\noindent As special cases, one has, for instance for the Kantorovich
distances between Poisson and exponential distributions (on $\NNO$ and
$\pR$):
\[ \begin{array}{rclcrcl}
d\big(\poisson[\lambda_{1}], \poisson[\lambda_{2}]\big)
& = &
\big|\, \lambda_{1} - \lambda_{2} \,\big|
& \qquad\qquad &
d\big(\Exp[\lambda_{1}], \Exp[\lambda_{2}]\big)
& = &
\big|\, \frac{1}{\lambda_{1}} - \frac{1}{\lambda_{2}} \,\big|.
\end{array} \]

\noindent This equation~\eqref{ContinuousCase} is very powerful.
However, the isometry results for draw distributions in this paper are
not instances of~\eqref{ContinuousCase} since these draw distributions
are not distributions on real numbers, but on multisets.

Another question that may be asked is if the (isometry) results for
draw distributions, with respect to the Kantorovich distance, also
hold for Kullback-Leibler divergence (see~\cite{CoverT06} for
extensive information). This is largely an open question. What we
can offer is the following result for multinomial distributions.
We write $\DKL$ for this Kullback-Leibler divergence and use
its definition implicitly in the proof below.

\begin{prop}
\label{KLdivMultinomialProp}
Let distributions $\omega, \omega'\in \Dst(X)$ be given with a
draw-size number $K\in\NNO$.  Then:
\[ \begin{array}{rcl}
\DKL\Big(\multinomial[K](\omega), \, \multinomial[K](\omega')\Big)
& = &
K\cdot\DKL\big(\omega, \omega'\big).
\end{array} \]
\end{prop}

\begin{myproof}
We may assume $\supp(\omega) \subseteq \supp(\omega')$ and write
$\supp(\omega') = \{x_{1}, \ldots, x_{n}\} \subseteq X$. 
\begin{equation*}
\begin{array}[b]{rcl}
\DKL\Big(\multinomial[K](\omega), \, \multinomial[K](\omega')\Big)
& = &
\displaystyle\sum_{\varphi\in\Mlt[K](X)} \,
   \multinomial[K](\omega)(\varphi) \cdot 
   \ln\left(\frac{\multinomial[K](\omega)(\varphi)}
      {\multinomial[K](\omega')(\varphi)}\right)
\\[+1em]
& = &
\displaystyle\sum_{\varphi\in\Mlt[K](X)} \,
   \multinomial[K](\omega)(\varphi) \cdot 
   \ln\left(\frac{\coefm{\varphi}\cdot \prod_{i} \omega(x_{i})^{\varphi(x_{i})}}
      {\coefm{\varphi}\cdot \prod_{i} \omega'(x_{i})^{\varphi(x_{i})}}\right)
\\[+1em]
& = &
\displaystyle\sum_{\varphi\in\Mlt[K](X)} \,
   \multinomial[K](\omega)(\varphi) \cdot 
   \ln\left({\textstyle{\displaystyle\prod}_{i}\,}
      \left(\frac{\omega(x_{i})}{\omega'(x_{i})}\right)^{\varphi(x_{i})}\right)
\\[+1em]
& = &
\displaystyle\sum_{\varphi\in\Mlt[K](X)} \,
   \multinomial[K](\omega)(\varphi) \cdot 
   {\textstyle{\displaystyle\sum}_{i}\,} \varphi(x_{i}) \cdot
      \ln\left(\frac{\omega(x_{i})}{\omega'(x_{i})}\right)
\\[+1em]
& = &
\displaystyle {\textstyle{\displaystyle\sum}_{i}\,}
   \left(\sum_{\varphi\in\Mlt[K](X)} \,
   \multinomial[K](\omega)(\varphi) \cdot \varphi(x_{i})\right) \cdot
      \ln\left(\frac{\omega(x_{i})}{\omega'(x_{i})}\right)
\\[+1em]
& = &
\displaystyle {\textstyle{\displaystyle\sum}_{i}\,}
   K\cdot\omega(x_{i}) \cdot
      \ln\left(\frac{\omega(x_{i})}{\omega'(x_{i})}\right)
   \qquad\mbox{by Proposition~\ref{FlrnMulnomProp}~\eqref{FlrnMulnomPropElt1}}
\\[+0.6em]
& = &
K\cdot\DKL\big(\omega, \omega'\big). \hspace*{\fill} \parbox[t]{0pt}{ \qedsymbol}
\end{array}
\end{equation*}
\end{myproof}


%
%
%
%
%
%

\subsection*{Acknowledgments}

Thanks are due to the anonymous reviewers for their careful reading
and detailed comments, leading to many points of improvement.

\bibliographystyle{alphaurl}


\bibliography{draw-isometries-journal.bib}

\appendix

\section{Proof of the equations in~\eqref{DstMetricEqn}}\label{DstApp}


%
%

We recall the three formulations~\eqref{DstMetricEqn} of the
Kantorovich distance.
\begin{equation}
\label{DstMetricMongeEqn}
\begin{array}{rcl}
d\big(\omega, \omega'\big)
\hspace*{\arraycolsep}\coloneqq\hspace*{\arraycolsep}
\displaystyle\bigwedge_{\tau\in\decouple^{-1}(\omega, \omega')} \tau \models d_{X}
& \smash{\stackrel{(*)}{=}} &
\displaystyle\bigvee_{p, \, p'\in \Obs(X), \, p\oplus p' \,\leq\, d_{X}} 
   \, \omega\models p \,+\, \omega'\models p' 
\\[+1.4em]
& \smash{\stackrel{(**)}{=}} &
\displaystyle\bigvee_{q\in \FactSh(X)} \big|\, \omega\models q \,-\,
   \omega'\models q \,\big|.
\end{array}
\end{equation}

\noindent The observation $p\oplus p' \in \Obs(X\times X)$ is defined
as $(p\oplus p')(x,x') = p(x) + p'(x')$. The minimum given by the meet
$\bigwedge$ exists since there is at least one coupling of
$\omega,\omega'$, namely the product $\omega\otimes\omega'$.  The two
joins $\bigvee$ are also taken over non-empty domains since one can
instantiate with the zero observation / factor $\zero$.

We first concentrate on equation~$(*)$
in~\eqref{DstMetricMongeEqn}. The inequality $(\geq)$ is easy: let
$p,p'\in\Obs(X)$ satisfy $p\oplus p' \leq d_{X}$. Then for each
coupling $\tau\in\Dst(X\times X)$ of $\omega,\omega'$ one has:
\[ \begin{array}{rcl}
\tau \models d_{X}
\hspace*{\arraycolsep}\geq\hspace*{\arraycolsep}
\tau \models p \oplus p' 
& = &
\displaystyle \sum_{x,x'\in X} \, \tau(x,x') \cdot (p \oplus p')(x,x')
\\[+1em]
& = &
\displaystyle \sum_{x,x'\in X} \, \tau(x,x') \cdot \big(p(x) + p'(x')\big)
\\[+1em]
& = &
\displaystyle \sum_{x,x'\in X} \, \tau(x,x') \cdot p(x) + 
   \sum_{x,x'\in X} \, \tau(x,x') \cdot p'(x')
\\[+0.8em]
& = &
\displaystyle \sum_{x\in X} \,
   \left(\sum_{x'\in X} \, \tau(x,x')\right) \cdot p(x) + 
   \sum_{x'\in X} \,\left(\sum_{x\in X} \, \tau(x,x')\right) \cdot p'(x')
\\[+1.2em]
& = &
\displaystyle \sum_{x\in X} \, \omega(x) \cdot p(x) + 
   \sum_{x'\in X} \,\omega'(x') \cdot p'(x')
\\
& = &
\omega\models p \,+\, \omega'\models p'.
\end{array} \]

\noindent As a result, since this inequality holds for all couplings
$\tau$,
\[ \begin{array}{rcl}
\displaystyle\bigwedge_{\tau\in\decouple^{-1}(\omega, \omega')} \tau \models d_{X}
& \,\geq\, &
\omega\models p \,+\, \omega'\models p'.
\end{array} \]

\noindent And since the latter holds for all $p,p'$ we get the
$(\geq)$ part of $\smash{\stackrel{(*)}{=}}$.
\[ \begin{array}{rcl}
\displaystyle\bigwedge_{\tau\in\decouple^{-1}(\omega, \omega')} \tau \models d_{X}
& \,\geq\, &
\displaystyle\bigvee_{p,p'\in \Obs(X)\text{ with } p\oplus p' \leq d_{X}} 
   \omega\models p \,+\, \omega'\models p'.
\end{array} \]

\noindent This inequality shows in particular that this join $\bigvee$
exists.


For the $(\leq)$ part of $\smash{\stackrel{(*)}{=}}$
in~\eqref{DstMetricMongeEqn} we have to do more work and exploit
Farkas Lemma, which we formulate first. It captures a separation
property that plays a central role in the duality theorem in linear
programming, see \textit{e.g.}~\cite{MatouekG06} for further details
and discussion.


\auxproof{
One writes a matrix $A\in\R^{m\times n}$ and $A^{T}\in\R^{n\times m}$ as:
\[ \begin{array}{rclcrcl}
A
& = &
\begin{pmatrix}
A_{11} & \cdots & A_{1n}
\\
\vdots & & \vdots
\\
A_{m1} & \cdots & A_{mn}
\end{pmatrix}
& \qquad\mbox{and}\qquad &
A^{T}
& = &
\begin{pmatrix}
A_{11} & \cdots & A_{m1}
\\
\vdots & & \vdots
\\
A_{1n} & \cdots & A_{mn}
\end{pmatrix}
\end{array} \]

\noindent They yield linear functions $A \colon \R^{n} \rightarrow
\R^{m}$ and $A^{T} \colon \R^{m} \rightarrow \R^{n}$. Explicitly, for
$v\in\R^{n}$ and $w\in \R^{m}$,
\[ \begin{array}{rclcrcccl}
A\cdot v
& = &
\begin{pmatrix}
\sum_{j} A_{1j}\cdot v_{j}
\\
\vdots
\\
\sum_{j} A_{mj}\cdot v_{j}
\end{pmatrix}
& \qquad &
A^{T}\cdot w
& = &
w^{T}\cdot A
& = &
\Big(\sum_{i} w_{i}\cdot A_{i1}, \, \ldots \, , \sum_{i} w_{i}\cdot A_{in}\Big).
\end{array} \]
}


\begin{lem}[Farkas Lemma]
\label{FarkasLem}
For a matrix $A\in\R^{m\times n}$ and a vector $b\in\R^{m}$ there
are equivalences of the following form.
\begin{enumerate}
\item \label{FarkasLemEq} $\exin{v}{\big(\nnR\big)^{n}}{A\cdot v = b}
  \;\Longleftrightarrow\; \neg\exin{u}{\R^{m}}{u^{T}\cdot A \geq \zero
    \mbox{ and } u^{T}\cdot b < 0}$.

\item \label{FarkasLemLeqPos} $\exin{v}{\big(\nnR\big)^{n}}{A\cdot v \leq b}
  \;\Longleftrightarrow\; \neg\exin{u}{\big(\nnR\big)^{m}}{u^{T}\cdot A
    \geq \zero \mbox{ and } u^{T}\cdot b < 0}$.

\end{enumerate}




\end{lem}

\begin{myproof}
A (short) proof of the first item may be found
in~\cite{Kager24}. Next, \cite[Prop.~6.4.3]{MatouekG06} contains a
derivation of the second item from the first. \QED

\auxproof{
One extends the matrix $A\in\R^{m\times n}$ to $A' = A\mid I \in
\R^{m\times (n+m)}$ by appending the identity matrix $I\in\R^{m\times
  m}$. For the direction $(\Rightarrow)$ in
item~\eqref{FarkasLemLeqPos}, let $A\cdot v \leq b$ for
$v\in\big(\nnR\big)^{n}$ and also, towards a contradiction, suppose we
have $u\in \big(\nnR\big)^{m}$ satisfying $u^{T}\cdot A \geq \zero
\mbox{ and } u^{T}\cdot b < 0$. From $A\cdot v \leq b$ we deduce that
there is a vector $w\in\big(\nnR\big)^{m}$ with $A\cdot v + w = b$ and
thus $A'\cdot(v,w) = b$. Using implication $(\Rightarrow)$
in~\eqref{FarkasLemEq} we know that there is no vector $x\in\R^{m}$
with $x^{T}\cdot A' \geq \zero$ and $x^{T}\cdot b < 0$. But we do have
$u\in \big(\nnR\big)^{m}$ with $u^{T}\cdot A \geq \zero$ and
$u^{T}\cdot b < 0$. The inequality $u^{T}\cdot A \geq \zero$ yields
$u^{T} \cdot A' \geq \zero$, since $u\geq\zero$. Hence taking $x = u$
gives a contradiction.

For $(\Leftarrow)$ in item~\eqref{FarkasLemLeqPos} we use the same
matrix $A'=A\mid I$ and prove the contrapositive. If there is no
$v\in\big(\nnR\big)^{n}$ with $A\cdot v \leq b$, then there are no
$v\in\big(\nnR\big)^{n}$, $w\in\big(\nnR\big)^{m}$ with $A'\cdot (v,w)
= b$. Item~\eqref{FarkasLemEq} then gives a vector $u\in\R^{m}$ with
$u\cdot A' \geq \zero$ and $u^{T}\cdot b < 0$. From $u\cdot A' \geq
\zero$ we obtain $u\cdot A \geq \zero$ and $u \geq \zero$. \QED
}
\end{myproof}

\auxproof{
\begin{myproof}
For the (only if)-part, assume $\vec{\lambda} \in \R^{m}$ as
indicated, whereas, towards a contradiction, that
$\mathsl{FS}\neq\emptyset$, say with $\vec{x}\in\mathsl{FS}$. Then,
since the $x_{j} \geq 0$,
\[ \begin{array}{rcl}
{\displaystyle\sum}_{i,j} \, \lambda_{i}\cdot A_{ij}\cdot x_{j}
& \leq &
0.
\end{array} \]

\noindent But also:
\[ \begin{array}{rcccccl}
{\displaystyle\sum}_{i,j} \, \lambda_{i}\cdot A_{ij}\cdot x_{j}
& = &
{\displaystyle\sum}_{i}\, \lambda_{i} \cdot
   {\displaystyle\sum}_{j}\, A_{ij}\cdot x_{j}
& \geq &
{\displaystyle\sum}_{i}\, \lambda_{i} \cdot b_{i}
& > &
0.
\end{array} \]

\noindent This is impossible. Hence $\mathsl{FS}=\emptyset$.

For the (if)-part assume $\mathsl{FS} = \emptyset$. We consider the
following cone in $\R^{m}$.
\[ \begin{array}{rcccl}
C
& \coloneqq &
\downset\set{A\cdot\vec{x}}{\vec{x}\in\big(\nnR\big)^{n}}
& = &
\setin{\vec{z}}{\R^m}{\exin{\vec{x}}{\big(\nnR\big)^{n}}{
   \vec{z} \leq A\cdot\vec{x}}}.
\end{array} \]

\noindent It is convex and closed. By assumption $b\not\in C$. By the
hyperplane separation theorem there is a non-zero vector
$\vec{\lambda}\in\R^{m}$ and a number $r\in\R$ such that:
\[ \begin{array}{rclcrcl}
{\displaystyle\sum}_{i} \, \lambda_{i}\cdot y_{i}
& \leq &
r, \quad\mbox{for all $\vec{y}\in C$}
& \qquad\mbox{and}\qquad &
{\displaystyle\sum}_{i} \, \lambda_{i}\cdot b_{i}
& > &
r.
\end{array} \]

\noindent By taking $\zero\in C$ we get $0\leq r$. We show that we may
as well take $r=0$. Let $r > 0$; we show that $\sum_{i}
\lambda_{i}\cdot y_{i} \leq 0$ for all $\vec{y}\in C$. If not, say
$\vec{z} \in C$ satisfies $s\coloneqq \sum_{i} \lambda_{i}\cdot z_{i}
> 0$. Take a scalar $t > \frac{r}{s} > 0$. Then $t\cdot\vec{z} \in C$,
and:
\[ \begin{array}{rcccccccl}
{\displaystyle\sum}_{i}\, \lambda_{i} \cdot (t\cdot \vec{z})_{i}
& = &
{\displaystyle\sum}_{i}\, \lambda_{i} \cdot t\cdot z_{i}
& = &
t\cdot {\displaystyle\sum}_{i}\, \lambda_{i} \cdot z_{i}
& = &
t \cdot s
& > &
r.
\end{array} \]

\noindent This is impossible. When $r > 0$ and $\sum_{i}
\lambda_{i}\cdot b_{i} > r$, then certainly $\sum_{i} \lambda_{i}\cdot
b_{i} > 0$. Thus we are done. \QED
\end{myproof}
}
\vspace{0.5\baselineskip}
\noindent 
\vspace{0.25\baselineskip}
We are going to apply this lemma in the Kantorovich context, in order
to prove the $(\leq)$ part of $\smash{\stackrel{(*)}{=}}$
in~\eqref{DstMetricMongeEqn}. Let $\omega, \omega'\in\Dst(X)$ be
given, where $X = \{x_{1}, \ldots, x_{n}\}$. We take $m = 2n$ in
Lemma~\ref{FarkasLem} with vector $b \in \R^{2n}$ given by the
sequence of probabilities:
\[ \begin{array}{rcl}
b
& \coloneqq &
\big(\omega(x_{1}), \ldots, \omega(x_{n}), \omega'(x_{1}), \ldots, 
   \omega'(x_{n})\big).
\end{array} \]

\noindent For convenience we write $t$ for the minimum validity
$\bigwedge_{\tau\in\decouple^{-1}(\omega, \omega')} \tau \models
d_{X}$. 

We use the matrix $A \in \R^{2n\times n^{2}}$ that captures
marginalisation, when applied to distributions, giving a mapping
$\Dst(X\times X) \rightarrow \Dst(X)\times\Dst(X)$.
\[ \begin{array}{rcl}
A
& \coloneqq &
\underbrace{\left(\begin{array}{ccccccccccccc}
1 & 1 & \cdots & 1 & 0 & 0 & \cdots & 0 & \cdots \cdots & 0 & 0 & \cdots & 0
\\[-0.5em]
0 & 0 & \cdots & 0 & 1 & 1 & \cdots & 1 & \cdots \cdots & 0 & 0 & \cdots & 0
\\[-0.5em]
& & \vdots & & & & \vdots & & & & & \vdots &
\\[-0.5em]
0 & 0 & \cdots & 0 & 0 & 0 & \cdots & 0 & \cdots \cdots & 1 & 1 & \cdots & 1
\\[-0.5em]
1 & 0 & \cdots & 0 & 1 & 0 & \cdots & 0 & \cdots \cdots & 1 & 0 & \cdots & 0 
\\[-0.5em]
0 & 1 & \cdots & 0 & 0 & 1 & \cdots & 0 & \cdots \cdots & 0 & 1 & \cdots & 0 
\\[-0.5em]
& & \vdots & & & & \vdots & & & & & \vdots &
\\[-0.5em]
0 & 0 & \cdots & 1 & 0 & 0 & \cdots & 1 & \cdots \cdots & 0 & 0 & \cdots & 1
\end{array}\right)}_{\text{$n^2$ columns}}
\begin{array}{c}
{\left.\begin{array}{c}
\mbox{}
\\[+1.5em]
\mbox{}
\end{array}\right\}\text{\scriptsize $n$ rows}}
\\
{\left.\begin{array}{c}
\mbox{}
\\[+1.5em]
\mbox{}
\end{array}\right\}\text{\scriptsize $n$ rows}}
\end{array}
\end{array} \]



\noindent We turn the matrix $A\in\R^{2n\times n^{2}}$ into a matrix
$A'\in \R^{(4n+1)\times n^{2}}$ in the following manner.
\begin{itemize}
\item On top of $A$ we add the row $d_{X}(x_{1}, x_{1}), d(x_{1},
  x_{2}), \ldots, d_{X}(x_{n}, x_{n})$ containing the $n^2$ distance
  values.

\item Below $A$ we add the matrix $-A$, with all numbers in $A$
  negated.
\end{itemize}

\newcommand{\tightdots}{...\ } 

\noindent For each $\varepsilon\!>\!0$ we form a similar sequence
$b_{\varepsilon}\!\in\!\R^{4n+1}$ of the form $(t-\varepsilon,
b_{1}, \tightdots, b_{2n}, -b_{1}, \tightdots,$$ -b_{2n})$. In this way the
following statement holds.
\[ \neg\exin{v}{\big(\nnR\big)^{n^2}}{
   A'\cdot v \leq b_{\varepsilon}}. \]

\noindent Indeed, $v = (v_{1}, \ldots, v_{n^2})$ with $A'\cdot v \leq
b_{\varepsilon}$ would give a multiset $\sigma\in\Mlt(X\times X)$,
namely $\sigma = \sum_{i,j} v_{(i-1)*n+j}\ket{x_{i}, x_{j}}$. It is a
coupling of $\omega,\omega'$ since $A\cdot v = b = (\omega, \omega')$
via the inequalities $A\cdot v \leq b$ and $-A\cdot v \leq
-b$. Therefore, $\sigma$ is a distribution. Moreover, it satisfies
$\sigma\models d_{X} \leq t - \varepsilon$. The latter contradicts the
minimality of $t$.

By applying item~\eqref{FarkasLemLeqPos} of Farkas
Lemma~\ref{FarkasLem} we get a vector $u\in\R^{4n+1}$ with $u \geq
\zero$ satisfying:
\begin{equation}
\label{FarkasLambdaEqn}
\begin{array}{rcl}
\displaystyle u_{0}\cdot d_{X} +
   (u_{1},\ldots,u_{2n})\cdot A -
      (u_{2n+1},\ldots,u_{4n})\cdot A
& \geq &
\zero
\\[+0.2em]
\displaystyle u_{0}\cdot (t-\varepsilon) +
   (u_{1},\ldots,u_{2n})\cdot b -
      (u_{2n+1},\ldots,u_{4n})\cdot b
& < &
0.
\end{array}
\end{equation}

\noindent We first cover the case $u_{0} = 0$. The vector $u$ is then
a solution in~\eqref{FarkasLambdaEqn} for $\varepsilon = t$, so that
applying the direction $(\Leftarrow)$ in
Lemma~\ref{FarkasLem}~\eqref{FarkasLemLeqPos} gives a coupling
$\sigma$ with $\sigma\models d_{X} = 0$. This gives $t = 0$.
But then we are done, since trivially:
\[ \begin{array}{rcl}
\displaystyle\bigvee_{p,p'\in \Obs(X)\text{ with } p\oplus p' \leq d_{X}} 
   \omega\models p \,+\, \omega'\models p'
& \geq &
\omega\models \zero \,+\, \omega'\models \zero
\\[-0.5em]
& = &
0
\hspace*{\arraycolsep}=\hspace*{\arraycolsep}
t
\hspace*{\arraycolsep}=\hspace*{\arraycolsep}
\displaystyle\bigwedge_{\tau\in\decouple^{-1}(\omega, \omega')} \tau \models d_{X}.
\end{array} \]

We may now assume $u_{0} > 0$, so we can form two observations
$p,p'\in\Obs(X)$ via the definitions:
\[ \begin{array}{rclcrcl}
p(x_{i})
& \coloneqq &
\displaystyle\frac{u_{2n+i} - u_{i}}{u_0}
& \qquad\mbox{and}\qquad &
p'(x_{i})
& \coloneqq &
\displaystyle\frac{u_{3n+i} - u_{n+i}}{u_0}.
\end{array} \]

\noindent The first of the inequalities~\eqref{FarkasLambdaEqn}
translates to $p(x_{i}) + p'(x_{j}) \leq d_{X}(x_{i}, x_{j})$ for each
$i,j$, that is, to $p\oplus p' \leq d_{X}$. The second inequality
gives us $\omega\models p + \omega'\models p' > t -
\varepsilon$. We thus get:
\[ \begin{array}{rcl}
\displaystyle\bigvee_{p,p'\in \Obs(X)\text{ with } p\oplus p' \leq d_{X}} 
   \omega\models p \,+\, \omega'\models p'
& > &
t - \varepsilon.
\end{array} \]

\noindent Since this holds for each $\varepsilon > 0$ we obtain:
\[ \begin{array}{rcccl}
\displaystyle\bigvee_{p,p'\in \Obs(X)\text{ with } p\oplus p' \leq d_{X}} 
   \omega\models p \,+\, \omega'\models p'
& \geq &
t
& = &
\displaystyle\bigwedge_{\tau\in\decouple^{-1}(\omega, \omega')} \tau \models d_{X}.
\end{array} \]

\noindent This proves the $(\leq)$ part of $\smash{\stackrel{(*)}{=}}$
in~\eqref{DstMetricMongeEqn}.

We now turn to the second equation $\smash{\stackrel{(**)}{=}}$
in~\eqref{DstMetricMongeEqn}, again starting from two distributions
$\omega,\omega'\in\Dst(X)$. Its proof is more elementary.  For
$(\geq)$, let $q\in\FactSh(X)$ be given. Without loss of generality we
assume $\omega\models q \,\leq\, \omega'\models q$. We take as
observations $p = -q$ and $p'=q$. Then, because $q$ is short:
\[ \begin{array}{rcccccccl}
(p \oplus p')(x,x')
& = &
p(x) + p'(x')
& = &
q(x') - q(x)
& \leq &
\big| \, q(x') - q(x) \, \big|
& \leq &
d_{X}(x,x').
\end{array} \]

\noindent Moreover,
\[ \begin{array}{rcl}
\big|\, \omega\models q \,-\, \omega'\models q \,\big|
& = &
\omega' \models q - \omega\models q
\\
& = &
\omega\models p \,+\, \omega'\models p' 
\\
& \leq &
\displaystyle\bigvee_{p,p'\in \Obs(X)\text{ with } p\oplus p' \leq d_{X}} 
   \omega\models p \,+\, \omega'\models p'.
\end{array} \]

\noindent Since this holds for all $q\in\FactSh(X)$ we get:
\[ \begin{array}{rcl}
\displaystyle\bigvee_{q\in \FactSh(X)} \big|\, \omega\models q \,-\,
   \omega'\models q \,\big|
& \,\leq\, &
\displaystyle\bigvee_{p,p'\in \Obs(X)\text{ with } p\oplus p' \leq d_{X}} 
   \omega\models p \,+\, \omega'\models p'.
\end{array} \]


For $(\leq)$ let $p,p'\in\Obs(X)$ be given with $p\oplus p' \leq
d_{X}$. We form $q\colon X \rightarrow \R$ as:
\[ \begin{array}{rcl}
q(x)
& \coloneqq &
\displaystyle\bigwedge_{y\in X} \, d_{X}(x,y) - p(y)
\end{array} \]

\noindent We make a few points explicit.
\begin{itemize}
\item By using $x$ in place of $y$ in the range of the above meet
  $\bigwedge$ we get $q(x) \leq -p(x)$. Further, since $p(x) + p'(x')
  \leq d_{X}(x,x')$ we have $p'(x') \leq d_{X}(x,x') - p(x) \leq
  q(x')$. Thus, $p\leq -q$ and $p'\leq q$.

\item Next, $q$ is short. Indeed, for $x,x' \in X$, and
arbitrary $y\in X$ we have, by the triangular inequality,
\[ \begin{array}{rcccl}
q(x)
& \leq &
d_{X}(x,y) - p(y)
& \leq &
d_{X}(x,x') + d_{X}(x',y) - p(y).
\end{array} \]

\noindent Since this holds for each $y$ we get:
\[ \begin{array}{rcl}
q(x)
& \leq &
\displaystyle\bigwedge_{y\in X} \, d_{X}(x,x') + d_{X}(x',y) - p(y)
\\[+1em]
& = &
d_{X}(x,x') + \displaystyle\bigwedge_{y\in X} \, d_{X}(x',y) - p(y)
\hspace*{\arraycolsep}=\hspace*{\arraycolsep}
d_{X}(x,x') + q(x').
\end{array} \]

\noindent Thus: $q(x) - q(x') \leq d_{X}(x,x')$. By symmetry also:
$q(x') - q(x) \leq d_{X}(x,x')$, and thus $\big|\,q(x) - q(x')\,\big|
= \max\big(q(x) - q(x'),q(x') - q(x)\big) \leq d_{X}(x,x')$.

\item Let $M$ be the minimum value of $q$, which exists, since we
  assume that $X$ is a finite set. We take $q' = q + M\cdot
  \one$. This is a factor, that is, a function $X \rightarrow \nnR$
  taking non-negative values. Moreover, $q'$ is short since $q$ is
  short.
\end{itemize}

\noindent We can now put things together. 
\[ \begin{array}{rcll}
\omega\models p \,+\, \omega'\models p'
& \leq &
\omega\models -q \,+\, \omega'\models q
   & \mbox{since $p\leq -q$ and $p'\leq q$}
\\
& \leq &
\big|\, \omega\models q \,-\, \omega'\models q \,\big|
\\[+0.2em]
& = &
\big|\, \omega\models q' \,-\, \omega'\models q' \,\big|
   & \mbox{since }q' = q + M\cdot \one
\\
& \leq &
\displaystyle\bigvee_{q\in \FactSh(X)} \big|\, \omega\models q \,-\,
   \omega'\models q \,\big|. \quad
\end{array} \]

\noindent Because this holds for all $p,p'$ we get, as required,
the inequality $(\leq)$ for $\smash{\stackrel{(**)}{=}}$
in~\eqref{DstMetricMongeEqn}:
\[ \begin{array}{rcl}
\displaystyle\bigvee_{p,p'\in \Obs(X)\text{ with } p\oplus p' \leq d_{X}} 
   \omega\models p \,+\, \omega'\models p'
& \,\leq\, &
\displaystyle\bigvee_{q\in \FactSh(X)} \big|\, \omega\models q \,-\,
   \omega'\models q \,\big|.
\end{array} \]

\auxproof{
\subsection{Proof in the other direction}


Consider the sum $\oplus = A^{T}$ of observables that we used before
as a (linear) function $\oplus \colon \R^{2n} \rightarrow
\R^{n^2}$. Let $\omega,\omega'\in\Dst(X)$ have Kantorovich
distance $t \coloneqq d(\omega,\omega') \in \nnR$. We consider the
linear map $B\in\R^{(1+n^{2})\times 2n}$ in:
\[ \xymatrix@R-2.2pc{
\R^{2n}\ar[rr]^-{B} & & \R \times \R^{n^2}
\\
v\ar@{|->}[rr] & & 
   \big(-v^{T}\cdot b, \, A^{T}\cdot v\big)
} \]

\noindent Thus, the matrix $B$ has $b$ on top of $A^{T}$.

We claim that:
\[ \exin{v}{\big(\nnR\big)^{2n}}{B\cdot v \leq (-t,d_{X})}. \]

\noindent If not, then by Farkas Lemma item~\eqref{FarkasLemLeqPos}
there is a pair $(r, u) \in \R\times\R^{n^2}$ with
$(r,u)\geq(0,\zero)$ such that $(r,u)^{T}\cdot B \geq 0$ and
$(r,u)^{T}\cdot (-t,d_{X}) < 0$. The former means that
$\marg{u}{1,0}(x_{i}) \geq r\cdot \omega(x_{i})$ and
$\marg{u}{0,1}(x_{i}) \geq r\cdot \omega'(x_{i})$. The latter means $u
\models d_{X} < r\cdot t$.

We distinguish two (exclusive) cases and show that are both impossible.
\begin{itemize}
\item Let $r=0$. Then $\marg{R}{1,0} = \zero = \marg{R}{0,1}$ so that
  $R = \zero$. But then we cannot have $R \models d_{X} < 0$.

\auxproof{
We assume an optimal coupling $\tau\in\Dst(X\times X)$
and consider $\tau' = \tau + R \in \Mlt(X)$. This is a coupling of
$\omega,\omega'$ and satisfies:
\[ \begin{array}{rcccccl}
\tau + R \models d_{X}
& = &
\tau \models d_{X} + R \models d_{X}
& < &
w + r\cdot w
& = &
w.
\end{array} \]

\noindent This is impossible, because $w$ is the least value.
}

\item We thus have $r > 0$, so that we can form $R' = \frac{1}{r}\cdot
  R \in \R^{n^2}$, still with $R' \geq \zero$. Moreover,
  $\marg{R'}{1,0} = \omega$ and $\marg{R'}{0,1} = \omega'$, giving:
\[ \begin{array}{rcccccl}
R' \models d_{X}
& = &
\frac{1}{r} \cdot \big(R \models d_{X}\big)
& < &
\frac{1}{r} \cdot r\cdot w
& = &
w.
\end{array} \]

\noindent This is impossible, because $w$ is the least value.
\end{itemize}
}

\section{Proof of the equations in~\eqref{MltMetricEqn}}\label{MltApp}

We start with a basic result about coupling and
decoupling~\eqref{MltDecoupleDiag} of multisets.

\begin{lem}
\label{DecoupleLem}
Consider sets $X,Y$ and a number $K\in\NNO$.
\begin{enumerate}
\item \label{DecoupleLemTriangle} There is a commuting triangle of the
  form:
\begin{equation}
\label{DecoupleAccDiag}
\vcenter{\xymatrix@R-0.8pc@C-1pc{
X^{K}\times Y^{K}\ar@/_3ex/@{->>}[dr]_-{\acc\times\acc}
   \ar[r]^-{\zip}_-{\cong} &
   (X\times Y)^{K}\ar@{->>}[r]^-{\acc} & 
   \Mlt[K](X\times\rlap{$Y)$}\ar@/^3ex/[dl]^-{\decouple}
\\
& \Mlt[K](X)\times\Mlt[K](Y)
}}
\end{equation}

\noindent As a result, the decouple function $\decouple =
\tuple{\Mlt(\pi_{1}), \Mlt(\pi_{2})} \colon \Mlt[K](X\times
Y)\rightarrow \Mlt[K](X) \times \Mlt[K](Y)$ is surjective.

\item \label{DecoupleLemInv} There is the following equality of
  subsets of $\Mlt[K](X\times Y)$, for multisets
  $\varphi\in\Mlt[K](X)$ and $\psi\in\Mlt[K](Y)$.
\[ \begin{array}{rcl}
\decouple^{-1}(\varphi,\psi)
& = &
\set{\acc\big(\zip(\vec{x},\vec{y})\big)}{\vec{x}\in\acc^{-1}(\varphi),
   \vec{y}\in\acc^{-1}(\psi)\,}.
\end{array} \]

\noindent In particular, the subset $\decouple^{-1}(\varphi,\psi)
\subseteq \Mlt[K](X\times Y)$ is finite.

\item If $\tau = \acc\big(\zip(\vec{x}, \vec{y})\big) \in
  \Mlt[K](X\times X)$, where $X$ is a metric space, then:
\[ \begin{array}{rcl}
\flrn(\tau) \models d_{X}
& \,=\, &
\frac{1}{K}\cdot d_{X^{K}}\big(\vec{x}, \vec{y}\big).
\end{array} \]

\end{enumerate}
\end{lem}

\begin{myproof}
\begin{enumerate}
\item Via the naturality of $\acc \colon (-)^{K} \Rightarrow \Mlt[K]$,
  we get for $\vec{x}\in\acc^{-1}(\varphi)$ and
  $\vec{y}\in\acc^{-1}(\psi)$,
\[ \begin{array}{rcl}
\decouple\Big(\acc\big(\zip(\vec{x},\vec{y})\big)\Big)
& = &
\tuple{\, \Mlt(\pi_{1})\Big(\acc\big(\zip(\vec{x},\vec{y})\big)\Big), \;
   \Mlt(\pi_{2})\Big(\acc\big(\zip(\vec{x},\vec{y})\big)\Big) \,}
\\
& = &
\tuple{\, \acc\Big((\pi_{1})^{K}\big(\zip(\vec{x},\vec{y})\big)\Big), \;
   \acc\Big((\pi_{2})^{K}\big(\zip(\vec{x},\vec{y})\big)\Big) \,}
\\
& = &
\tuple{ \, \acc\big(\pi_{1}(\vec{x},\vec{y})\big), \,
   \acc\big(\pi_{2}(\vec{x},\vec{y})\big) \,}
\\
& = &
\tuple{\, \acc(\vec{x}), \, \acc(\vec{y}) \,}
\\
& = &
\tuple{\, \varphi, \, \psi \,}.
\end{array} \]

\noindent This allows us to show that $\decouple \colon
\Mlt[K](X\times Y)\rightarrow \Mlt[K](X) \times \Mlt[K](Y)$ is a
surjective function: let a pair
$\tuple{\varphi,\psi}\in\Mlt[K](X)\times\Mlt[K](Y)$ be given. Using
the surjectivity of accumulation we can find $\vec{x}\in X^{K}$ and
$\vec{y}\in Y^{K}$ with $\acc(\vec{x}) = \varphi$ and $\acc(\vec{y}) =
\psi$. The above argument shows that $\chi \coloneqq
\acc\big(\zip(\vec{x},\vec{y})\big) \in \Mlt[K](X\times Y)$ satisfies
$\decouple(\chi) = \tuple{\acc(\vec{x}), \acc(\vec{y})} =
\tuple{\varphi,\psi}$.

\item The previous point gives the inclusion $(\supseteq)$. For
  $(\subseteq)$, let $\chi\in\decouple^{-1}(\varphi,\psi)$ be given.
  There is a sequence $\vec{z}\in (X\times Y)^{K}$ with $\acc(\vec{z})
  = \chi$. Write $\vec{z} = \zip(\vec{x}, \vec{y})$, for $\vec{x} =
  (\pi_{1})^{K}(\vec{z})$ and $\vec{y} = (\pi_{2})^{K}(\vec{z})$.  We
  claim that $\acc(\vec{x}) = \varphi$ and $\acc(\vec{y}) = \psi$. We
  elaborate only the first equation, since the second one is obtained
  in the same way. By assumption, $\chi$ is coupling of $\varphi,
  \psi$, so:
\[ \begin{array}{rcccccl}
\varphi
& = &
\Mlt(\pi_{1})(\chi)
& = &
\Mlt(\pi_{1})\big(\acc(\zip(\vec{x}, \vec{y}))\big)
& = &
\acc(\vec{x}).
\end{array} \]

\noindent The last equation is obtained as in the previous point.

\item Let $\tau = \acc\big(\zip(\vec{x}, \vec{y})\big) \in
  \Mlt[K](X\times X)$, for $\vec{x}, \vec{y} \in X^{K}$. Then:
\[ \begin{array}[b]{rcl}
\flrn(\tau) \models d_{X}
& = &
\frac{1}{K} \cdot \displaystyle \sum_{(u,v)\in X^{2}} \,
   \acc\big(\zip(\vec{x}, \vec{y})\big)(u,v) \cdot d_{X}(u,v)
\\
& = &
\frac{1}{K}\cdot\displaystyle \sum_{1\leq i\leq K} \, d_{X}(x_{i}, y_{i})
\;\hspace*{\arraycolsep}\smash{\stackrel{\eqref{ProductMetricEqn}}{=}}\hspace*{\arraycolsep}\;
\textstyle\frac{1}{K}\cdot d_{X^{K}}\big(\vec{x}, \vec{y}\big).
\end{array} \eqno{\QEDbox} \]

\end{enumerate}
\end{myproof}

\begin{thm}
Equation~\eqref{MltMetricEqn} holds, that is, for a metric space $X$
with multisets $\varphi,\varphi'\in\Mlt[K](X)$,
\[ \begin{array}{rcl}
\displaystyle\bigwedge_{\tau\in\decouple^{-1}(\varphi, \varphi')}
   \flrn(\tau) \models d_{X}
& \,=\, &
\frac{1}{K}\cdot\displaystyle\bigwedge_{\vec{x}\in\acc^{-1}(\varphi), \,
   \vec{y}\in\acc^{-1}(\varphi')} d_{X^{K}}\big(\vec{x}, \vec{y}\big)
\end{array} \]
\end{thm}

\begin{myproof}
By Lemma~\ref{DecoupleLem}. For $(\leq)$ we notice that for each pair
of vectors $\vec{x}\in\acc^{-1}(\varphi),
\vec{y}\in\acc^{-1}(\varphi')$ we have that $\tau =
\acc\big(\zip(\vec{x},\vec{y})\big)$ is a coupling of
$\varphi,\varphi'$ with $\flrn(\tau) \models d_{X} \,=\,
\frac{1}{K}\cdot d_{X^{K}}(\vec{x}, \vec{y})$. For $(\geq)$ we notice
that each coupling $\tau$ is of the form $\tau =
\acc\big(\zip(\vec{x},\vec{y})\big)$, for some vectors
$\vec{x}\in\acc^{-1}(\varphi), \vec{y}\in\acc^{-1}(\varphi')$. Then
again, $\flrn(\tau) \models d_{X} \,=\, \frac{1}{K}\cdot
d_{X^{K}}(\vec{x}, \vec{y})$.
\end{myproof}\QED

\end{document}